\documentclass[twocolumn]{aastex631}

\usepackage{amsmath}
\usepackage{graphicx}
\usepackage{booktabs}
\usepackage{bm}
\usepackage{xspace}
\usepackage{color}
\usepackage{hyperref}

\newcommand{\lya}{Ly\ensuremath{\alpha}\xspace}

\newcommand{\hi}{H~{\sc i}\xspace}

\newcommand{\oiii}{[O~{\sc iii}]\xspace}

\newcommand{\ha}{H\ensuremath{\alpha}\xspace}
\newcommand{\hb}{H\ensuremath{\beta}\xspace}

\newcommand{\kms}{\ensuremath{\mathrm{km~s^{-1}}}\xspace}

\newcommand{\msun}{\ensuremath{M_\odot}\xspace}

\newcommand{\zspec}{\ensuremath{z_{\rm spec}}\xspace}
\newcommand{\zphot}{\ensuremath{z_{\rm phot}}\xspace}

\newcommand{\mstar}{\ensuremath{M_\star}\xspace}

\newcommand{\sfr}{\ensuremath{\mathrm{SFR}}\xspace}

\newcommand{\ebv}{\ensuremath{E(B\!-\!V)}\xspace}

\newcommand{\spitzer}{{\it Spitzer}\xspace}
\newcommand{\herschel}{{\it Herschel}\xspace}
\newcommand{\wise}{{\it WISE}\xspace}
\newcommand{\galex}{{\it GALEX}\xspace}

\providecommand{\okina}{\textquoteleft}

\shorttitle{CHEF I. Multiwavelength Master Catalog}
\shortauthors{Liang et al.}

\graphicspath{{./}{figures/}}

\newcommand{\unitcolhead}[1]{\multicolumn{1}{c}{\vrule depth 4pt height 8pt width 0pt\relax#1}\ignorespaces}
\newcommand{\unittwocolhead}[1]{\multicolumn{2}{c}{\vrule depth 4pt height 8pt width 0pt\relax#1}\ignorespaces}

\begin{document}

\title{The \textit{Cosmic Himalayas} Extended Field (CHEF). I.\ A Multiwavelength Master Catalog Centered on an Extreme Quasar Concentration and Its Extended Environment}

\correspondingauthor{Yongming Liang}
\email{ymliang@icrr.u-tokyo.ac.jp}

\author[0000-0002-2725-302X]{Yongming Liang}
\affiliation{Institute for Cosmic Ray Research, The University of Tokyo, 5-1-5 Kashiwanoha, Kashiwa, Chiba 277-8582, Japan}
\affiliation{National Astronomical Observatory of Japan, 2-21-1 Osawa, Mitaka, Tokyo 181-8588, Japan}

\author[0000-0002-1049-6658]{Masami Ouchi}
\affiliation{National Astronomical Observatory of Japan, 2-21-1 Osawa, Mitaka, Tokyo 181-8588, Japan}
\affiliation{Institute for Cosmic Ray Research, The University of Tokyo, 5-1-5 Kashiwanoha, Kashiwa, Chiba 277-8582, Japan}

\author[0000-0001-5804-1428]{Sebastiano Cantalupo}
\affiliation{Department of Physics, University of Milan Bicocca, Piazza della Scienza 3, I-20126 Milano, Italy}
\affiliation{INAF--Osservatorio Astronomico di Trieste, Via G. B. Tiepolo 11, 34143 Trieste, Italy}

\author[0000-0003-3310-0131]{Xiaohui Fan}
\affiliation{Steward Observatory, University of Arizona, 933 N. Cherry Avenue, Tucson, AZ 85721, USA}

\author[0000-0001-8467-6478]{Zheng Cai}
\affiliation{Department of Astronomy, Tsinghua University, Beijing 100084, China}

\author[0000-0001-6251-649X]{Mingyu Li}
\affiliation{Department of Astronomy, Tsinghua University, Beijing 100084, China}

\author[0000-0003-3954-4219]{Nobunari Kashikawa}
\affiliation{Department of Astronomy, School of Science, The University of Tokyo, 7-3-1 Hongo, Bunkyo-ku, Tokyo, 113-0033, Japan}

\author[0000-0002-5011-5178]{Masayuki Tanaka}
\affiliation{National Astronomical Observatory of Japan, 2-21-1 Osawa, Mitaka, Tokyo 181-8588, Japan}

\author[0000-0002-1319-3433]{Hidenobu Yajima}
\affiliation{Center for Computational Sciences, University of Tsukuba, 1-1-1 Tennodai, Tsukuba, Ibaraki 305-8577, Japan}

\author[0000-0001-7457-8487]{Kentaro Nagamine}
\affiliation{Department of Earth and Space Science, Osaka University, 1-1 Machikaneyama, Toyonaka, Osaka 560-0043, Japan}

\author[0009-0005-1273-0742]{Yuto Kuwayama}
\affiliation{Department of Earth and Space Science, Osaka University, 1-1 Machikaneyama, Toyonaka, Osaka 560-0043, Japan}

\author[0000-0002-5712-6865]{Yuri Oku}
\affiliation{Department of Earth and Space Science, Osaka University, 1-1 Machikaneyama, Toyonaka, Osaka 560-0043, Japan}

\author[0000-0002-5045-6052]{Keita Fukushima}
\affiliation{Institute for Data Innovation in Science, Seoul National University, Seoul 08826, Republic of Korea}

\author[0000-0002-7464-7857]{Hyunbae Park}
\affiliation{Center for Theoretical Physics of the Universe, Institute for Basic Science (IBS), Republic of Korea}

\author[0000-0003-4442-2750]{Rhythm Shimakawa}
\affiliation{Waseda Institute for Advanced Study (WIAS), Waseda University, 1-21-1 Nishi-Waseda, Shinjuku, Tokyo 169-0051, Japan}

\author[0000-0002-2993-1576]{Tadayuki Kodama}
\affiliation{Astronomical Institute, Tohoku University, 6-3 Aramaki, Aoba-ku, Sendai, Miyagi 980-8578, Japan}

\author[0000-0001-9687-4973]{Naveen Reddy}
\affiliation{Department of Physics and Astronomy, University of California, Riverside, USA}

\author[0000-0002-4834-7260]{Charles Steidel}
\affiliation{Cahill Center for Astronomy and Astrophysics, California Institute of Technology, USA}

\author[0000-0002-9453-0381]{Kei Ito}
\affiliation{Cosmic Dawn Center (DAWN), Denmark}
\affiliation{DTU Space, Technical University of Denmark, Elektrovej 327, DK2800 Kgs. Lyngby, Denmark}

\author[0000-0002-6047-430X]{Yuichi Harikane}
\affiliation{Division of Physics and Astronomy, Graduate School of Science, Kyoto University, Japan}

\author[0009-0004-4332-9225]{Tomokazu Kiyota}
\affiliation{Astronomical Science Program, Graduate Institute for Advanced Studies, SOKENDAI, 2-21-1 Osawa, Mitaka, Tokyo 181-8588, Japan}
\affiliation{National Astronomical Observatory of Japan, 2-21-1 Osawa, Mitaka, Tokyo 181-8588, Japan}

\author[0000-0002-1199-6523]{Dongsheng Sun}
\affiliation{Center for Computational Sciences, University of Tsukuba, 1-1-1 Tennodai, Tsukuba, Ibaraki 305-8577, Japan}

\author[0000-0002-2597-2231]{Kazuhiro Shimasaku}
\affiliation{Department of Astronomy, School of Science, The University of Tokyo, 7-3-1 Hongo, Bunkyo-ku, Tokyo, 113-0033, Japan}

\author[0000-0003-2984-6803]{Masafusa Onoue}
\affiliation{Waseda Institute for Advanced Study (WIAS), Waseda University, 1-21-1 Nishi-Waseda, Shinjuku, Tokyo 169-0051, Japan}

\author[0000-0001-9452-0813]{Takuma Izumi}
\affiliation{National Astronomical Observatory of Japan, 2-21-1 Osawa, Mitaka, Tokyo 181-8588, Japan}

\author[0000-0002-8863-888X]{Andrea Travascio}
\affiliation{INAF--Osservatorio Astronomico di Trieste, Via G. B. Tiepolo 11, 34143 Trieste, Italy}
\affiliation{Department of Physics, University of Milan Bicocca, Piazza della Scienza 3, I-20126 Milano, Italy}

\author[0000-0001-5060-1398]{Martin Elvis}
\affiliation{Center for Astrophysics $|$ Harvard \& Smithsonian, 60 Garden St., Cambridge, MA 02138, USA}

\author[0000-0002-3554-3318]{Giuseppina Fabbiano}
\affiliation{Center for Astrophysics $|$ Harvard \& Smithsonian, 60 Garden St., Cambridge, MA 02138, USA}

\author[0000-0002-9593-8274]{Weichen Wang}
\affiliation{Chinese Academy of Sciences South America Center for Astronomy (CASSACA), NAOC, Beijing 100101, China}
\affiliation{Centro Conjunto China-Chile de Astronomía (CCJCA), Departamento de Astronomía, Universidad de Chile, Santiago, Chile}

\author[0000-0003-2273-9415]{Haibin Zhang}
\affiliation{National Astronomical Observatory of Japan, 2-21-1 Osawa, Mitaka, Tokyo 181-8588, Japan}

\author[0000-0001-9299-5719]{Khee-Gan Lee}
\affiliation{Kavli IPMU (WPI), The University of Tokyo, 5-1-5 Kashiwanoha, Kashiwa, Chiba 277-8583, Japan}

\author[0000-0003-0111-8249]{Yunjing Wu}
\affiliation{Kavli IPMU (WPI), The University of Tokyo, 5-1-5 Kashiwanoha, Kashiwa, Chiba 277-8583, Japan}

\author[0000-0002-7738-6875]{J. Xavier Prochaska}
\affiliation{UCO/Lick Observatory, University of California, 1156 High Street, Santa Cruz, CA 95064, USA}

\author[0000-0002-0000-6977]{John Silverman}
\affiliation{Kavli IPMU (WPI), The University of Tokyo, 5-1-5 Kashiwanoha, Kashiwa, Chiba 277-8583, Japan}

\author[0000-0001-7943-0166]{Fangxia An}
\affiliation{Yunnan Observatories, Chinese Academy of Sciences, Kunming 650011, People's Republic of China}

\author[0000-0002-0673-0632]{Hisakazu Uchiyama}
\affiliation{Department of Advanced Sciences, Faculty of Science and Engineering, Hosei University, Japan}
\affiliation{National Astronomical Observatory of Japan, 2-21-1 Osawa, Mitaka, Tokyo 181-8588, Japan}

\author[0009-0007-0864-7094]{Junya Arita}
\affiliation{Department of Astronomy, School of Science, The University of Tokyo, 7-3-1 Hongo, Bunkyo-ku, Tokyo, 113-0033, Japan}

\author[0000-0002-9373-3865]{Xin Wang}
\affiliation{School of Astronomy and Space Science, University of Chinese Academy of Sciences (UCAS), Beijing 100049, China}
\affiliation{National Astronomical Observatories, Chinese Academy of Sciences, Beijing 100101, China}
\affiliation{Institute for Frontiers in Astronomy and Astrophysics, Beijing Normal University, Beijing 102206, China}

\author[0000-0001-6503-8315]{Takanobu Kirihara}
\affiliation{Kitami Institute of Technology, 165 Koen-cho, Kitami, Hokkaido 090-8507, Japan}

\author[0000-0002-9888-6895]{Chenghao Zhu}
\affiliation{Institute for Cosmic Ray Research, The University of Tokyo, 5-1-5 Kashiwanoha, Kashiwa, Chiba 277-8582, Japan}

\author[0000-0002-9634-2923]{Zhenya Zheng}
\affiliation{Shanghai Astronomical Observatory, Chinese Academy of Sciences, China}

\author[0000-0001-7832-5372]{Benjamin Horowitz}
\affiliation{Kavli IPMU (WPI), The University of Tokyo, 5-1-5 Kashiwanoha, Kashiwa, Chiba 277-8583, Japan}

\author[0000-0002-4314-5686]{Dongdong Shi}
\affiliation{Center for Fundamental Physics, School of Mechanics and Optoelectronic Physics, Anhui University of Science and Technology, Huainan, Anhui 232001, China}

\author[0000-0002-4193-2539]{Denis Burgarella}
\affiliation{Aix Marseille Univ, CNRS, CNES, LAM, Marseille, France}

\author[0009-0003-7910-0265]{Hao Zhang}
\affiliation{Aix Marseille Univ, CNRS, CNES, LAM, Marseille, France}

\author[0000-0002-3134-9526]{Jiaqi Lin}
\affiliation{Shanghai Astronomical Observatory, Chinese Academy of Sciences, China}

\begin{abstract}
We present the multiwavelength master catalog of the \textit{Cosmic Himalayas} Extended Field (CHEF), about $1~\mathrm{deg}^2$ of deep imaging around the \textit{Cosmic Himalayas}.
This concentration of 11 quasars at $z=2.16$--$2.20$ is the densest quasar overdensity in the $\sim$10,000~deg$^2$ SDSS footprint.
Twenty-four bands, from \galex\ far-ultraviolet to JCMT/SCUBA-2 850~\micron, come from public surveys of SDSS Stripe~82 and our CFHT, Subaru, CTIO, and JCMT observations.
The \galex\ and $u$ bands bracket the Lyman break at $z\approx2.2$, and the $J$ and $K$ bands the Balmer break.
Three narrowbands isolate \lya\ and \ha\ at the structure redshift, and the 850~\micron\ map traces dust-obscured star formation.
Sources detected independently in optical, near-infrared, and narrowband images are cross-matched into $456{,}513$ objects with uniform photometry and measured completeness.
The optical catalog reaches $5\sigma$ depths of about 26~mag in $ugri$.
The catalog gives photometric redshifts, rest-frame colors, structure-membership classes, and, for $382{,}799$ objects, stellar masses and star-formation rates.
The photometric redshifts scatter by $3\%$ in $1+z$ for spectroscopic galaxies at $z<1.6$ and by $10\%$, with $22\%$ outliers, at $z\approx2$ for COSMOS2020 galaxies degraded to our depths.
A compilation of $6{,}237$ spectra gives secure redshifts for $4{,}102$ sources, including 24 \ha\ emitters newly confirmed with Keck/MOSFIRE.
Value-added samples list \ha\ and \lya\ emitters, color-selected star-forming and passive galaxies, submillimeter sources, and spectroscopic quasars.
Outside the structure, the field's galaxies, active nuclei, and dusty sources from low redshift to $z>3$ form a protocluster-field counterpart to blank-field surveys.
The survey page will host the products.\footnote{\url{https://cosmic-himalayas.github.io/chef/}}
\end{abstract}


\section{Introduction} \label{sec:intro}

At ``cosmic noon'' ($z\sim2$--$3$, near the peak of the cosmic star-formation history; \citealt{MadauDickinson2014}), the co-evolution of galaxies, supermassive black holes, and the intergalactic medium (IGM) depends on environment.
The morphology--density \citep{Dressler1980} and star-formation--density \citep{Gomez2003} relations established locally are expected to emerge, or even reverse, in the overdense progenitors of present-day clusters \citep{Overzier2016}.
Testing this expectation requires the stellar masses (\mstar), star-formation rates (\sfr), and dust attenuation of galaxies in and around such structures.
At these redshifts the rest-frame ultraviolet (UV), which traces unobscured star formation, is observed in the optical.
The Balmer/4000~\AA\ break and the rest-frame optical continuum, which constrain \mstar, are observed in the near-infrared, and the dust-reprocessed light is observed in the far-infrared and submillimeter.
Master catalogs that combine such imaging with photometric redshifts (\zphot) and spectral energy distribution (SED) fits exist for the deep extragalactic fields \citep{Skelton2014,Laigle2016,Weaver2022,Shuntov2025}.
Identifying the members of a structure at $z\sim2$ requires more than such catalogs.
Spectroscopic surveys are sparse at these redshifts, and broadband \zphot\ are too imprecise to isolate a structure.
The emission-line galaxies of a structure are therefore selected with narrowband filters tuned to its redshift \citep{Sobral2013,Shimakawa2018,Harikane2019,Kikuta2023,Umeda2025}.
Known structures at $z\approx2$--$2.5$ with deep multiwavelength photometry include the $z=2.095$ cluster in COSMOS found in a near-infrared medium-band catalog \citep{Spitler2012,Yuan2014} and the Hyperion proto-supercluster at $z=2.45$ \citep{Cucciati2018}.
The radio-galaxy fields PKS~1138$-$262 and USS~1558$-$003 have been mapped with \ha\ narrowbands \citep{Koyama2013,Hayashi2012}.
Quasar-selected structures at these redshifts include those of \citet{Hennawi2015} and \citet{Cai2017}.

The BOSSJ0210 field contains an extreme quasar-selected structure.
Within the Mapping the Most Massive Overdensity Through Hydrogen with Subaru (MAMMOTH-Subaru) project \citep{Cai2016,Cai2017}, \citet{Liang2021} imaged fields selected for coherently strong \lya\ absorption in SDSS quasar spectra.
One of these fields coincided with an extreme quasar concentration.
\citet{Liang2025} named this concentration of 11 quasars at $z=2.16$--$2.20$ the \textit{Cosmic Himalayas} and established that it is, to our knowledge, the densest quasar overdensity ($17\sigma$) in the $\sim$10,000~deg$^2$ SDSS footprint.
In the CROCODILE cosmological hydrodynamic simulations, the counts of quasars in cells follow a heavy-tailed distribution, under which a concentration like the \textit{Cosmic Himalayas} is rare but expected in $\Lambda$CDM \citep{Kuwayama2026}.
Such peaks therefore sample a tail that moderate overdensities do not constrain, and the galaxies and gas around them have to be measured where the peaks occur.
The quasar concentration coincides with neither the galaxy density peaks traced by 452 \lya\ emitters (LAEs; Subaru/HSC NB387) nor the extrema of a three-dimensional IGM \hi\ tomographic map.
It lies instead near the border between opaque and transparent IGM volumes \citep{Liang2025}.
A companion paper (CHEF~II) shows that the \ha\ emitters of the structure confirmed with Keck/MOSFIRE spectroscopy form two kinematically distinct groups about $2{,}000~\kms$ apart, which may be two structures along the line of sight, one structure seen in part, or a pair that will merge.
Interpreting these differences between the quasars, the galaxies, and the gas requires more than galaxy counts.
\citet{Liang2025} located the galaxies and estimated ultraviolet star-formation rates for the \lya\ emitters of the two density peaks, but did not measure their stellar masses and dust attenuation, which the panchromatic catalog assembled here supplies.

This paper is the first of the \textit{Cosmic Himalayas} Extended Field (CHEF) series on deep multiwavelength observations of the \textit{Cosmic Himalayas} and the $\approx1~\mathrm{deg}^2$ around the structure.
It presents the master catalog, CHEF2026, named like COSMOS2020 \citep{Weaver2022} after the year of its construction, and defines the samples on which the later papers of the series build.
The survey covers the structure and its surroundings and has uniform photometry from the far-ultraviolet to the submillimeter.
The catalog can support a growing set of studies, among them the following.
\begin{itemize}\setlength{\itemsep}{0pt}\setlength{\parsep}{0pt}\setlength{\topsep}{2pt}
\item \textit{The structure and its members.}
Companion papers measure the kinematics of the structure from Keck/MOSFIRE spectra of the \ha\ emitters (CHEF~II) and the selection, overdensity, and luminosity function of the \ha\ emitters of the NB2083 and NB2096 images (CHEF~III).
The NB387 image allows the \lya\ escape fraction of the \ha\ emitters to be measured at their positions and from their comparison with the LAEs.
\item \textit{Galaxy populations and environment.}
The stellar masses, star-formation rates, and rest-frame colors of the fitted galaxies, together with the color-selected star-forming and passive galaxies, can trace how galaxy properties change from the structure to the surrounding field.
\item \textit{Obscured star formation and active nuclei.}
The SCUBA-2 sources, the active galactic nuclei (AGN) selected by their \wise\ colors, and the quasars locate obscured star formation and AGN activity around the quasar concentration, and ALMA and \textit{Chandra} observations can extend this census (Section~\ref{sec:summary_points}).
\item \textit{Gas and simulations.}
\lya\ nebulae around the quasars and galaxies and the IGM tomography of \citet{Liang2025} can trace the gas of the structure, and cosmological simulations of such environments, such as CROCODILE \citep{Kuwayama2026}, can provide the expectations for its galaxies, quasars, and gas.
\item \textit{Legacy value.}
Beyond the structure, the catalog can serve studies of galaxies, AGN, and dusty sources from low redshift to $z>3$, and it can provide a protocluster field against which blank-field surveys are compared.
\end{itemize}

Section~\ref{sec:obs} describes the survey design and the imaging, with the astrometry, point-spread functions (PSFs), depths, and masks of each band.
Section~\ref{sec:catalog} describes the source detection, completeness, photometry, cross-matching, and classification, and Section~\ref{sec:validation} validates the photometry.
Section~\ref{sec:zphot} presents the spectroscopic redshifts in the field, the photometric redshifts, and the SED fits.
Section~\ref{sec:science} defines the value-added samples (the narrowband emitters, the color-selected and infrared-selected samples, the submillimeter sources, and the structure membership).
Section~\ref{sec:summary} describes the data products and summarizes the catalog, and Appendices~\ref{sec:app_datamodel} and \ref{sec:app_vi} give the catalog data model and the visual-inspection procedure.

Throughout, we adopt the flat $\Lambda$CDM cosmology of \citet{Planck2020}, with $H_0=67.66~\kms~\mathrm{Mpc^{-1}}$ ($h=0.6766$) and $\Omega_m=0.30966$ (the \texttt{Planck18} parameters of \texttt{Astropy}; \citealt{Astropy2022}).
Magnitudes are AB \citep{Oke1974}.
Spectral-feature wavelengths are in vacuum.
We adopt a \citet{Chabrier2003} initial mass function (IMF) for all stellar-population quantities.
We quote the scatter of residuals as the normalized median absolute deviation (NMAD).
\section{Survey Design and Data} \label{sec:obs}

\subsection{Field and wavelength coverage} \label{sec:survey}

\begin{figure*}
\centering
\includegraphics[width=\textwidth]{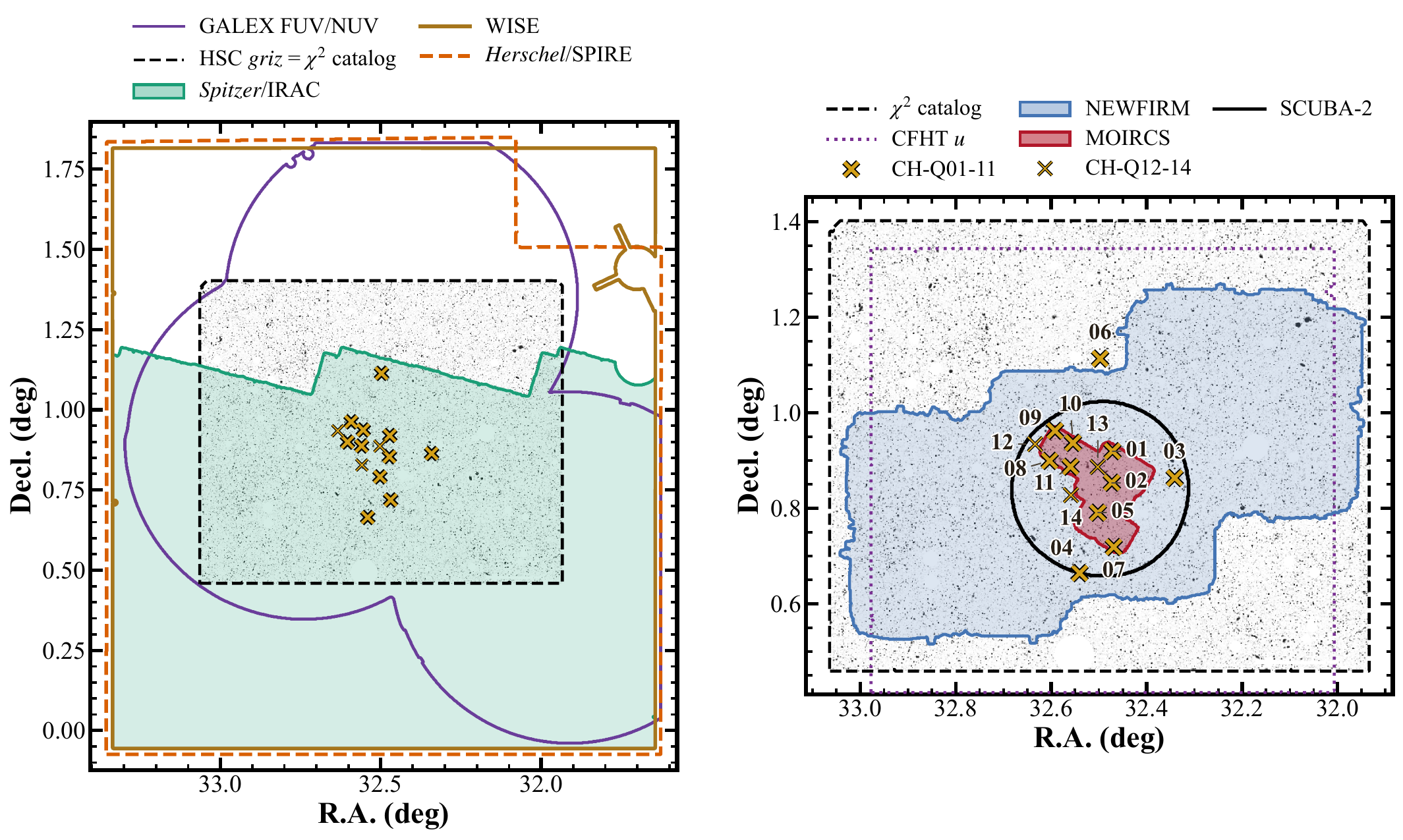}
\caption{Coverage of the imaging, over the HSC $r$ image.
\textit{Left:} the coverage of \galex\ FUV and NUV (purple), \spitzer/IRAC (shaded teal), \wise\ W1--W4 (brown), and \herschel/SPIRE (orange, dashed).
The black dashed line is the HSC $griz$ footprint, which is also that of the catalog of detections on the $\chi^2$ image of these bands (Section~\ref{sec:catalog_det}).
\textit{Right:} within the $\chi^2$ catalog footprint, the coverage of CFHT $u$ (dotted), NEWFIRM (shaded blue), MOIRCS (shaded red), and SCUBA-2 (black).
Each outline marks the coverage of at least one band of the instrument, and Table~\ref{tab:bands} gives the per-band areas.
Instrument colors follow Figure~\ref{fig:filterset}.
Crosses mark the quasars of the structure, labeled by number in the right panel: thick crosses CH-Q01 to CH-Q11 of \citet{Liang2025} and thin crosses the associated quasars CH-Q12 to CH-Q14.
R.A.\ increases to the left.}
\label{fig:footprints}
\end{figure*}

\textit{Field.}
The survey targets a single field, CHEF (the BOSSJ0210 field of MAMMOTH-Subaru), which contains the \textit{Cosmic Himalayas} quasar overdensity at $z\approx2.18$.
The field lies in SDSS Stripe~82 \citep{Annis2014}, where public imaging from the ultraviolet to the far-infrared is available.
The field is centered at ($32\fdg49$, $+0\fdg88$; J2000) on the Subaru/HSC NB387 mosaic of \citet{Liang2021}.
The effective area of this mosaic after masking is $1.31~\mathrm{deg}^2$.
The catalog is constructed on the grid of HSC-SSP tract 9491, which covers the central $1.065~\mathrm{deg}^2$, and its coordinates are on the \textit{Gaia} DR3 frame (Section~\ref{sec:bandqa_astrom}).
At $z=2.18$ the NB387 area corresponds to a square $\approx110$~cMpc on a side.
It contains the quasar overdensity, the two LAE density peaks northwest and southeast of the overdensity \citep{Liang2025}, and the surrounding field used as an internal density baseline.
In this catalog the quasars of the structure are named CH-Q01 to CH-Q14, with redshifts from ALMA CO(4--3), Keck/MOSFIRE, the first data release of the Dark Energy Spectroscopic Instrument (DESI DR1), SDSS DR16Q, and SDSS DR18 in this order of precedence (Section~\ref{sec:specz}).
CH-Q01 to CH-Q11 are the 11 quasars of \citet{Liang2025}.
CH-Q12 to CH-Q14 are DESI DR1 quasars that share a friends-of-friends group in comoving three-dimensional space with at least one of CH-Q01 to CH-Q11 at a linking length of 21~cMpc, as they do at every length from 20 to 22~cMpc.
The 34 quasars whose \lya\ forest (1040--1185~\AA\ in the rest frame) covers the structure, at $2.262\leq z\leq2.717$, are named CH-BGQ01 to CH-BGQ34.
Figure~\ref{fig:footprints} shows where each instrument has data and where CH-Q01 to CH-Q14 lie.

\begin{deluxetable*}{llcDcDD}[t]
\tablecaption{Imaging bands of CHEF2026.
\label{tab:bands}}
\tablewidth{0pt}
\tablehead{
\colhead{Band} & \colhead{Instrument} & \colhead{$\lambda_c$} &
\twocolhead{FWHM} &
\colhead{$5\sigma$ depth} & \twocolhead{Area} &
\twocolhead{$A_\lambda/$} \\
\unitcolhead{} & \unitcolhead{} & \unitcolhead{(\micron)} &
\unittwocolhead{(\arcsec)} & \unitcolhead{(AB mag or mJy)} & \unittwocolhead{(deg$^2$)} & \unittwocolhead{$E(B-V)$}
}
\decimals
\startdata
FUV          & \galex          & 0.154 & 4.2       & 24.50     & 2.14 & 7.836 \\
NUV          & \galex          & 0.232 & 5.3       & 23.25     & 2.15 & 7.009 \\
$u$          & CFHT/MegaCam    & 0.368 & 0.67      & 26.30     & 0.87 & 4.684 \\
NB387        & Subaru/HSC      & 0.386 & 1.18      & 24.68     & 1.31 & 4.538 \\
$g$          & Subaru/HSC      & 0.477 & 0.83      & 26.12     & 1.02 & 3.585 \\
$g$ (deep)   & Subaru/HSC      & 0.477 & 0.76      & 26.66     & 0.96 & 3.585 \\
$r$          & Subaru/HSC      & 0.623 & 0.54      & 25.81     & 1.02 & 2.542 \\
$i$          & Subaru/HSC      & 0.770 & 0.56      & 25.64     & 1.03 & 1.818 \\
$z$          & Subaru/HSC      & 0.892 & 0.65      & 24.92     & 1.04 & 1.429 \\
$y$          & Subaru/HSC      & 1.003 & 0.56      & 24.17     & 1.03 & 1.216 \\
$J_X$        & CTIO/NEWFIRM    & 1.25  & 1.35      & 23.17     & 0.51 & 0.800 \\
NB2096       & CTIO/NEWFIRM    & 2.091 & 1.27      & 22.22     & 0.51 & 0.364 \\
NB2083       & Subaru/MOIRCS   & 2.084 & 0.61      & 23.21     & 0.031 & 0.366 \\
$K_s$        & Subaru/MOIRCS   & 2.147 & 0.57      & 23.50     & 0.030 & 0.354 \\
$K_{Xs}$     & CTIO/NEWFIRM    & 2.146 & 1.21      & 23.31     & 0.51 & 0.355 \\
3.6~\micron & \spitzer/IRAC   & 3.55  & 1.91      & 21.46     & 1.96 & 0.181 \\
4.5~\micron & \spitzer/IRAC   & 4.49  & 1.79      & 21.54     & 1.79 & 0.124 \\
W1           & \wise           & 3.4   & 6.1       & 19.51     & 3.06 & 0.195 \\
W2           & \wise           & 4.6   & 6.4       & 19.17     & 3.09 & 0.119 \\
W3           & \wise           & 12.1  & 9.1       & 16.79     & 2.54 & 0 \\
W4           & \wise           & 22.2  & 13.2      & 14.46     & 2.54 & 0 \\
250~\micron & \herschel/SPIRE & 250   & 18.2      & 74~mJy   & 3.16 & 0 \\
350~\micron & \herschel/SPIRE & 350   & 25.2      & 73~mJy   & 3.16 & 0 \\
500~\micron & \herschel/SPIRE & 500   & 36.3      & 83~mJy   & 3.18 & 0 \\
850~\micron & JCMT/SCUBA-2    & 850   & 13        & 10.0~mJy & 0.106 & 0 \\
\enddata
\tablecomments{$\lambda_c$: central wavelength (for $u$ and the narrowbands, the mean wavelength of the full-system curve).
The narrowband FWHMs are $55.5$~\AA\ (NB387), $281$~\AA\ (NB2083), and $203$~\AA\ (NB2096).
The deep $g$ stack, a second image in the $g$ band, is not used in the fits.
This stack and the \spitzer/MIPS 24~\micron\ image, whose sources in the catalog of \citet{SEIP2020} are priors of the far-infrared deblending and whose fluxes enter the SED fits (Sections~\ref{sec:catalog_phot} and~\ref{sec:sed_cigale}), are not counted among the 24 bands.
FWHM: native PSF FWHM, the half-maximum width of a stack of isolated \textit{Gaia} stars from $u$ to $K_{Xs}$ and in IRAC (Section~\ref{sec:bandqa_psf}), the PSF model of the fits for \galex\ \citep{Morrissey2007}, the AllWISE PSF for \wise\ W1 and W2 \citep{Wright2010}, and the beam of the deblending for W3, W4, SPIRE, and SCUBA-2.
The MOIRCS and NEWFIRM PSFs vary over $0\farcs47$--$0\farcs63$ and $1\farcs1$--$2\farcs0$, respectively, across their mosaics.
$5\sigma$ depth: median over the field of the local $5\sigma$ point-source limit on the image used for each band's photometry (Section~\ref{sec:bandqa_depth}), in a $2\arcsec$ aperture from $u$ to $K_{Xs}$, for the fitted PSF-total flux for \galex\ and IRAC, and in the standard AllWISE apertures on the images for \wise.
SPIRE: $5\sigma$ total (instrumental and confusion) noise, from the clipped pixel scatter of the maps.
SCUBA-2: depth in the uniform-noise area ($0.082~\mathrm{deg}^2$; Section~\ref{sec:vas_submm}).
Area: effective area after masking (Section~\ref{sec:bandqa_mask}).
$A_\lambda/E(B-V)$: the \citet{Fitzpatrick1999} law ($R_V=3.1$) integrated through each passband for a 7000~K blackbody, per unit $E(B-V)$ on the scale of \citet{SchlaflyFinkbeiner2011}.
Beyond $3.3~\micron$ the law is continued as $\lambda^{-1.61}$, and beyond $10~\micron$ the coefficient is zero.}
\end{deluxetable*}

\begin{figure*}
\centering
\includegraphics[width=0.95\textwidth]{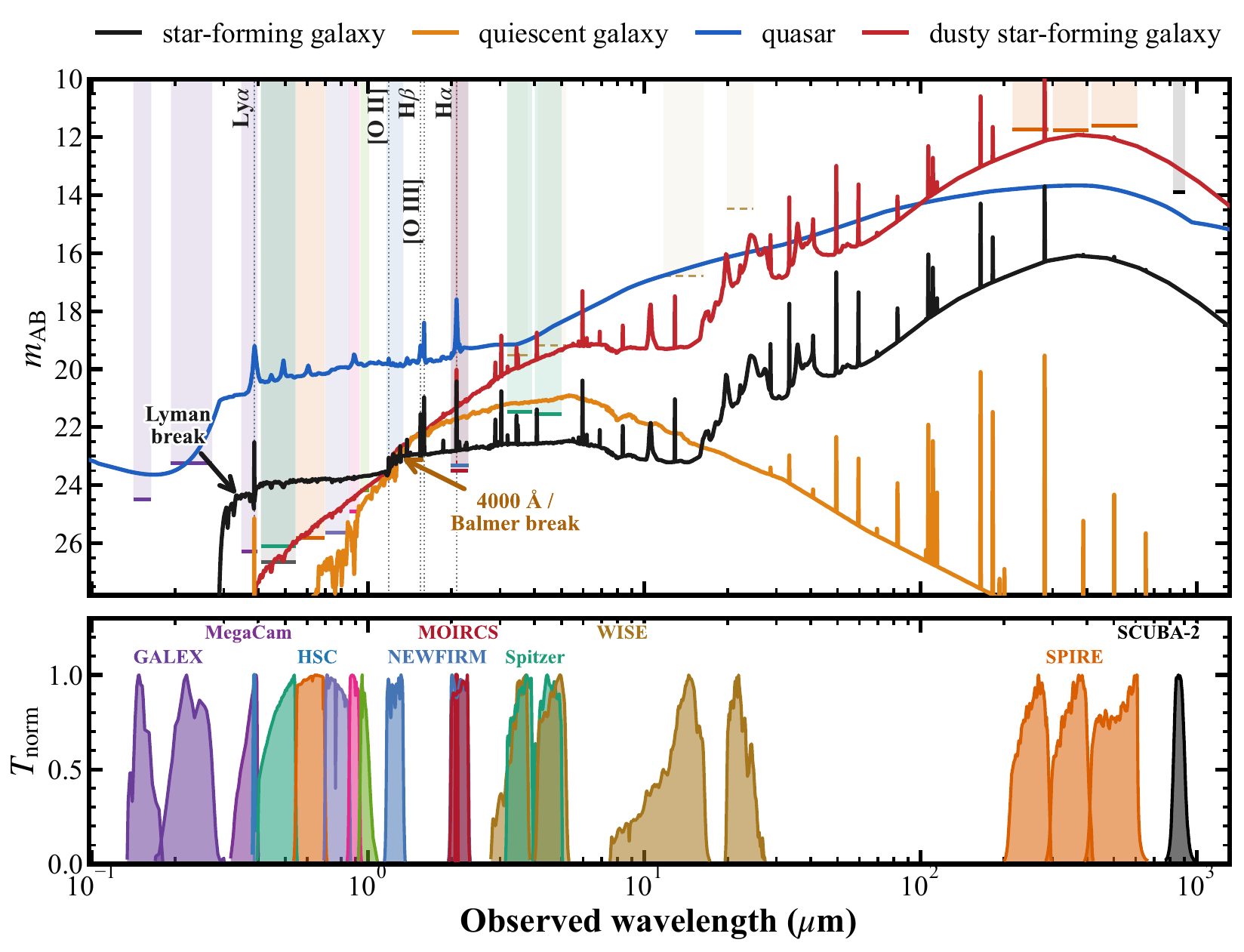}
\caption{Spectral coverage and depth of the survey.
\textit{Top:} four model SEDs at $z=2.18$ over the $5\sigma$ depth of each band (Table~\ref{tab:bands}), drawn as a shaded column that reaches down to the limit, so an object is detected where its SED lies above the bottom of a column.
Dashed limits mark the \wise\ bands, and the SPIRE and SCUBA-2 depths are converted to AB magnitudes.
The Lyman and 4000~\AA/Balmer breaks and the rest-frame emission lines at $z=2.18$ are marked.
The galaxy SEDs are illustrative \texttt{CIGALE} models built with the modules of the catalog SED fits (Section~\ref{sec:sed_cigale}) and delayed-$\tau$ star-formation histories.
They are a star-forming galaxy on the main sequence of \citet{Speagle2014} ($\mstar=10^{10}~\msun$, $\sfr=27~\msun~\mathrm{yr^{-1}}$, stellar $A_V=0.53$~mag), an unattenuated quiescent galaxy ($\mstar=10^{11}~\msun$), and a dusty star-forming galaxy ($\mstar=2\times10^{11}~\msun$, $\sfr=1{,}000~\msun~\mathrm{yr^{-1}}$, stellar $A_V=3.0$~mag, beyond the fitted grid), whose far-infrared emission follows from energy balance with the \citet{Dale2014} templates.
The narrow spikes are nebular emission lines.
The quasar is the unreddened type-1 template of the quasar fits (\citealt{Salvato2009}; Section~\ref{sec:zphot_z}) at $i=20$, and all four include the intergalactic absorption of \citet{Meiksin2006}.
\textit{Bottom:} transmission of every band, normalized to unit peak, colored by instrument.
The narrowband curves include the atmosphere, and the deep $g$ stack uses the HSC $g$ filter.}
\label{fig:filterset}
\end{figure*}

\textit{Wavelength coverage.}
The bands are chosen to bracket the spectral features that constrain \zphot\ and \mstar\ at the structure redshift (Figure~\ref{fig:filterset}; Table~\ref{tab:bands}).
At $z\approx2.2$ the Lyman break ($\approx0.29~\micron$) lies between \galex\ and CFHT $u$, and HSC $griz$ sample the rest-UV continuum.
The Balmer/4000~\AA\ break ($\approx1.2$--$1.3~\micron$) lies between HSC $y$ on its blue side and the rest-frame optical bands MOIRCS $K_s$ and NEWFIRM $K_{Xs}$ on its red side, and NEWFIRM $J_X$ covers the break.
\spitzer/IRAC and \wise\ W1 and W2 sample the rest-frame near-infrared, and \wise\ W3 and W4, \herschel/SPIRE, and JCMT/SCUBA-2 sample the dust emission of obscured sources.
The dusty model of Figure~\ref{fig:filterset}, a galaxy forming $1{,}000~\msun~\mathrm{yr^{-1}}$ at $z=2.18$, has $21$~mJy at 850~\micron, above the $5\sigma$ depth of SCUBA-2, and $46$--$59$~mJy in the SPIRE bands, below their depths (Table~\ref{tab:bands}).
Three narrowbands are tuned to the structure redshift.
NB387 isolates \lya\ at $z=2.177\pm0.023$.
The passbands of NB2083 (MOIRCS) and NB2096 (NEWFIRM) shift to the blue with the angle of incidence on the filter, which grows with the distance from the center of each pointing.
Between the centers and the corners of the pointings, the shift is about 110~\AA\ for NB2083 (from the MOIRCS filter-shift calculator of the Subaru Telescope) and 260~\AA\ for NB2096 \citep{Nakajima2012}.
Averaged over the field, NB2083 and NB2096 isolate \ha\ at $z=2.152$--$2.195$ and $2.165$--$2.199$, respectively (half maximum).

\subsection{Imaging} \label{sec:obs_imaging}

\textit{Ultraviolet and optical.}
\label{sec:obs_galex}\label{sec:obs_cfht}\label{sec:obs_hsc} The \galex\ FUV ($1539$~\AA) and NUV ($2316$~\AA) images are those of the Medium Imaging Survey \citep{Martin2005}, retrieved from MAST with their response and flag maps.
The CFHT/MegaCam $u$ image (filter MP9302) combines 20 exposures of 570--600~s ($11.6$~ks) taken on 2024 September 25--27 (UT; CFHT program 24BS07; PI: M.~Li).
The exposures were coadded with \texttt{SWarp} \citep{Bertin2002} after each CCD was registered astrometrically to the reference frame.
The $u$ zero point is set on SDSS Stripe~82 standard stars \citep{Thanjavur2021} and on synthetic photometry from the \textit{Gaia} DR3 XP spectra \citep{DeAngeli2023,Montegriffo2023}, which agree to $0.001$~mag, with a combined $1\sigma$ uncertainty of $0.041$~mag.
The HSC $grizy$ images are the Wide-layer coadds of tract 9491 from the third public data release of the HSC-SSP \citep[PDR3;][]{Miyazaki2018,Komiyama2018,Kawanomoto2018,Furusawa2018,Bosch2018,Aihara2022}.
The NB387 image and a deep $g$ stack ($4\times600$~s) were taken in 2018 January in the Subaru queue program S17B-041 for MAMMOTH-Subaru and keep the calibration of \citet{Liang2021}.
The deep $g$ stack is the continuum image of the NB387 selection.
Together with the NB387 image, it defines the area in which the LAEs are matched, and its photometry at the $\chi^2$ positions is part of the catalog.
Elsewhere in this paper, $g$ denotes the HSC-SSP image.

\textit{Near-infrared.}
\label{sec:obs_moircs}\label{sec:obs_newfirm} MOIRCS on Subaru \citep{Suzuki2008} imaged four pointings on the structure ($0.03~\mathrm{deg}^2$) in $K_s$ ($2.4$--$2.7$~ks per pointing) and NB2083 ($8.6$--$9.0$~ks) on 2024 September 21--23 (UT; Subaru program S24B-135; PI: Y.~Liang).
The data were reduced with \texttt{MCSRED2} \citep{Tanaka2011} and coadded with \texttt{SWarp}.
The MOIRCS photometric calibration is set per detector on unsaturated, isolated stars against the mean of three $K$-band references: UKIDSS \citep{Lawrence2007}, VHS \citep{McMahon2013}, and our NEWFIRM $K_{Xs}$ image.
The calibration uncertainty is $0.03$--$0.04$~mag per detector.
NEWFIRM on the Blanco 4-m telescope at CTIO \citep{Probst2008} imaged three pointings ($0.51~\mathrm{deg}^2$) in $J_X$ ($7.2$~ks per pointing), $K_{Xs}$ ($18.7$--$21.9$~ks), and NB2096 ($20.9$--$22.5$~ks).
The observations were taken on 14 nights between 2024 September 13 and 2025 December 10 (UT; NOIRLab programs 2024B-182778 and 2025B-897187; PI: Y.~Liang).
The data were reduced with a custom pipeline that masks detector persistence and rejects artifacts in each exposure.
Because the dither pattern repeats, the persistent images of bright stars would stack into spurious NB2096 sources at fixed positions on the sky.
Every pixel that was bright in either of the two preceding exposures of the night and is not bright in the exposure itself is therefore masked in each NB2096 exposure, together with a margin of three pixels.
The artifact rejection changes the $2\arcsec$ fluxes of sources by about $1\%$ or less.
The zero points of every pointing and detector are tied to UKIDSS and VHS stars at the mean of the two surveys' levels, with uncertainties of $0.008$~mag ($J_X$), $0.009$~mag ($K_{Xs}$), and $0.014$~mag (NB2096).

\textit{Mid-infrared to submillimeter.}
\label{sec:obs_irac}\label{sec:obs_mips}\label{sec:obs_wise}\label{sec:obs_hers}\label{sec:obs_jcmt} The \spitzer/IRAC 3.6 and 4.5~\micron\ images \citep{Werner2004,Fazio2004} are the combined atlas tiles of the SpIES program \citep{Timlin2016} from the NASA/IPAC Infrared Science Archive.
The SpIES coverage ends at $\delta\approx+1\fdg0$--$1\fdg2$, so the northern part of the catalog tract has no IRAC data (Figure~\ref{fig:footprints}).
The \wise\ W1 and W2 photometry is taken from the AllWISE catalog \citep{Wright2010,Cutri2013}, and the W3 and W4 fluxes are measured on the AllWISE images.
The \herschel/SPIRE 250, 350, and 500~\micron\ maps come from the \herschel\ Stripe 82 Survey \citep[HerS;][]{Viero2014} in the HerMES DR4 release \citep{Oliver2012}.
The JCMT/SCUBA-2 850~\micron\ map \citep{Holland2013} combines ten observations with the PONG900 pattern ($6.75$~h of scanning in total) taken on six nights between 2023 August 21 and September 12 (UT; JCMT program M23BP028; PI: Y.~Liang).
It was reduced with \texttt{SMURF} \citep{Chapin2013}, calibrated with a flux conversion factor \citep{Dempsey2013} of $495$~Jy~pW$^{-1}$~beam$^{-1}$ \citep{Mairs2021}, and matched-filtered.

\subsection{Astrometry} \label{sec:bandqa_astrom}

All images except those of NEWFIRM are registered to the astrometric frame of the HSC-SSP.
Stars matched between each image and the HSC reference show residual offsets of $\lesssim0\farcs06$ per axis in the HSC images, $\lesssim0\farcs08$ in MOIRCS, $\lesssim0\farcs17$ in IRAC, and $0\farcs1$--$0\farcs2$ in CFHT $u$, and these offsets are removed before photometry.

\begin{figure}[t]
\centering
\includegraphics[width=\columnwidth]{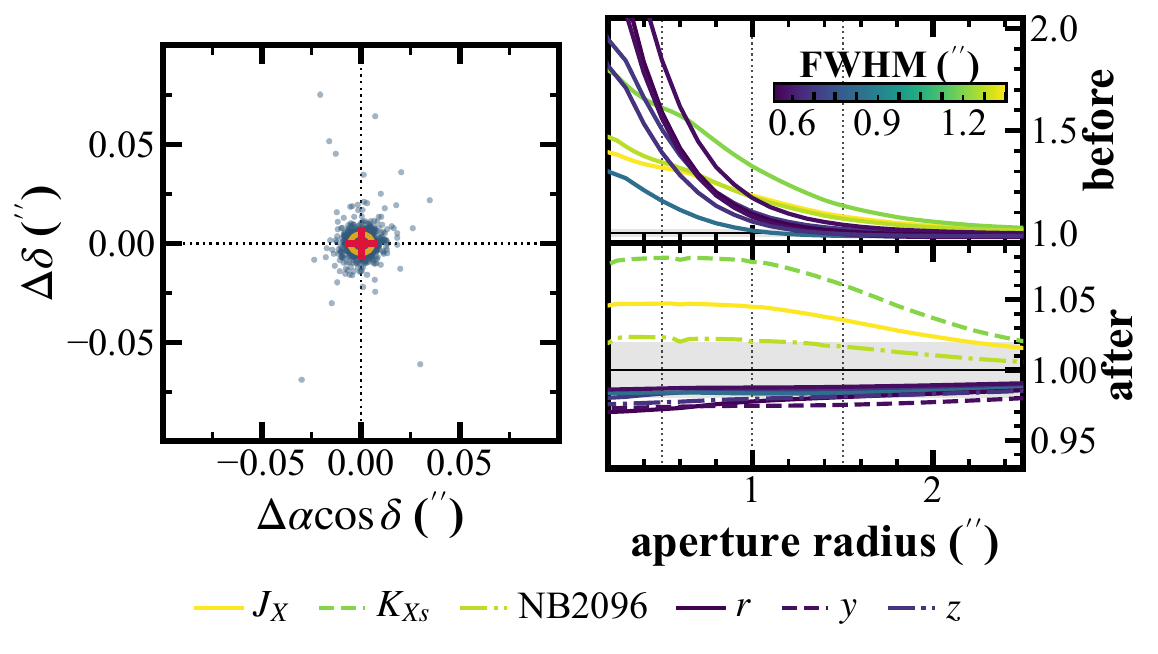}
\caption{Astrometry and PSF homogenization.
\textit{Left:} positions of the $\chi^2$ catalog relative to \textit{Gaia} DR3 (catalog minus \textit{Gaia}) for the 730 isolated stars unsaturated in $g$, $r$, $i$, and $z$ (Section~\ref{sec:bandqa_astrom}).
The red cross marks the per-axis median, and the gold ellipse marks the per-axis scatter (NMAD).
\textit{Right:} the curve of growth of each band divided by that of its target, on the stars not used to build the kernels, before matching (top, colored by native FWHM) and after convolution to the $1\farcs05$ target ($u$ and $grizy$) or to the local NEWFIRM target (bottom), so that a matched band is a horizontal line at unity.
Dotted lines mark the radii of apertures of 1, 2, and 3\arcsec\ diameter, and the gray band is $\pm2\%$.}
\label{fig:astrompsf}
\end{figure}

We compare the positions of the $\chi^2$ catalog with \textit{Gaia} DR3 \citep{GaiaMission2016,GaiaDR3}.
The position of a star on the $\chi^2$ image refers to the $\mathrm{S/N}^2$-weighted mean epoch of its $g$, $r$, $i$, and $z$ exposures, to which we propagate each \textit{Gaia} position with its proper motion (2017.9--2018.6 for the comparison stars).
For 730 isolated, point-like stars at $19<G<20.8$ that are unsaturated in all four bands, the $\chi^2$ positions measured on the HSC-SSP PDR3 coadds have median offsets of $-0\farcs005$ in right ascension and $+0\farcs016$ in declination.
We find the same declination offset for these stars in the positions of the PDR3 catalog \citep{Aihara2022}.
The catalog coordinates remove these median offsets, and for the same stars they agree with \textit{Gaia} to better than $0\farcs001$ in the median, with a per-axis scatter of $0\farcs006$ and $0\farcs005$, respectively (Figure~\ref{fig:astrompsf}, left).
The correction changes the catalog coordinates only, and the photometry is measured on the astrometric solution of each image.
Brighter stars are saturated in the HSC Wide images, and at $17<G<18$ their catalog positions have median offsets of $+0\farcs007$ in right ascension and $+0\farcs010$ in declination.
The positions of the MOIRCS $K_s$, NB2083, and NB387 detections, measured on their native grids, are corrected to \textit{Gaia} in the same way, for NB387 with a declination term that varies across the field, and on the stars of Figure~\ref{fig:astrompsf} (left) with an NB387 detection the catalog NB387 positions have a median offset of $-0\farcs008$ in declination.
The NEWFIRM pointings are each registered to \textit{Gaia} and resampled onto a common projection for the field coadds.
The median offset of every combination of pointing and detector is at most $0\farcs02$ in right ascension and $0\farcs03$ in declination, with larger offsets in one corner of the central pointing (Section~\ref{sec:known_issues}).
The catalog positions of the $K_{Xs}$ and NB2096 detections follow this solution without a further correction.

\subsection{Point-spread functions and homogenization}
\label{sec:bandqa_psf}\label{sec:catalog_psf}

Per-band PSF models are built from isolated, unsaturated \textit{Gaia} DR3 stars, split at random into two halves.
One half builds the PSFs and matching kernels, and the other half validates them.
Matched-aperture colors require a common PSF, but convolving every band to the broadest PSF would discard the resolution of HSC and MOIRCS.
We therefore homogenize only within groups of similar PSF and measure the bands with broad PSFs by prior-based PSF fitting (\galex\ and IRAC) or deblending (\wise\ W3 and W4, SPIRE, and SCUBA-2; Section~\ref{sec:catalog_phot}).
CFHT $u$ and HSC $grizy$ are matched on the tract grid to a target of $1\farcs051$ FWHM (the $g$ PSF broadened by a Gaussian).
The kernels, from regularized Fourier division \citep[cf.][]{Aniano2011}, vary across the field.
MOIRCS $K_s$ and NB2083 are matched to the same target with kernels that vary across the pointings.
The NEWFIRM bands are matched to a common local target, $\max(1\farcs60,\ \mathrm{FWHM_{max}}+0\farcs10)$, where $\mathrm{FWHM_{max}}$ is the broadest of their native PSFs at that position.
Fourier division of their critically sampled PSFs is unstable, so the NEWFIRM kernels are Gaussians solved on a Moffat model of the local PSF.

On the validation stars the matched FWHMs reach their targets within $0\farcs03$.
The encircled energy within apertures of $1\arcsec$, $2\arcsec$, and $3\arcsec$ diameter matches the target to $0.023$ or better in the optical bands (Figure~\ref{fig:astrompsf}, right) and in MOIRCS.
In the $2\arcsec$ aperture the encircled energy of NEWFIRM $K_{Xs}$ and $J_X$ stays $7.7\%$ and $4.5\%$ above the target.
The forced photometry at the near-infrared and narrowband positions therefore converts each aperture flux to a total with the encircled energy of the band's own local PSF (Section~\ref{sec:catalog_phot}).
The PSF models, the kernels, and their validation are provided with the catalog.

\subsection{Depths and uncertainties} \label{sec:bandqa_depth}

\begin{figure*}
\centering
\includegraphics[width=\textwidth]{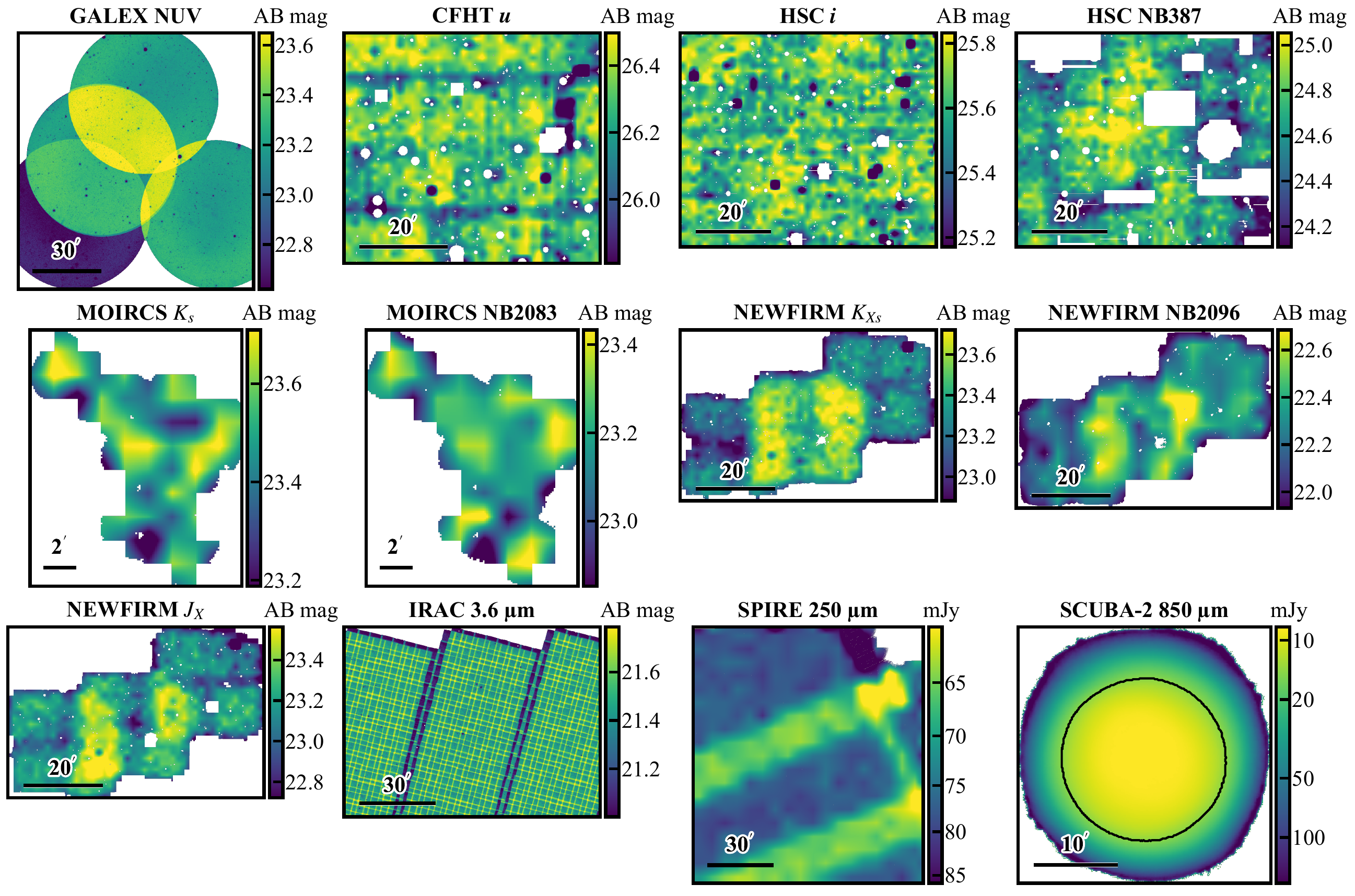}
\caption{Spatial $5\sigma$ point-source depth in twelve bands.
Table~\ref{tab:bands} gives the depth definition of each band.
Each panel has its own field of view, angular scale bar, and color scale.
Yellow is deeper.
Depths are in AB magnitudes, except for SPIRE 250~\micron\ and SCUBA-2 850~\micron, which are in mJy.
The black line on the SCUBA-2 map outlines the uniform-noise area of Table~\ref{tab:bands}.
The NB2096 map shows the depth of the image on which the \ha\ emitters are selected (Section~\ref{sec:vas_hae}).
The catalog NB2096 photometry reaches $22.22$~mag (Table~\ref{tab:bands}).
Blank pixels are outside the coverage or contain too few empty apertures, mostly around bright stars.}
\label{fig:depthmaps}
\end{figure*}

For the bands measured in apertures, the $5\sigma$ point-source depth is measured from the clipped scatter of fluxes in source-free apertures within cells across the image used for photometry.
Figure~\ref{fig:depthmaps} maps the depth for representative bands, and Table~\ref{tab:bands} gives its median over the field.
The completeness and the narrowband selections use the local depth (Sections~\ref{sec:catalog_compl} and~\ref{sec:vas_hae}).
Resampling, coaddition, and PSF matching correlate neighboring pixels, so the pixel variance underestimates the error of an aperture flux.
The aperture and Kron flux uncertainties of the catalog are therefore the \texttt{SExtractor} errors multiplied by a factor $\sigma_{\rm scale}$, set per band, image, and aperture to reproduce the empty-aperture scatter and provided with the catalog.
In the $2\arcsec$ aperture, the factor is $1.15$--$1.22$ on the native HSC images, whose noise is only weakly correlated, and $3.6$--$8.5$ on the resampled NEWFIRM images and the PSF-matched images.
Resampling and the matching convolution lower the pixel variance but leave the noise of a $2\arcsec$ sum almost unchanged.

\subsection{Masks and effective areas} \label{sec:bandqa_mask}\label{sec:bandqa_area}

Each band carries a mask with two components: a footprint that trims the low-coverage rim of the mosaic, and bad regions around bright stars and image artifacts (saturation trails, ghosts, diffraction spikes, and halos).
The bright-star radii are measured on each band's own image for every \textit{Gaia} DR3 star and capped by a magnitude--radius relation of the form used for HSC-SSP \citep{Coupon2018}, so that the deep optical images receive larger masks than the shallower infrared images.
Every mask was inspected visually over the whole mosaic and corrected by hand where needed.
Table~\ref{tab:bands} gives the effective areas after masking.
The area unmasked in all of $g$, $r$, $i$, and $z$ (the four-band area), which defines \texttt{optical\_coverage\_flag}$~=0$, is $1.001~\mathrm{deg}^2$.
The $\chi^2$ detection covers $1.045~\mathrm{deg}^2$.
Sources outside a band's effective area have no flux in that band and enter the SED fits as missing data.
The masks are provided with the catalog.

\section{Catalog Construction} \label{sec:catalog}

The bands of this survey span a factor of $\approx65$ in PSF FWHM, from $0\farcs54$ (HSC $r$) to $35\arcsec$ (SPIRE 500~\micron), and no single image detects every population at $z\sim2.2$.
We therefore separate detection from measurement.
Sources are detected independently in six images (Section~\ref{sec:catalog_det}), each band is measured at the resolution its data allow (Section~\ref{sec:catalog_phot}), and the detected samples are linked by cross-matching (Section~\ref{sec:catalog_assoc}).

\subsection{Source detection} \label{sec:catalog_det}

Sources are detected with \texttt{SExtractor} \citep{Bertin1996} in six images, each chosen for its target population and each with its own footprint and selection function:
\begin{enumerate}
\item \textit{Optical $\chi^2$}: the sum $\chi^2=\sum_b f_b^2 w_b$ \citep{Szalay1999} of the native HSC $g$, $r$, $i$, and $z$ images on the tract grid, where $w_b$ is the inverse variance of band $b$ (zero in masked pixels).
This sum sets a statistical threshold without favoring blue or red continua.
Its detections form the general rest-UV continuum sample, which we call the $\chi^2$ catalog.
For most objects, every optical and near-infrared band is measured at these positions.
The $y$ band, the shallowest HSC band, would carry about $1.3\%$ of the $\mathrm{S/N}^2$ weight of a source with a flat $f_\nu$ spectrum, so it is not part of the $\chi^2$ image but is measured at its positions with the other bands.
\item \textit{MOIRCS $K_s$}: a near-infrared selection over the MOIRCS pointings.
\item \textit{NEWFIRM $K_{Xs}$}: a near-infrared selection over the NEWFIRM area, which recovers red and optically faint sources below the $\chi^2$ threshold.
\item \textit{NB387}: the counterparts of the published \lya\ emitters at $z\approx2.18$ (Section~\ref{sec:vas_lae}).
\item \textit{NB2083}: \ha\ emitters at $z\approx2.18$ over the MOIRCS area.
\item \textit{NB2096}: \ha\ emitters at $z\approx2.18$ over the NEWFIRM area, kept separate from NB2083 because the filter, PSF, depth, and footprint differ.
\end{enumerate}
Sources in the near-infrared and narrowband images are detected on the native grids, where the noise is least correlated.
The $K_{Xs}$, NB2083, and NB2096 detections use a noise-map weight that makes the threshold a constant multiple of the measured local noise.
For $K_s$, $K_{Xs}$, NB2083, and NB2096, the threshold and minimum area are the combination that maximizes the injection--recovery completeness (Section~\ref{sec:catalog_compl}) at a false-detection fraction $F\leq0.5\%$, where $F$ is the number of detections on the negated image divided by the number on the real image, and the \texttt{SExtractor} configuration of each detection image is provided on request.
Table~\ref{tab:compl} gives the number of detections in each image.

\textit{Significance of the $\chi^2$ detections.}
The $\chi^2$ image cannot be negated, because $\chi^2$ is positive by construction.
Its threshold ($1.5$ times the background rms of the filtered image) admits noise peaks, and $18$--$19\%$ of the $\chi^2$ detections are spurious (Section~\ref{sec:catalog_compl}).
Every $\chi^2$ detection therefore carries its peak significance on the filtered image, \texttt{peak\_snr}, and the sign statistic $L=\sum_b (K*f_b\sqrt{w_b})$ at that peak.
The statistic $L$ is the sum of the filtered, noise-normalized band fluxes, where $K$ is the detection filter.
Because $\chi^2$ does not change when every band flips sign, a noise peak has $L<0$ half of the time, whereas none of the $26{,}198$ injected point sources recovered at $\texttt{peak\_snr}\geq4.5$ does.
The $285{,}537$ detections with $\texttt{peak\_snr}\geq4.5$ and $L>0$ have a false-detection fraction of $\approx0.3\%$.
These detections form the recommended sample for optical selections (Section~\ref{sec:catalog_flow}) and the optical priors of the \galex\ and IRAC photometry (Section~\ref{sec:catalog_phot}).
A near-star flag marks the detections that lie within 1.3 times the radius of the masked halo of the four $G<10$ stars ($130$--$260\arcsec$) or within $25\arcsec$ of a star with $13.5\leq G<15$.
Around these stars the $\chi^2$ image also picks up fragments of the stellar halos.

\subsection{Detection completeness and contamination}
\label{sec:catalog_compl}\label{sec:val_compl}\label{sec:val_spurious}

We measure the completeness of every detection image by injecting artificial sources and recovering them with the detection settings of the catalog.
The exception is NB387, whose \lya\ emitters keep the selection function of \citet{Liang2021}.
Most injected sources are Moffat profiles with the local FWHM and $\beta$ of the band's PSF, and $10$--$15\%$ are exponential disks convolved with that PSF.
The sources are placed at least two FWHMs from every real detection and from each other, so the completeness is that of isolated sources.
For the $\chi^2$ image each source has one of three UV slopes ($\beta_{\rm UV}=-2.3$, $-2$, and $0$) across $g$, $r$, $i$, and $z$, and the $\chi^2$ image is rebuilt from the injected bands.
In the $K_s$, $K_{Xs}$, NB2083, and NB2096 images, the completeness of isolated point sources is a single function of $m-m_{\rm det}(\mathbf{x})$.
Here $m_{\rm det}(\mathbf{x})$ is the total magnitude at which a point source of the local PSF has a filtered peak equal to the detection threshold, given the local noise of the detection weight map and the detection filter.
The function is $C=c_{\max}/\{1+\exp[(m-m_{\rm det}-d_{50})/s]\}$ with $c_{\max}=0.985$--$0.990$, $d_{50}=-0.09$ to $-0.20$~mag, and $s=0.14$--$0.18$~mag.
For the $\chi^2$ image the corresponding quantity is the expected filtered peak $\sum_b f_b^2 w_b \max(K*P_b^2)$, where $P_b$ is the PSF of band $b$.
This peak depends on the colors of the source.
The completeness of the $\chi^2$ image is a cumulative Gaussian in the logarithm of the expected peak.
We provide maps of these detection limits and of the 50, 80, and $90\%$ completeness magnitudes.
To build a random catalog for a given luminosity function, points are drawn uniformly over the detection area, at least $2\arcsec$ from its edge, and each point is kept with probability $C$ at a magnitude drawn from that luminosity function.
Spurious detections are counted on the negated images for the near-infrared and narrowband detections.
For the $\chi^2$ image we use two estimates: a pure-noise $\chi^2$ image built from the measured per-band noise and pixel autocorrelation, and the count of detections with $L<0$, which gives $N_{\rm false}\approx2~N(L<0)$.

\begin{figure}[t]
\centering
\includegraphics[width=\columnwidth]{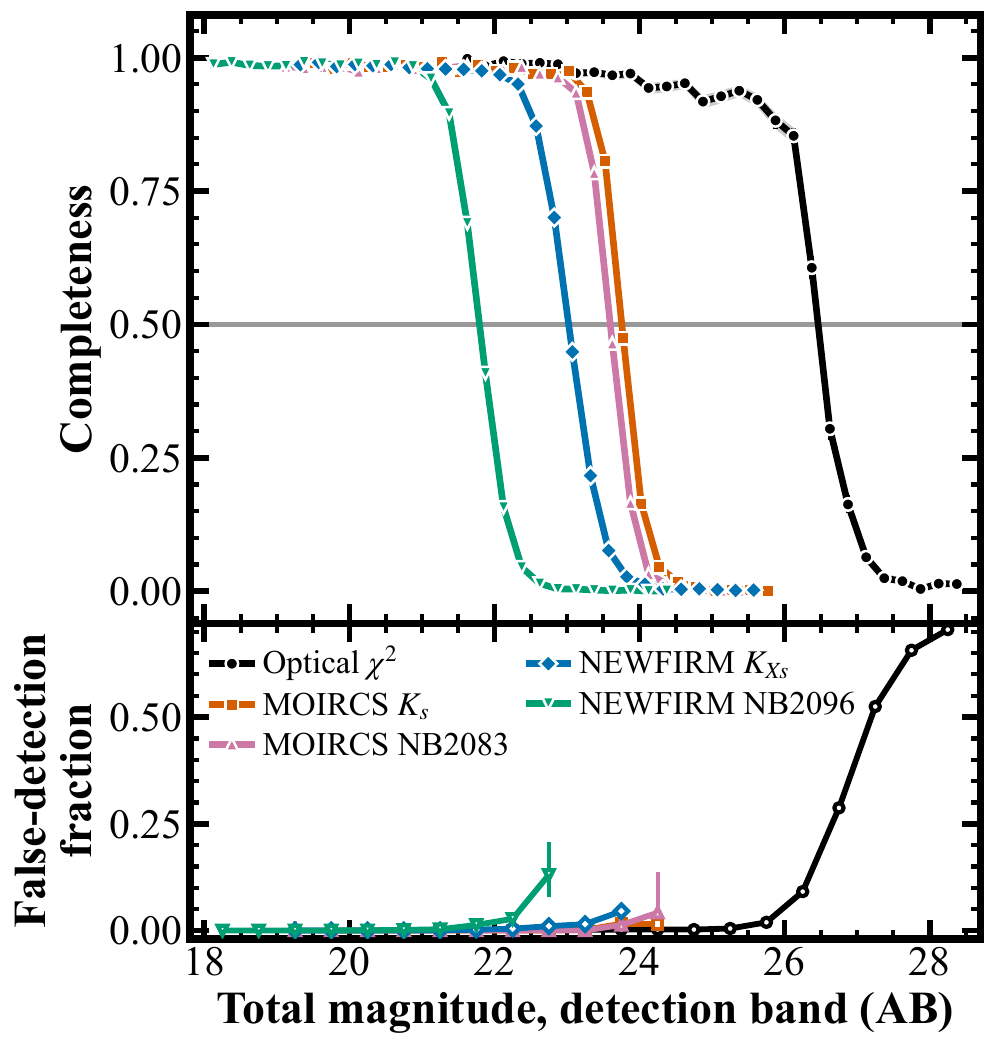}
\caption{Completeness (top; filled symbols, with $68\%$ Wilson intervals) and false-detection fraction (bottom; open symbols, with $68\%$ Poisson intervals) of the detection images against total magnitude in the detection band.
The five detection images are the optical $\chi^2$ image (black circles), MOIRCS $K_s$ (orange squares) and NB2083 (pink triangles), and NEWFIRM $K_{Xs}$ (blue diamonds) and NB2096 (green inverted triangles).
This magnitude is the injected total magnitude for the completeness (for the $\chi^2$ image, the $i$ magnitude of a flat-$f_\nu$ source) and the measured Kron magnitude for the false detections.
The gray line marks $50\%$ completeness.
The false-detection fraction is shown where a bin contains at least 20 real detections.
The $\chi^2$ image cannot be negated, so its false-detection fraction is twice the fraction of detections with $L<0$, each placed at the $i$ Kron magnitude of its absolute flux (Section~\ref{sec:catalog_compl}).
The intervals of this fraction come from a bootstrap over detections.}
\label{fig:completeness}
\end{figure}

Table~\ref{tab:compl} and Figure~\ref{fig:completeness} give the results.
The $\chi^2$ catalog is $50\%$ complete at $i=26.46$ for a flat-$f_\nu$ spectrum and $93\%$ complete at the local $5\sigma$ depth in $i$.
The spurious fraction falls from $72\%$ at $\texttt{peak\_snr}=3.0$--$3.5$ to $5\%$ at $4.5$--$5.0$ and to $0.3\%$ or less above $6$.
Placing each detection with $L<0$ at the $i$ Kron magnitude of its absolute flux, we find a false-detection fraction below $2\%$ at $i<26$, which rises at fainter magnitudes (Figure~\ref{fig:completeness}).
Where fewer than four bands have data and within $2\arcsec$ of the coverage edge, the $50\%$ completeness magnitude lies $0.28$ and $0.56$~mag brighter, respectively, than the model predicts from the expected peak.
The $K_s$ and NB2083 samples are $96\%$ complete at their local $5\sigma$ depth.
The NEWFIRM $K_{Xs}$ and NB2096 samples are only $70\%$ and $68\%$ complete at that depth, because the correlated noise of the NEWFIRM coadds makes the detection threshold shallower than the $2\arcsec$ aperture depth.

\begin{deluxetable}{lrDccc}
\tabletypesize{\footnotesize}
\setlength{\tabcolsep}{3pt}
\tablecaption{Detection completeness and contamination of each detection image.
\label{tab:compl}}
\tablewidth{0pt}
\tablehead{\colhead{Detection image} & \colhead{$N_{\rm det}$} &
\twocolhead{Area} & \colhead{$F$} & \colhead{$C(m_{5\sigma})$} & \colhead{$m_{50}$} \\
\unitcolhead{} & \unitcolhead{} & \unittwocolhead{(deg$^2$)} & \unitcolhead{(\%)} &
\unitcolhead{} & \unitcolhead{(AB mag)}}
\decimals
\startdata
Optical $\chi^2$ & 425{,}670 & 1.045 & 18--19\tablenotemark{a} & 0.93 & 26.46 \\
MOIRCS $K_s$     & 3{,}498   & 0.030 & 0.23 & 0.956 & 23.76 \\
NEWFIRM $K_{Xs}$ & 39{,}081  & 0.51  & 0.44 & 0.701 & 23.03 \\
NB387            & 52{,}812  & 1.31  & 0.06\tablenotemark{b} & \nodata & \nodata \\
NB2083           & 3{,}222   & 0.029 & 0.13 & 0.964 & 23.60 \\
NB2096           & 17{,}545  & 0.51  & 0.40 & 0.681 & 21.80 \\
\enddata
\tablecomments{Area: detection area after masking.
$F$: false-detection fraction.
$C(m_{5\sigma})$: completeness of isolated sources at the local $5\sigma$ $2\arcsec$ depth (for $\chi^2$, the depth in $i$).
$m_{50}$: total magnitude of $50\%$ completeness (for $\chi^2$, in $i$ for a flat-$f_\nu$ spectrum).
NB387 keeps the selection function of \citet{Liang2021}.}
\tablenotetext{a}{Fraction of all $\chi^2$ detections that are spurious.
The fraction is $\approx0.3\%$ for the $285{,}537$ detections with $\texttt{peak\_snr}\geq4.5$ and $L>0$.}
\tablenotetext{b}{Ratio of negated to real detections brighter than the \lya\ selection limit.}
\end{deluxetable}

\subsection{Photometry} \label{sec:catalog_phot}

\textit{Matched apertures at the $\chi^2$ positions.}
The optical and near-infrared bands are measured with \texttt{SExtractor} in dual-image mode, with the $\chi^2$ image as the detection image.
Each band is resampled onto the tract grid after its frame offset is removed (Section~\ref{sec:bandqa_astrom}).
Each band except NEWFIRM is measured on its native and on its PSF-homogenized image, and both measurements are in the catalog.
We measure $2\arcsec$ and $3\arcsec$ aperture fluxes for colors, \texttt{MAG\_AUTO} \citep{Kron1980} fluxes, and a total flux.
The dual-image mode gives photometry in $ugrizy$, MOIRCS $K_s$ and NB2083 (matched to $1\farcs05$), and NEWFIRM $J_X$, $K_{Xs}$, and NB2096 (native) at all $425{,}670$ positions.
The total flux of a band is its $2\arcsec$ aperture flux on the PSF-homogenized image multiplied by the ratio of the Kron flux on the native $i$ image (Kron factor $2.5$) to the $2\arcsec$ flux on the homogenized $i$ image.
The same factor scales every band and its uncertainty.
The colors of a source are therefore those of the $2\arcsec$ aperture, and its normalization is that of the $i$-band Kron aperture.
The catalog total fluxes, \texttt{f\_X}, are the fluxes entering the fits, except in the photometric-redshift and SED fits of 118 extended galaxies (Section~\ref{sec:sed_cigale}).
They are corrected for Galactic extinction.
No correction is made for the light outside the Kron aperture, which is about $0.07$~mag for a point source.
The median correction from the $2\arcsec$ aperture to total is $0.22$~mag for point sources and $0.33$~mag for galaxies at $i=24$.
Sources without a usable ratio take the median factor, $0.30$~mag: the $38\%$ without a measured $i$-band Kron flux and the $8\%$ whose ratio lies outside 1--5 or whose $i$ aperture is saturated or masked.

Before the scaling to total flux, the NEWFIRM $2\arcsec$ aperture fluxes, measured on the native images, are multiplied by the ratio of the encircled energies of the homogenized $i$-band PSF and of the local NEWFIRM PSF in that aperture.
The median correction is $0.19$~mag in $J_X$ and $0.10$~mag in $K_{Xs}$, and it ranges over $0.08$--$0.34$ and $-0.02$ to $0.15$~mag between the three pointings.
The deep $g$ stack is measured at the same positions on its native image and on its image matched to $1\farcs05$.
NB387, whose $1\farcs2$ PSF is broader than the target, is measured at its own detections, with $u$ and the deep $g$ smoothed to the NB387 PSF for the \lya\ colors ($g-\mathrm{NB387}$ in the NB387 selection of \citealt{Liang2021}).

\textit{Forced photometry at the near-infrared and narrowband positions.}
The $K_s$ and $K_{Xs}$ detections are measured at their own positions in eleven bands ($u$, $grizy$, $K_s$, NB2083, $J_X$, $K_{Xs}$, NB2096), and the NB2083 and NB2096 detections in $g$, $i$, and $K_s$ or $K_{Xs}$, respectively.
Each position is carried into the frame of every band through the stellar offsets between the two frames.
The flux is measured there in a $2\arcsec$ aperture with a local background and a local empty-aperture uncertainty and is converted to a point-source total with the encircled energy of the band's PSF at that position.
Neighbors are flagged but not subtracted.
For point sources the forced $K_{Xs}$ total is $0.02$~mag brighter than \texttt{MAG\_AUTO} in the median.
The forced photometry provides the colors used in the \ha\ selection (Section~\ref{sec:vas_hae}) and the photometry of the $K_{Xs}$ catalog and of the $426$ $K_s$ detections without a $\chi^2$ counterpart.
The narrowband samples also carry line fluxes in the $2\arcsec$ selection aperture, without an aperture correction.
The point-source correction would be $0.10$~mag in NB2083, measured on the native image, and $0.66$~mag in NB2096, measured on the images matched to $1\farcs6$.

\textit{Prior-based photometry.}
The PSFs of \galex\ ($4\farcs2$--$5\farcs3$) and IRAC ($1\farcs8$--$1\farcs9$) are too broad for PSF homogenization, so these bands are measured by fixed-position PSF fitting at $288{,}858$ priors: the $285{,}537$ significant $\chi^2$ detections and $3{,}321$ $K_s$ and $K_{Xs}$ detections without a significant $\chi^2$ counterpart.
Sources are grouped into blends by their separation relative to the PSF and fitted with a free local background.
The catalog fluxes come from one joint linear fit of all groups with \texttt{SciPy} \citep{Virtanen2020} without a non-negativity constraint.
The fluxes of faint sources can therefore be negative, and their uncertainties include the pixels and fluxes shared with neighbors.
The PSF is a Moffat profile for \galex\ and, for IRAC, one empirical circularized median stack of about 200 isolated stars per channel, used over the whole field.
To tie the IRAC fluxes to the IRAC absolute calibration, we divide them by $k=1.079$ (ch1) and $1.148$ (ch2).
The factor $k$ is the median ratio, over isolated stars, of the flux fitted with the unit-normalized PSF to the flux in the $12\farcs2$-radius calibration aperture.
\galex, which has too few isolated stars, keeps $k=1$.
For both \galex\ and IRAC, we correct the fluxes for an additive bias and rescale their errors, both measured from injected sources per band and environment (isolated, in a group, or at its boundary), and add the remaining residual in quadrature as a systematic error.
On the half of the injections not used to derive these corrections, the corrected fluxes of sources injected at 3 and 10 times the local noise have a median bias of $-0.08$ to $+0.20$ times the statistical error and a scatter of $0.78$--$1.00$ times the catalog uncertainty.
Isolated IRAC sources agree with the SpIES catalog \citep{Timlin2016} to $0.036$ (ch1) and $0.016$~mag (ch2) in the median.
The IRAC fit residuals yield $190{,}963$ further IRAC-only positions (local maxima at $\mathrm{S/N}\geq4$ more than 1.5 PSF FWHMs from every prior), which are measured in IRAC and provided on request.

\textit{\wise.}
The W1 and W2 fluxes are the profile-fit photometry of the AllWISE catalog \citep{Cutri2013}.
Measurements with photometric quality A--C, no contamination flag, and $\mathrm{S/N}\geq2$ are matched one-to-one within $2\arcsec$ to the IRAC priors and the IRAC-only positions.
The other priors within half a beam FWHM of an AllWISE source receive its flux as an upper limit, and isolated non-detections receive a $3\sigma$ upper limit from the local noise measured in empty apertures.
W1 and W2 do not enter the photometric-redshift fits, and the SED fits use them only outside the IRAC footprint. W3 and W4 are measured with the far-infrared bands.

\textit{Mid-infrared, far-infrared, and submillimeter.}
The \wise\ W3 and W4, MIPS 24~\micron, SPIRE, and SCUBA-2 fluxes are deblended by non-negative least squares within each beam group, at the positions of the IRAC priors, the $\chi^2$ detections below the significance cut, and the MIPS 24~\micron\ sources of the \spitzer\ Enhanced Imaging Products \citep{SEIP2020}.
A prior stays free when its flux predicted by an SED fit to the broad bands reaches the local point-source noise, with at most eight free priors per beam, the brightest predicted first, and the other priors are fixed at their predicted fluxes and subtracted.
The flux of a free prior is the median of a residual bootstrap, corrected for the bias measured on point sources injected into the maps, and its uncertainty is the bootstrap width rescaled with the same injections, so that it includes confusion noise.
A source is detected where its $\mathrm{S/N}$ reaches the larger of 3 and the 99th percentile of the $\mathrm{S/N}$ of sources injected at zero flux, between 3.0 and 4.1 depending on band and environment.
An undetected free prior receives a $3\sigma$ upper limit, and a fixed or merged prior, or an object without a prior, receives the summed flux of the free priors within one beam FWHM and of the residual at its position, plus three times its uncertainty.
At 850~\micron, the candidate counterpart with the lowest chance probability of each peak of the blind catalog (Section~\ref{sec:vas_submm}) takes the deboosted flux of the peak, and the other objects within one beam FWHM receive that flux as an upper limit.

\textit{Uncertainties and limits.}
Aperture and Kron flux uncertainties are the weight-map uncertainties rescaled by the empty-aperture factor $\sigma_{\rm scale}$ of Section~\ref{sec:bandqa_depth} (for Kron fluxes, that of the band's native $2\arcsec$ aperture), and a total flux keeps the signal-to-noise ratio of its $2\arcsec$ aperture.
In the \galex, optical, near-infrared, and IRAC bands, non-detections keep their measured flux and uncertainty for the SED fits \citep[cf.][]{Whitaker2011}, and in the \wise, MIPS, SPIRE, and SCUBA-2 bands they enter as upper limits.
Below $2\sigma$, the catalog magnitude is the local $2\sigma$ limit, and the row carries a limit flag.
The uncertainties of the prior-based bands are the fit covariance, and those of the deblended bands are the injection-calibrated bootstrap widths.

\textit{Galactic extinction.}
Apart from the total fluxes, the catalog fluxes are not corrected for Galactic extinction.
Each source's reddening is the \citet{SchlegelFinkbeiner1998} value scaled by $0.86$ \citep{SchlaflyFinkbeiner2011}, and it ranges over $E(B-V)=0.022$--$0.030$ across the field.
The \ha\ selections and the SED fits correct each source with its own reddening, and the reddening and the coefficients are provided with the catalog.

\subsection{Cross-matching and catalog format} \label{sec:catalog_assoc}

One near-infrared, IRAC, or far-infrared source can correspond to several optical components, so the catalog is built from linked tables of objects, detections, measurements, and associations.
The associations are likelihood-ratio matches \citep{SutherlandSaunders1992,Salvato2018}.
Each detection is compared with the detections of the denser image within a search radius of five times the combined positional uncertainty, capped at $0\farcs5$ to $1\farcs5$ depending on the pair of images.
The likelihood ratio is $LR=f(r)~q(m)/n(m)$.
Here $f(r)$ is a circular Gaussian whose width combines the positional uncertainties of the two images, and $n(m)$ is the surface density of the candidates per magnitude.
The magnitude distribution of true counterparts, $q(m)$, is measured as the excess of candidates within the search radii over the background density $n(m)$.
The reliability of a candidate is $R=LR/(\sum LR+1-Q)$, where the sum runs over all candidates of the detection and $Q$ is the integral of $q(m)$.
The most reliable candidate is accepted if $R\geq0.5$, and further candidates with $R\geq0.2$ are kept as components of a blend.
Detections linked by accepted pairs form one object, which takes the position of the first available of its detections in the order MOIRCS $K_s$, NB2083, the $\chi^2$ image, NB387, NB2096, and $K_{Xs}$.
The $541{,}828$ detections of the six images form $456{,}513$ objects, of which $2{,}378$ are detected in $K_{Xs}$ alone.
Repeating the matching with one sample shifted by $1\arcmin$ gives a false-match fraction of $0.15\%$, and $2.2\%$ of the matches are one-to-many.
The catalog (Section~\ref{sec:products}) has one row for each of these objects.

\subsection{Source classification: star, galaxy, and quasar}
\label{sec:catalog_class}

\begin{figure}[t]
\centering
\includegraphics[width=\columnwidth]{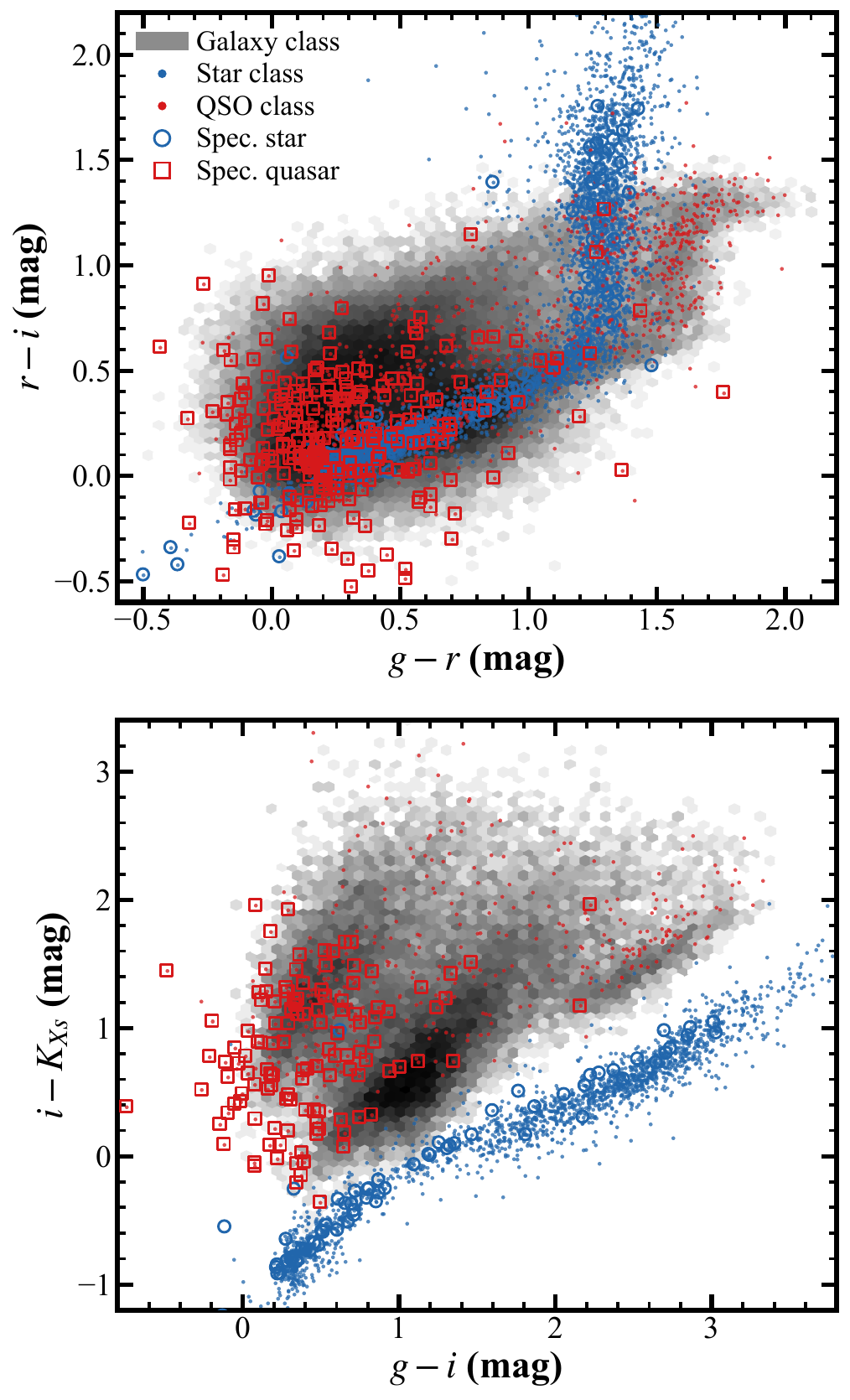}
\caption{Star (blue), galaxy (gray), and QSO (red) classes in two color--color planes, in total-flux colors corrected for Galactic extinction.
\textit{Top:} $r-i$ against $g-r$ for the $112{,}154$ objects with $\mathrm{S/N}>5$ in $g$, $r$, $i$, and $z$.
\textit{Bottom:} $i-K_{Xs}$ against $g-i$ for the $28{,}827$ of these objects with $K_{Xs}$ $\mathrm{S/N}>5$ in the NEWFIRM area.
Open circles mark spectroscopic stars, and open squares mark spectroscopic quasars.
The classes include the overrides by external evidence.}
\label{fig:sgqso}
\end{figure}

Because the field contains the densest known $z\sim2.2$ quasar overdensity, quasars are classified explicitly, so that they are neither lost nor allowed to bias the galaxy photometric redshifts.
Each source receives a class (star, galaxy, or QSO) from four inputs.
The first is the \texttt{class\_star} index of \texttt{SExtractor} in the HSC $i$ image, with point sources defined by \texttt{class\_star}$~>0.95$.
The second is the $\chi^2$ of the best stellar, galaxy, and quasar templates in the fourteen-band \texttt{LePhare} fit of Section~\ref{sec:zphot_z}, which fits the three template sets together for every fitted object, with the AGN and AGN--host templates of \citet{Salvato2009,Salvato2011} as the quasar set.
The third is the \wise\ $W1-W2$ color (Section~\ref{sec:catalog_phot}).
The fourth is external evidence: a significant \textit{Gaia} parallax or proper motion for a star, and a spectroscopic confirmation (Section~\ref{sec:specz}) for a QSO.
A point source is a star if its stellar $\chi^2$ is lowest.
A source is a QSO if it is compact (\texttt{class\_star}$~\geq0.8$) and its quasar $\chi^2$ is lower than its stellar $\chi^2$ and lower than its galaxy $\chi^2$ by more than $0.25$ per fitted band.
Its best quasar template must also contain an AGN component, which excludes the six pure host templates.
A source whose $W1-W2$ color meets the criterion of \citet{Assef2018} at $90\%$ reliability (Section~\ref{sec:vas_ir}) is also a QSO.
The remaining sources form the galaxy class, which can contain host-dominated AGN.
The \textit{Gaia} evidence enters the photometric class (\texttt{class\_photometric}), and the spectroscopic confirmation overrides it to give \texttt{class\_adopted}.
Before the spectroscopic override and measured against the spectroscopic classes, the star class has a purity of $100\%$ and a completeness of $96.0\%$ ($429$ stars).
The galaxy class has a purity of $96.1\%$ and a completeness of $98.9\%$ ($2{,}374$ galaxies).
The QSO class has a purity of $84.7\%$ and a completeness of $67.5\%$, measured on the $246$ spectroscopic quasars inside the four-band area with a consistent spectroscopic class.
Because the galaxies with spectra represent the field galaxies that can enter the QSO class only at $i<21$, the $214$ of the $1{,}415$ objects in the photometric QSO class at $i<21$ carry the flag \texttt{flag\_qso\_purity\_calibrated}.
Of the other $1{,}201$, fainter or without an $i$ measurement, $88$ are spectroscopic quasars, and \texttt{flag\_qso\_faint} marks the remaining $1{,}113$, whose purity is not measured.
Figure~\ref{fig:sgqso} shows the classes in two color--color planes.

\subsection{Quality flags and the recommended sample} \label{sec:catalog_flow}

\begin{deluxetable}{lp{1.45in}r}
\tabletypesize{\footnotesize}
\tablecaption{Selection of the recommended optical sample and its cleaner subsets from the $\chi^2$ detection list.
\label{tab:selflow}}
\tablewidth{0pt}
\tablehead{
\colhead{Step} & \colhead{Cut} & \colhead{$N$}
}
\startdata
Detection list    & \texttt{SExtractor} detections on the masked $\chi^2(griz)$ image & 425{,}670 \\
Significance      & $\texttt{peak\_snr}\geq4.5$ and $L>0$ (recommended sample) & 285{,}537 \\
Flag 0            & inside the four-band area, $\geq3$ unflagged bands & 222{,}241 \\
Not near a star   & no near-star flag (Section~\ref{sec:catalog_det}) & 217{,}042 \\
By class          & star (Section~\ref{sec:catalog_class}) & 4{,}637 \\
                  & QSO & 1{,}069 \\
                  & galaxy & 211{,}336 \\
\enddata
\tablecomments{Each step applies to the sample of the step above it, and the three class rows divide the last sample.}
\end{deluxetable}

Table~\ref{tab:selflow} lists the steps from the $\chi^2$ detection list to the recommended sample and to cleaner subsets of it, with the number of sources that pass each step.
Every $\chi^2$ detection also carries a quality flag.
Flag~0 marks detections with no masked pixel in any of $g$, $r$, $i$, and $z$, a valid $2\arcsec$ magnitude in at least three of $grizy$, and no \texttt{SExtractor} flag in at least three of $g$, $r$, $i$, and $z$.
The last condition excludes blended detections and detections with bright neighbors.
Flag~1 marks detections truncated by the edge of the image, and flag~2 marks the rest: detections measured in fewer than three bands, detections with a masked pixel, and detections with fewer than three unflagged bands, almost all of them blends.
Selections that must avoid blends and masked pixels can also require flag~0, and selections affected by bright-star halos can further require no near-star flag (Table~\ref{tab:selflow}).
The object table carries a separate \texttt{optical\_coverage\_flag} column that tests only the position.
Of the $456{,}513$ objects, $443{,}070$ lie inside the area covered by all of $g$, $r$, $i$, and $z$ (\texttt{optical\_coverage\_flag}$~=0$), and the others lie where only some (2) or none (3) of these bands have usable data.

\section{Photometric Validation} \label{sec:validation}

We validate the photometry against an external catalog and between measurement methods, and we report the blend fraction and the number counts.

\subsection{External and internal consistency}
\label{sec:val_ext}\label{sec:val_internal}\label{sec:app_extphot}

In nominal $2\arcsec$ apertures on the native images, our magnitudes of the sources matched uniquely within $1\arcsec$ with $\mathrm{S/N}\geq5$ are $0.008$--$0.011$~mag fainter in the median than those of the HSC-SSP PDR3 pipeline catalog \citep{Bosch2018,Aihara2022} at $19$--$22$~mag.
Most of the difference comes from our slightly smaller aperture ($1\farcs999$ against $2\farcs016$).
Toward the turnover of the matched-source counts, our local background subtraction removes a further $14$--$39$~nJy per aperture ($g$ to $y$).
In $12\arcsec$ apertures, unsaturated stars agree with their VHS predictions to $0.005$~mag in $K_{Xs}$, $0.010$~mag in $J_X$, $0.03$~mag in MOIRCS $K_s$, and $0.008$~mag in NB2096.
The catalog applies no further zero-point correction.
Within the catalog, the Kron minus matched-$2\arcsec$-aperture magnitude of point sources is $-0.14$ to $-0.23$~mag, depending on the band, and constant within $0.04$~mag from 19~mag to the depth in $r$, $i$, $z$, $y$, and $K_s$.
For the common $1\farcs05$ target PSF, the $2\arcsec$ aperture misses $0.30$~mag of point-source light, and the native-image Kron aperture recovers part of it, so the scaling of the total fluxes to the $i$-band Kron flux restores $0.23$~mag of this loss (Section~\ref{sec:catalog_phot}).

\textit{Uncertainties.}
For faint isolated sources, the catalog uncertainties are $0.94$--$1.07$ times the empty-aperture noise at their positions in the HSC, CFHT $u$, and NEWFIRM bands.
In MOIRCS $K_s$ and NB2083, whose single factor is set by the scatter over the whole mosaic, they are $1.25$ and $1.42$ times this noise.
Two pairs of images measure the same band: the HSC-SSP $g$ image and the deep $g$ stack, and MOIRCS $K_s$ and NEWFIRM $K_{Xs}$ in the MOIRCS area.
For isolated sources at $3\leq\mathrm{S/N}<20$ the normalized differences $(f_1-f_2)/\sqrt{\sigma_1^2+\sigma_2^2}$ have an NMAD of $0.84$ ($g$, $83{,}332$ sources) and $0.92\pm0.03$ ($K$, $1{,}290$ sources), against unity for correct uncertainties.
Faint sources below the detection limit add the same noise to both images and cancel in the difference, so these widths test only the part of the uncertainty that differs between the images.

\subsection{Blending} \label{sec:val_blend}

Sources closer together than the color aperture contaminate each other's photometry.
In the detection catalog, $1.5\%$ of the $\chi^2$ detections have another detection within $1\arcsec$, the radius of the $2\arcsec$ color aperture, and $22.3\%$ have one within $2\arcsec$.
To measure blends that the detection does not resolve, we injected $32{,}000$ galaxies with COSMOS2020 colors and sizes at random positions into the HSC images and repeated the extraction.
With the injections weighted to the field magnitude distribution at $20\leq i<26.5$, $23.0\%$ ($22.6$--$23.3\%$) of the injected galaxies are recovered in one segment together with an existing detection, so their photometry is that of the blend.

\subsection{Number counts} \label{sec:val_counts}

\begin{figure*}
\centering
\includegraphics[width=\textwidth]{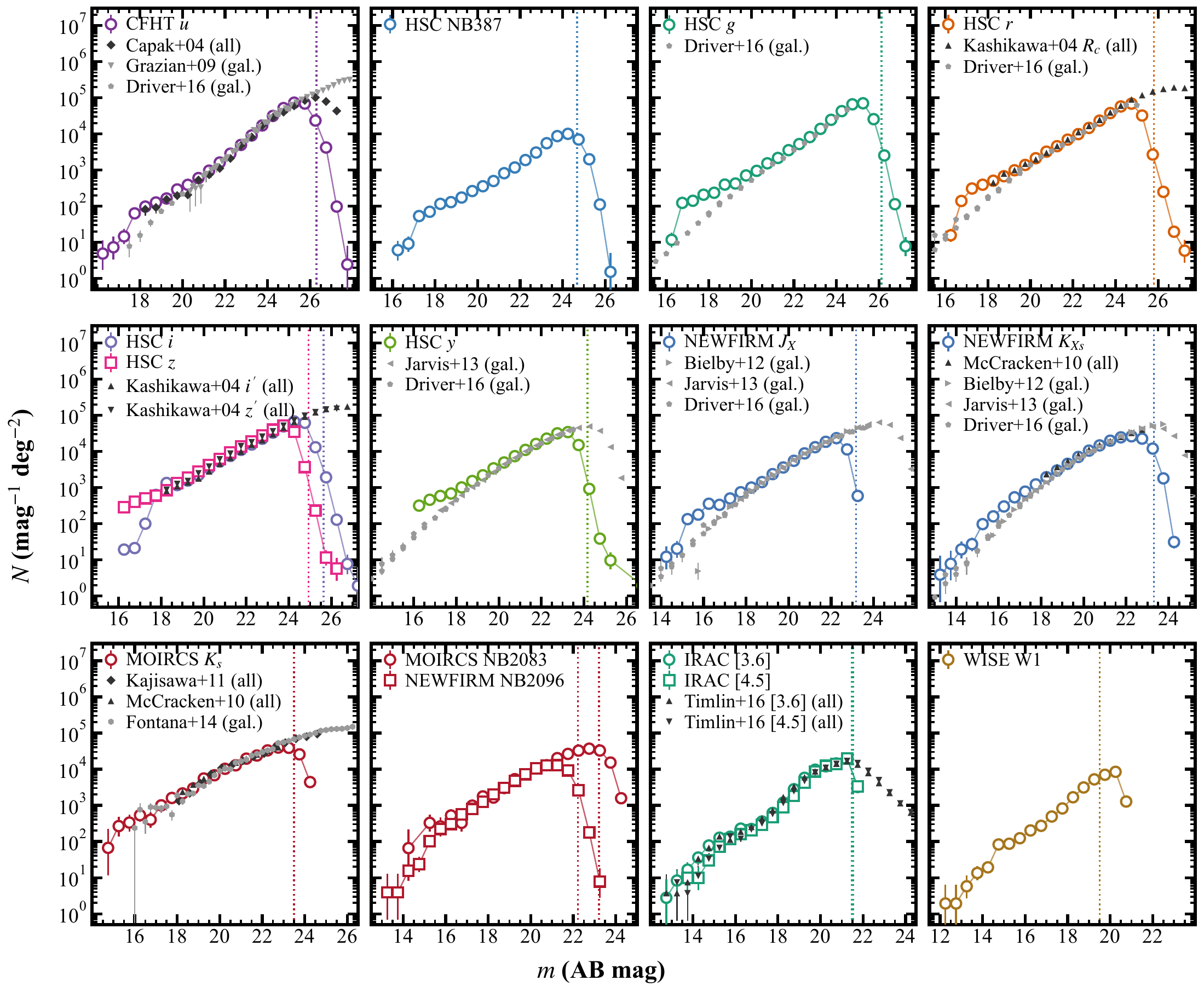}
\caption{Differential number counts at $\mathrm{S/N}>5$ in $0.5$~mag bins, per square degree of the area on which each band is counted, without star, completeness, or Galactic-extinction corrections (open colored symbols with Poisson $1\sigma$ intervals; \citealt{Gehrels1986}).
Dotted vertical lines mark the $5\sigma$ point-source depths of Table~\ref{tab:bands}.
Small filled gray symbols are comparison counts on the AB scale, labeled by first author and year.
Dark gray symbols are counts of all objects from \citet{Capak2004}, \citet{Kashikawa2004}, \citet{McCracken2010}, \citet{Kajisawa2011}, and \citet{Timlin2016}, and light gray symbols are galaxy counts, which exclude the stars that ours include, from \citet{Grazian2009}, \citet{Jarvis2013}, \citet{Driver2016}, \citet{Bielby2012}, and \citet{Fontana2014}.
The counts of \citet{Grazian2009} and \citet{Fontana2014} are corrected for incompleteness.
The IRAC comparison is the SpIES catalog counted on the same sky.
Every panel spans $12.5$~mag, so the count slopes can be compared between panels.}
\label{fig:numbercounts}
\end{figure*}

Figure~\ref{fig:numbercounts} shows the differential number counts in fifteen bands for measurements at $\mathrm{S/N}>5$, per square degree of the area on which each band is counted.
The counts use the $\chi^2$ catalog in $u$ and HSC $grizy$, the self-detected NB387, NB2083, and NB2096 samples, the $K_{Xs}$ detections in $J_X$ and $K_{Xs}$, the MOIRCS $K_s$ detections, the prior-based IRAC photometry, and the AllWISE W1 counterparts.
Where published counts in similar filters exist, ours agree with them to within $17\%$, summed over the magnitude ranges of each comparison (Figure~\ref{fig:numbercounts}).
No tabulated counts exist at these depths for the narrowbands or W1.
The counts turn over $0.2$--$1.4$~mag brighter than the point-source depths of Table~\ref{tab:bands}, because most counted sources are extended and an extended source reaches $\mathrm{S/N}=5$ at a brighter total magnitude than a point source.
The W1 counts continue past the W1 depth because their signal-to-noise ratio is that of the AllWISE profile fits.
The counts are not corrected for incompleteness.

\subsection{Known issues} \label{sec:known_issues}

The following properties of the catalog affect how some of its columns are used.
\begin{enumerate}
\item At the $\chi^2$ positions, the NEWFIRM aperture fluxes are measured on the native images, so colors between them and the PSF-matched bands should be formed from total fluxes.
\item In the northeast corner of the central NEWFIRM pointing (R.A.\ $32\fdg63$--$32\fdg71$, decl.\ $+1\fdg04$--$+1\fdg07$), stars are offset from \textit{Gaia} by up to $0\farcs36$ in $J_X$, $0\farcs32$ in NB2096, and $0\farcs10$ in $K_{Xs}$.
The NEWFIRM colors of sources there are therefore less certain than elsewhere.
The catalog flags the 929 objects in this region (\texttt{flag\_newfirm\_astrometry}).
\item The HSC $griz$ exposures were taken in 2016--2019 and the $u$ and near-infrared images in 2024--2025, so the photometry of variable sources mixes epochs.
The rest-frame ultraviolet variability of the \textit{Cosmic Himalayas} quasars on timescales of years \citep{Lin2026} is not included in the error floors of the fits (Section~\ref{sec:zphot_z}).
\end{enumerate}

\section{Photometric Redshifts and Physical Properties} \label{sec:zphot}

We fit the total fluxes of Section~\ref{sec:catalog} with two photometric-redshift codes, in the thirteen broad bands from \galex\ FUV to IRAC 4.5~\micron\ and in NB387, and fit the spectral energy distributions (SEDs) at fixed redshift.
We test the redshifts on field spectroscopic redshifts and on COSMOS2020 galaxies degraded to our depths, and the stellar masses and star-formation rates on the same galaxies, on far-infrared stacks, and against star-forming galaxies of COSMOS2020 and COSMOS2025.

\subsection{Spectroscopic-redshift compilation} \label{sec:specz}

\begin{deluxetable*}{lrrrcl}
\tabletypesize{\footnotesize}
\setlength{\tabcolsep}{4pt}
\tablecaption{Spectroscopic redshifts in the field.
\label{tab:specz}}
\tablewidth{0pt}
\tablehead{
\colhead{Source} & \colhead{$N_{\rm spec}$} & \colhead{$N_{\rm pass}$} &
\colhead{$N_{\rm obj}$} & \colhead{$z$ range} & \colhead{Content}
}
\startdata
DESI DR1      & 2{,}775 & 2{,}318 & 2{,}286 & $<0.01$--4.10 & galaxies, quasars, and stars \\
HETDEX        & 2{,}021 & 1{,}420 & 579     & 0.01--3.50    & single-line [O~{\sc ii}] and \lya\ emitters \\
SDSS DR18     & 1{,}221 & 891     & 872     & $<0.01$--5.73 & galaxies, quasars, and stars \\
SDSS DR16Q    & 160     & 160     & 158     & 0.25--3.16    & quasars \\
Keck/MOSFIRE  & 60      & 33      & 31      & 2.15--3.30    & \ha\ emitters, quasars, and \oiii\ emitters \\
\hline
Compilation   & 6{,}237 & 4{,}822 & \nodata & \nodata   & $4{,}102$ distinct sources with a secure redshift ($3{,}618$ extragalactic) \\
\enddata
\tablecomments{$N_{\rm spec}$: spectra in the field.
$N_{\rm pass}$: spectra that pass the quality criteria of their survey (Section~\ref{sec:specz}).
$N_{\rm obj}$: passing spectra within $1\arcsec$ of a catalog object.
Surveys observe some sources repeatedly and overlap one another, and the compilation also counts sources without a catalog object within $1\arcsec$, so the $N_{\rm obj}$ column does not sum to it.
The redshift range is that of the passing spectra, including stars at $z\approx0$.
Keck/MOSFIRE observed 58 targets and 2 serendipitous sources (Section~\ref{sec:specz}).}
\end{deluxetable*}

We compile the spectroscopic redshifts (\zspec) in the field from five sources (Table~\ref{tab:specz}): DESI DR1 \citep{DESI2025}, the SDSS DR18 spectroscopic catalog \citep{Almeida2023}, the SDSS DR16 quasar catalog \citep[DR16Q;][]{Lyke2020}, Public Source Catalog~2 \citep{MentuchCooper2026} of the Hobby--Eberly Telescope Dark Energy Experiment \citep[HETDEX;][]{Gebhardt2021}, and our Keck/MOSFIRE \citep{McLean2012} $K$-band spectroscopy of the \ha\ emitters.
A spectrum passes, and its redshift is secure, if it meets the criteria of its survey and class.
These are the DR1 validation criteria of its DESI target class, \texttt{ZWARNING}$~=0$ on a good SDSS DR18 plate, the final DR16Q quasar flag, the published HETDEX confidence thresholds for [O~{\sc ii}] and \lya\ emitters, and, for MOSFIRE, a line identification confirmed in CHEF~II.
Spectra are matched to the cross-matched objects (Section~\ref{sec:catalog_assoc}) within $1\arcsec$, keeping one secure redshift per object in the order of precedence ALMA CO(4--3), MOSFIRE, DESI, DR16Q, DR18, HETDEX [O~{\sc ii}], and HETDEX \lya.
The CO(4--3) redshifts are the systemic redshifts of CH-Q01, CH-Q09, and CH-Q11, observed in ALMA program 2025.1.01243.S and presented in a paper of the CHEF series in preparation.
For the broad-absorption-line quasar CH-Q08, whose DESI template fit misplaces the broad lines, we adopt the DR16Q catalog redshift \citep{Lyke2020}.
Of the $6{,}237$ spectra, $4{,}822$ pass, giving $4{,}102$ distinct sources with a secure redshift, of which $3{,}618$ are extragalactic and $3{,}024$ are fitted in Section~\ref{sec:zphot_z}.

The Keck/MOSFIRE program of 2025 December (U239; PI: N.~Reddy) targeted 54 narrowband-selected galaxies and 4 quasars.
Its spectra confirm 24 \ha\ emitters of the structure and 3 \oiii\ or \hb\ emitters at $z\approx3.2$--$3.3$, and measure the \ha\ lines of the 4 quasars.
The other 27 targets show a tentative \ha\ line (3) or no line (24), and two further galaxies on the slits are \oiii\ emitters at $z=3.10$ and $3.14$.
CHEF~II presents the spectra and their classification.

\subsection{Photometric-redshift methods} \label{sec:zphot_z}

We fit the total fluxes of the cross-matched objects (Section~\ref{sec:catalog_assoc}), corrected for Galactic extinction, in fourteen bands: \galex\ FUV and NUV, CFHT $u$, HSC $grizy$ and NB387, NEWFIRM $J_X$ and $K_{Xs}$, MOIRCS $K_s$, and IRAC 3.6 and 4.5~\micron.
The \galex\ and IRAC fluxes are prior-based PSF fits, and the NEWFIRM $J_X$ and $K_{Xs}$ total fluxes include an encircled-energy correction (Section~\ref{sec:catalog_phot}).
Every measured flux enters as a Gaussian measurement at its signed value, so faint and negative fluxes constrain the fit.
A band without a measurement, or measured on masked pixels (Section~\ref{sec:bandqa_mask}), is left out rather than set to zero.
Model fluxes use the photon-counting response of each band.
Of the $443{,}070$ objects inside the $griz$ area (Section~\ref{sec:catalog_flow}), we fit the $412{,}777$ observed in at least five bands.

\texttt{LePhare} \citep{Arnouts1999,Ilbert2006} provides the primary redshift, because it gives the lower outlier fraction on the field validation half ($2.0\%$ against $5.8\%$ for \texttt{EAZY}; Section~\ref{sec:zphot_acc}) and its configuration follows that of COSMOS2020, on which our degraded-photometry tests are built.
As in COSMOS2015 and COSMOS2020 \citep{Laigle2016,Weaver2022}, \texttt{LePhare} fits the COSMOS galaxy templates, with emission lines added, and a stellar library to every object, and, as in COSMOS2020, the same run fits the quasar and AGN--host templates of \citet{Salvato2009,Salvato2011}.
The galaxy templates use the Small Magellanic Cloud \citep[SMC;][]{Prevot1984} and \citet{Calzetti2000} attenuation laws at $\ebv=0$--$0.5$, the second also in two versions with a 2175~\AA\ bump.
The quasar templates use the SMC law over the same range, in steps of $0.05$.
The redshift grid runs from $z=0$ to $10$ in steps of $0.005$.
The galaxy templates are limited to absolute magnitudes $-24<M_g<-5$, which excludes bright quasars from the galaxy solutions at their redshifts.
The error floors are $0.02$~mag in the optical broad bands, $0.05$~mag in NB387, the near-infrared, the \galex\ bands, and IRAC 3.6~\micron, and $0.10$~mag at 4.5~\micron.
As in COSMOS2020, the catalog point estimate is the median of the profile likelihood, built from the minimum $\chi^2$ over templates at each redshift, with the 68\% and 95\% intervals of this likelihood.
\texttt{EAZY} \citep{Brammer2008} is the comparison code, with its 12 default templates built with the Flexible Stellar Population Synthesis code \citep{Conroy2009}, its template-error function, a $5\%$ error floor (after a $0.05$~mag floor in NB387), and the redshift range $0.01<z<10$.
Neither code applies a prior on apparent magnitude or on the redshift distribution.
The quasar templates give a second profile likelihood for every object, whose median is the quasar-template redshift.
On the 253 spectroscopic quasars, this redshift has an outlier fraction of $40\%$ and $\sigma_{\rm NMAD}=0.13$, against $66\%$ and $0.29$ for the galaxy templates.
The catalog gives this redshift for every fitted object as \texttt{z\_qso}, with its 68\% interval, for use with the QSO class (Section~\ref{sec:catalog_class}).

For \texttt{LePhare} we apply the template offsets of the iterative calibration \citep{Ilbert2006}, run on the field calibration half (Section~\ref{sec:zphot_acc}), because they lower its outlier fraction on the validation half in the fits of the thirteen broad bands.
They are $+0.07$ and $-0.20$~mag in \galex\ FUV and NUV, $-0.011$~mag in NB387, and at most $0.05$~mag in the optical and near-infrared bands and $0.06$~mag in IRAC.
\texttt{EAZY} uses no broadband offsets and scales NB387 by $0.995$.\label{sec:zphot_zp}

\subsection{Accuracy of the photometric redshifts} \label{sec:zphot_acc}

We quantify the accuracy with the residual $\delta=(\zphot-z_{\rm ref})/(1+z_{\rm ref})$ against a reference spectroscopic redshift $z_{\rm ref}$.
The bias $b$ is the median of $\delta$, the normalized median absolute deviation is $\sigma_{\rm NMAD}=1.4826\times\mathrm{median}(|\delta-b|)$, and the outlier fraction $\eta$ is the fraction of objects with $|\delta|>0.15$, counting failed fits as outliers.
Intervals are the 16th to 84th percentiles of $1{,}000$ bootstrap resamplings of the galaxies, except for field outlier fractions, which carry $68\%$ Wilson intervals.

\begin{figure*}
\centering
\includegraphics[width=0.82\textwidth]{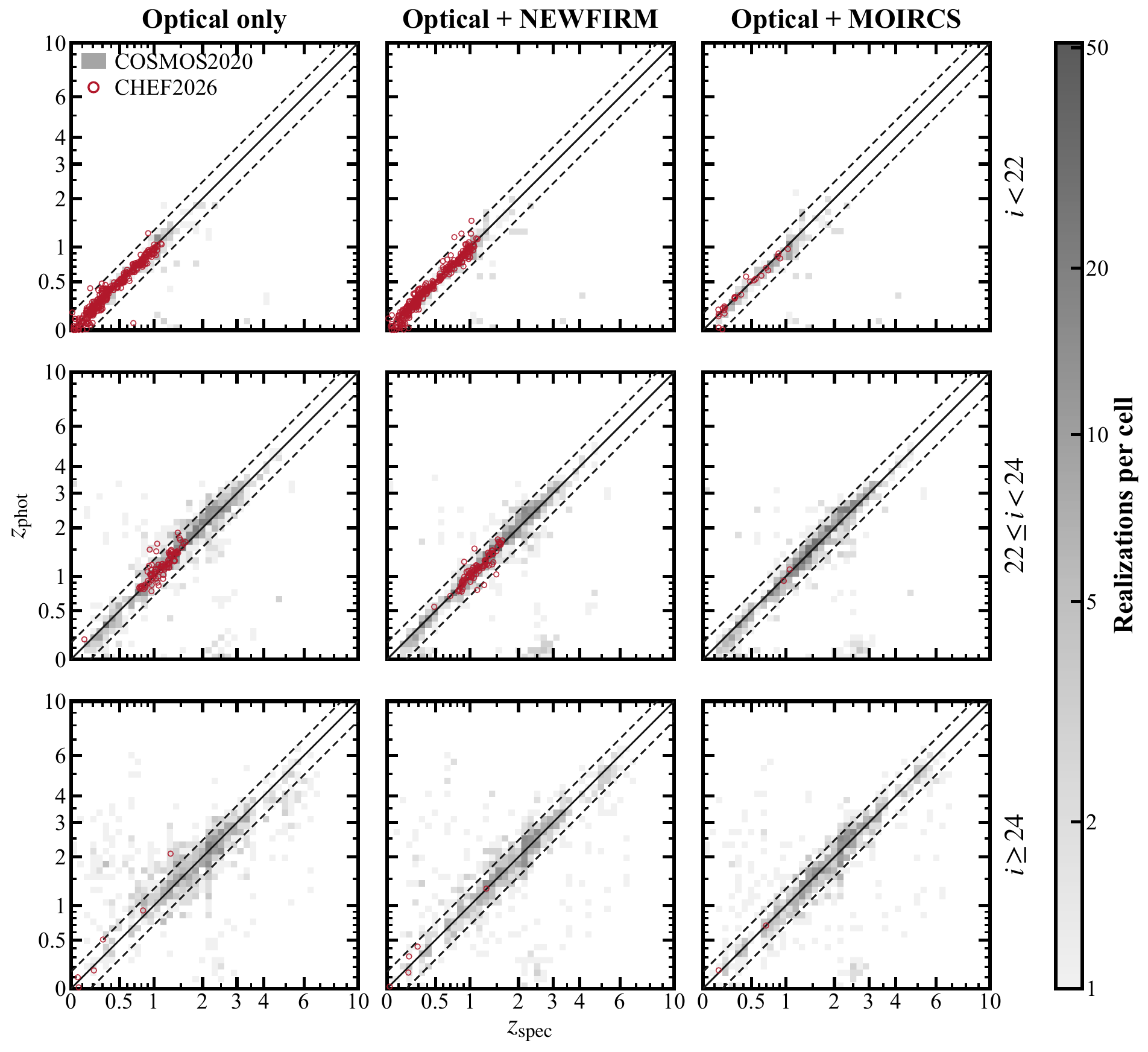}
\caption{\texttt{LePhare} photometric against spectroscopic redshift, in bins of total $i$ magnitude (rows) and near-infrared coverage (columns: optical only, optical plus NEWFIRM, optical plus MOIRCS), with both axes logarithmic in $1+z$.
The gray shading is the density, in cells uniform in $\log(1+z)$ with 45 per axis, of the COSMOS2020 galaxies of the validation part degraded to our depths and coverage.
Each galaxy has up to two realizations per coverage class.
The red circles are the 941 field galaxies of the validation half with a \texttt{LePhare} redshift, each in the panel of its own $i$ magnitude and coverage.
The 13 of them undetected in $i$ are in the bottom row.
The solid line is the one-to-one relation, and the dashed lines mark $|\delta|=0.15$.}
\label{fig:zphot_density}
\end{figure*}

\textit{Field spectroscopic galaxies.}
The DESI and SDSS galaxy redshifts of fitted objects give $1{,}880$ galaxies at $z=0.01$--$1.62$, split at random into calibration and validation halves of 938 and 942.
On the validation half (Figure~\ref{fig:zphot_density}), \texttt{LePhare} reaches $\eta=2.02\%$ ($1.61$--$2.53\%$), $\sigma_{\rm NMAD}=0.0298$, and $b=-0.002$, and \texttt{EAZY} $\eta=5.84\%$ ($5.12$--$6.65\%$), $\sigma_{\rm NMAD}=0.0329$, and $b=+0.008$.
These bright galaxies at $z<1.62$ test the low-redshift field, not $z\approx2$.
Over the field, the two codes agree within $|\Delta z|/(1+z)<0.15$ for $86.5\%$ of the objects with both redshifts.
The redshift differences of close pairs at $2\farcs5<\theta<15\arcsec$ \citep{QuadriWilliams2010}, measured relative to those of pairs at $2'$--$4'$, give $\sigma_z/(1+z)=0.13\pm0.05$ at $\zphot>1.5$ and $i<24$.
The degraded COSMOS2020 galaxies in the same ranges have $\sigma_{\rm NMAD}=0.064$ with $19\%$ outliers.
The close-pair width counts the moderate outliers that the NMAD down-weights, and at $24\leq i<25$ the two estimates agree at $0.10$.

\begin{figure*}
\centering
\includegraphics[width=0.96\textwidth]{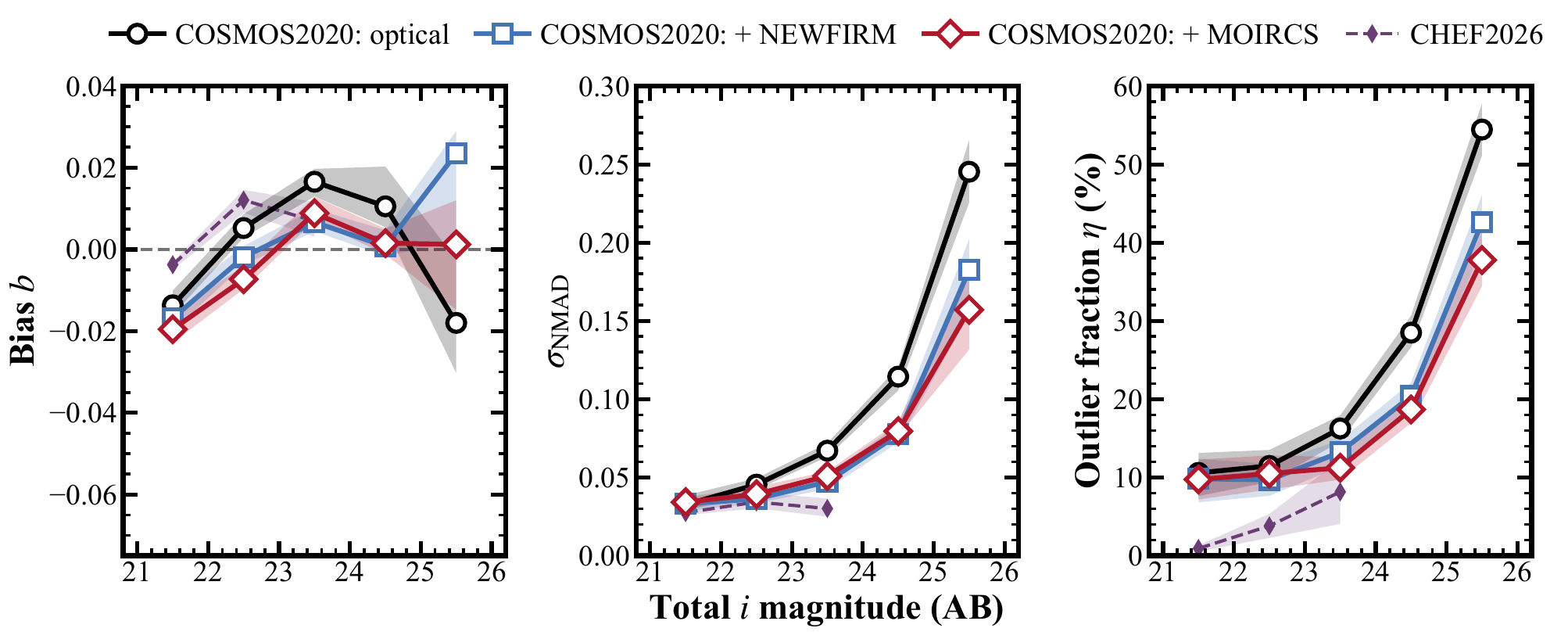}
\caption{Accuracy of the \texttt{LePhare} photometric redshifts against brightness for the COSMOS2020 galaxies of the validation part degraded to our depths (Section~\ref{sec:zphot_acc}).
From left to right, the panels show the bias $b$, the scatter $\sigma_{\rm NMAD}$, and the outlier fraction $\eta$ ($|\delta|>0.15$ or a failed fit) against the total $i$ magnitude of the COSMOS2020 galaxy, for optical only (black), optical plus NEWFIRM (blue), and optical plus MOIRCS (red) coverage.
Each galaxy has up to two noise realizations per class and equal weight.
The purple dashed curve shows the field galaxies of the validation half over all coverages, in magnitude bins with at least 20 galaxies.
The uncertainty bands are the 16th--84th percentiles of $1{,}000$ bootstrap resamplings of the galaxies.}
\label{fig:zphot_depmag}
\end{figure*}

\textit{COSMOS galaxies at the depth of this field.}
We test the accuracy at $z\approx2$ on COSMOS2020 galaxies \citep{Weaver2022} with secure spectroscopic redshifts, whose photometry is transformed to our filters and degraded to our depths.
We use COSMOS2020 rather than COSMOS2025 \citep{Shuntov2025} because it measures the same ground-based optical, near-infrared, and IRAC bands as ours, so its photometry maps directly onto our filters.
Our $u$, $grizy$, $J_X$, $K_s$ and $K_{Xs}$, IRAC, and \galex\ bands take the COSMOS2020 CFHT $u$, HSC $grizy$, UltraVISTA $J$ and $K_s$, IRAC, and \galex\ fluxes, respectively, and NB387 the flux of the CHORUS NB387 catalog of the COSMOS field \citep{Inoue2020}.
Where the passbands differ, each galaxy receives a color term from the templates that fit its COSMOS2020 photometry at its spectroscopic redshift, integrated through both measured passbands.
The fluxes are corrected for Galactic extinction and converted to our $2\arcsec$ apertures and $i$-band Kron normalization with an exponential profile convolved with each PSF.
Each galaxy is realized twice in each of three coverage classes (optical only, optical plus NEWFIRM, optical plus MOIRCS), each time with the band coverage, including \galex\ and IRAC, of a field object of that class within $0.25$~mag in total $i$.
Gaussian noise then raises the uncertainty of every band to the value our catalog gives a source of the same flux at that object's position, and the limit rule of Section~\ref{sec:catalog_phot} applies.
These galaxies are split at random, by galaxy and within bins of redshift and $i$ magnitude, into calibration ($60\%$), validation ($20\%$), and test ($20\%$) parts of $3{,}301$, $1{,}102$, and $1{,}100$ galaxies, with all realizations of a galaxy in one part.
We chose the fitting settings (code, apparent-magnitude priors, template set, treatment of flux and template errors, near-infrared aperture correction, redshift estimator, and band set) on the calibration and validation parts and applied them once to the test part.
The accuracies below are those of the catalog configuration (Section~\ref{sec:zphot_z}) on the $1{,}102$ validation galaxies.
Over all redshifts \texttt{LePhare} gives $\eta=19.3\%$ ($18.5$--$20.2\%$), $\sigma_{\rm NMAD}=0.0671$, and $b=+0.001$.
At $2\leq z<2.5$ (192 galaxies) it gives $\eta=21.7\%$ ($19.6$--$23.4\%$), $\sigma_{\rm NMAD}=0.0961$, and $b=+0.010$, and \texttt{EAZY} gives $\eta=23.5\%$ and $25.6\%$ in the two samples.
The test part, which played no role in the choice of the settings, was fitted once with the thirteen broad bands and the same settings but without the photon-counting responses and the \texttt{LePhare} template offsets of Section~\ref{sec:zphot_z}.
It gives \texttt{LePhare} $\eta=19.8\%$ ($18.9$--$20.6\%$) and $\sigma_{\rm NMAD}=0.066$ over all redshifts and $\eta=22.4\%$ ($20.6$--$24.5\%$) and $\sigma_{\rm NMAD}=0.110$ at $2\leq z<2.5$, within one percentage point in $\eta$ and $0.01$ in $\sigma_{\rm NMAD}$ of the validation part with the same thirteen bands.
The accuracy depends mainly on brightness (Figure~\ref{fig:zphot_depmag}).
The \texttt{LePhare} outlier fraction rises from $10.0\%$ at $i<22$ to $44.9\%$ at $i\geq25$, and $\sigma_{\rm NMAD}$ from $0.0330$ to $0.1904$.
In the thirteen-band fits, near-infrared coverage lowers $\eta$ from $23.1\%$ without it to $17.4\%$ with NEWFIRM and $15.7\%$ with MOIRCS.
Adding IRAC and \galex\ to the nine optical and near-infrared bands lowers it by $2.1$ percentage points ($1.8$--$2.5$), against $1.0$ and $0.8$ points for IRAC and \galex\ alone.
Figure~\ref{fig:zphot_density} shows how the redshifts of faint galaxies fail.
Without near-infrared bands they are compressed toward $\zphot\approx1$--$2$, and with them a group at $i\geq22$ and $z_{\rm ref}=2$--$2.7$ falls onto a low-redshift alias at $\zphot<0.5$.

The outlier fraction falls steeply with the number $n_3$ of the thirteen broad bands in which an object has $\mathrm{S/N}\geq3$ in its signed flux, from $57\%$ with none to $7.7\%$ with twelve.
Each \texttt{LePhare} redshift carries a reliability class based on $n_3$: good ($n_3\geq7$), medium ($5$--$6$), or low ($\leq4$), with COSMOS outlier fractions of $8$--$14\%$, $21$--$26\%$, and $40$--$57\%$.
Of the fitted objects, $59{,}846$ are good, $100{,}318$ medium, and $252{,}423$ ($61\%$) low, and 190 have no valid \texttt{LePhare} likelihood.
Only 3 of the 942 field validation galaxies are in the low class, so its outlier fraction comes from COSMOS.
The nominal $68\%$ and $95\%$ \texttt{LePhare} likelihood intervals contain the reference redshift for $53.3\%$ and $80.3\%$ of the $1{,}102$ COSMOS validation galaxies, so they are too narrow.
The catalog therefore also gives intervals calibrated by temperature scaling of the likelihood, fitted on the field calibration half and so independent of COSMOS.
These contain the reference redshift for $72.8\%$ ($71.9$--$73.6\%$) and $93.0\%$ ($92.3$--$93.6\%$) of the same galaxies, and the calibrated \texttt{EAZY} intervals for $69.0\%$ and $91.1\%$.
The width of the nominal intervals still ranks the redshifts.
With $w=(z_{84}-z_{16})/[2(1+\zphot)]$, the outlier fraction of these galaxies rises from $9$--$17\%$ at $w<0.1$ to $77\%$ at $w\geq0.5$.
At $2\leq z<2.5$, one galaxy in five is a \texttt{LePhare} outlier, and the scatter in redshift, $(1+z)\,\sigma_{\rm NMAD}$, is three times the half-width of the photometric-redshift window of the structure ($2.08<z<2.28$; Section~\ref{sec:catalog_member}).
Galaxies with most of their likelihood in this window are therefore only photometric-redshift candidates, and narrowband emitters and spectroscopic redshifts trace the structure.

\begin{figure*}[t]
\centering
\includegraphics[width=0.9\textwidth]{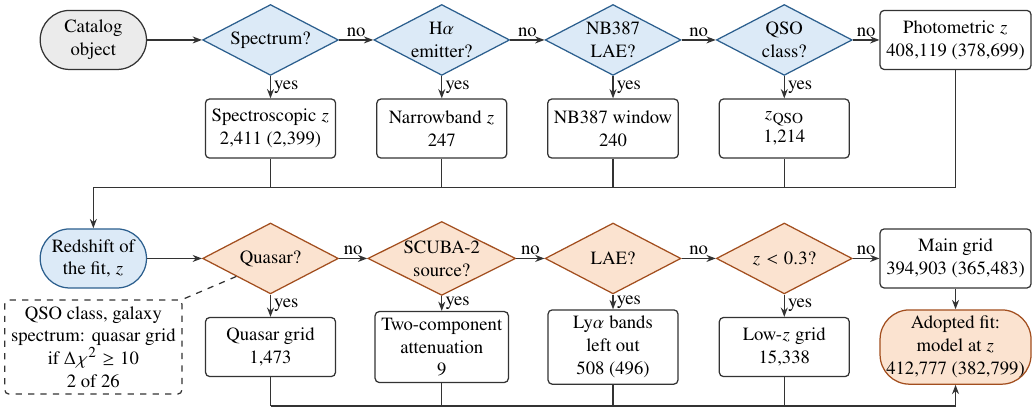}
\caption{Choice of the adopted SED fit of each catalog object.
The first row sets the redshift of the fit, in this order of precedence: a secure spectroscopic redshift, the narrowband redshift of an emitter of the final \ha\ samples, the middle of the NB387 window for an NB387 \lya\ emitter, the quasar-template redshift for the QSO class, and otherwise the photometric redshift.
The second row sets the model from the population of the object by the first of these tests that it passes: a quasar spectrum or the QSO class, the counterpart of a SCUBA-2 source, a \lya\ emitter, and a fit redshift below 0.3.
An object of the QSO class with a galaxy spectrum or a narrow-line \lya\ spectrum (dashed box) is fitted with both grids and passes the first test only if the quasar grid lowers $\chi^2$ by at least 10, as 2 of the 26 such objects do, and the others continue with the galaxy tests.
A quasar stays on the quasar grid, also as a SCUBA-2 counterpart, and leaves out the \lya\ bands as a \lya\ emitter.
The adopted fit is this model at this redshift, and the other fits that apply to an object are released beside it as auxiliary fits.
Numbers are the objects and, in parentheses, the successful fits of each outcome, and the 546 spectroscopic stars and objects without a valid redshift are not fitted (Section~\ref{sec:sed_cigale}).}
\label{fig:sedrule}
\end{figure*}

\textit{Near-infrared objects and narrowband emitters.}
Of the catalog objects detected in MOIRCS $K_s$ or NEWFIRM $K_{Xs}$ but not in the $\chi^2$ image, 295 detected in $K_s$ and $2{,}193$ detected only in $K_{Xs}$ are fitted with the same settings on their forced photometry at the near-infrared position (Section~\ref{sec:catalog_phot}).
The catalog \texttt{LePhare} redshifts give $|\delta|>0.15$ for 5 of the 15 NB2083 \ha\ emitters with a secure galaxy redshift.
The 3 NB2096 emitters with such a redshift and a catalog fit are also NB2083 emitters, and none is an outlier.
The SED fits of the \ha\ emitters therefore use the redshift of their spectrum or, without one, an assumed narrowband redshift (Section~\ref{sec:sed_cigale} and Figure~\ref{fig:sedrule}).
Of the 294 distinct objects of the final \ha\ samples, 16 have no SED fit (10 without linked photometry and 6 outside the $griz$ area).

\begin{figure*}[t]
\centering
\includegraphics[width=0.995\textwidth]{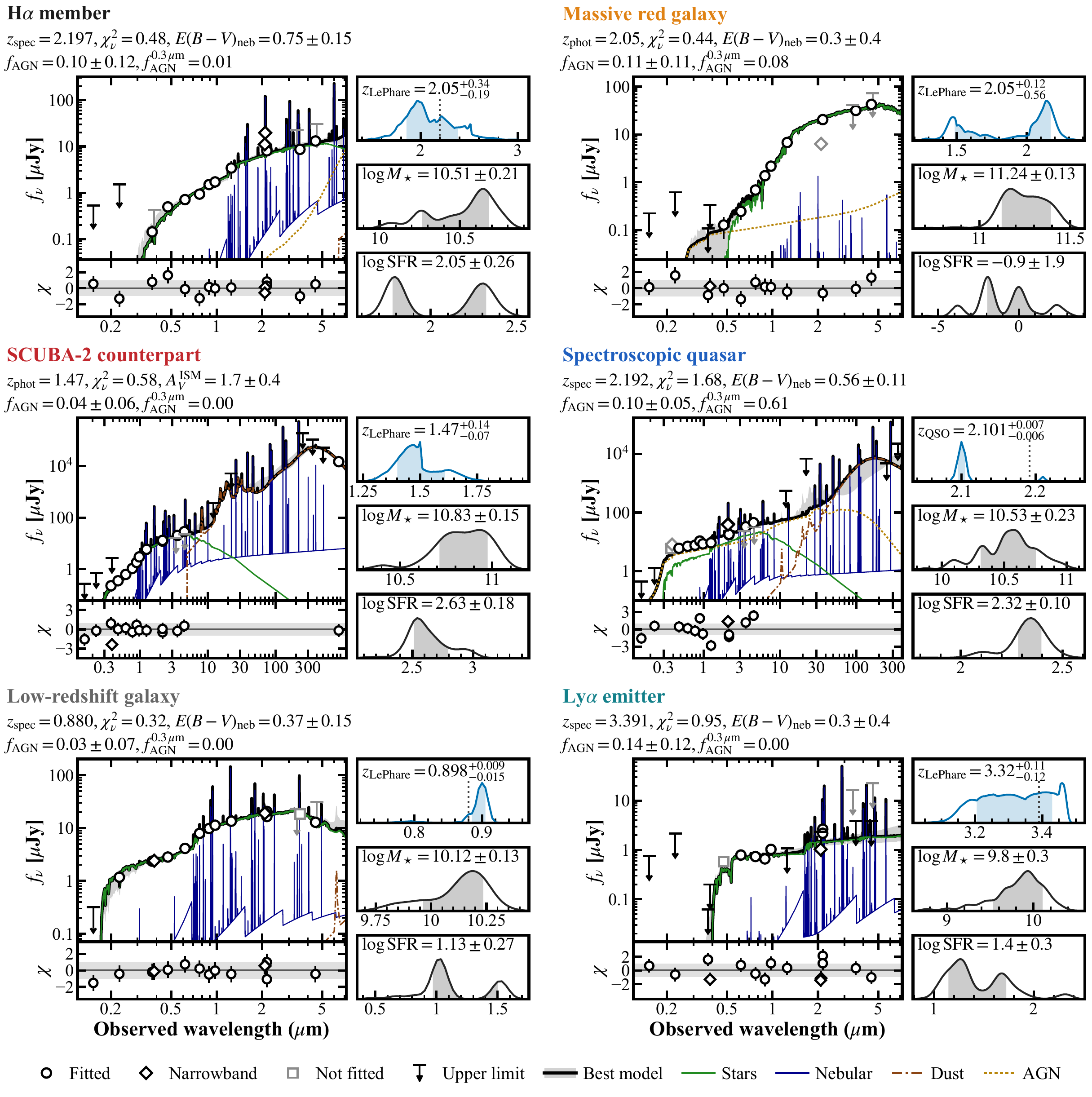}
\caption{\texttt{CIGALE} fits of six example objects: an \ha\ member, a massive red galaxy, the counterpart of the most significant SCUBA-2 source, a spectroscopic quasar, a low-redshift galaxy, and a HETDEX \lya\ emitter \citep{MentuchCooper2026}.
Each panel shows the adopted fit (Figure~\ref{fig:sedrule}): the quasar grid for the quasar, the two-component attenuation for the SCUBA-2 counterpart, the \lya\ treatment for the \lya\ emitter, and the main grid for the others.
Each panel ends at the longest band that constrains the fit.
The best model (black) and its components are drawn at the native \texttt{CIGALE} resolution, with the $68\%$ range of the posterior models in gray.
Gray symbols and arrows are not fitted (among them NB387, the \lya\ bands of the quasar and the \lya\ emitter, and masked IRAC 3.6~\micron).
Black arrows mark fitted bands below $\mathrm{S/N}=2$, drawn at $2\sigma$ while the fits use their signed fluxes, and the fitted upper limits of the mid- and far-infrared bands (Section~\ref{sec:catalog_phot}).
Lower panels show the residuals in units of the fitted error.
Right panels show the photometric-redshift distribution (quasar templates for the quasar) with the spectroscopic redshift dotted, and the posteriors of $\log\mstar$ and $\log\sfr$ with shaded $68\%$ ranges.
$A_V^{\rm ISM}$ is the $V$-band attenuation by the diffuse interstellar medium, $f_{\rm AGN}$ the fraction of the dust luminosity emitted by the AGN, and $f_{\rm AGN}^{0.3\,\mu{\rm m}}$ the AGN share of the model flux at rest-frame 0.3~\micron, which for the quasar comes mostly from the accretion disk.}
\label{fig:sedfit}
\end{figure*}

\begin{figure*}[t]
\centering
\includegraphics[width=\textwidth]{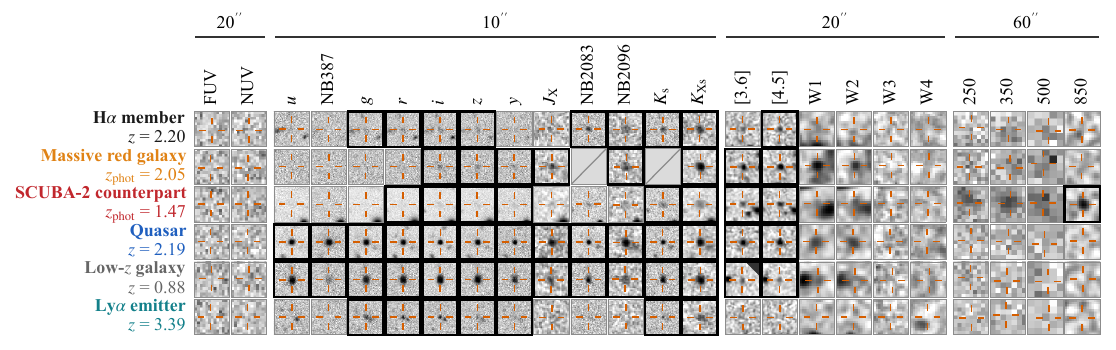}
\caption{Cutouts of the six objects of Figure~\ref{fig:sedfit} in the 24 bands, from \galex\ FUV to SCUBA-2 850~\micron.
Each stamp is cut from the image on which the catalog photometry was measured, without resampling, north up and east left.
The stamp side is $20\arcsec$ for \galex, $10\arcsec$ from $u$ to the $K$ bands, $20\arcsec$ for IRAC and \wise, and $60\arcsec$ for SPIRE and SCUBA-2.
The gray scale is an asinh stretch per stamp, from $2\sigma$ below the local sky to the 99.5th percentile.
Orange ticks mark the catalog position, placed with each band's own astrometry.
Black frames mark bands with catalog $\mathrm{S/N}\geq5$, and gray cells with a diagonal mark bands without data at the position.
Row labels give the class and redshift in the colors of Figure~\ref{fig:sedfit}.
The redshifts of the massive red galaxy and of the SCUBA-2 counterpart are photometric, and MOIRCS has no data at the position of the massive red galaxy.
A dark corner marks a band in which the object lies in the image mask (IRAC 3.6~\micron\ of the low-redshift galaxy).
That measurement is kept in the catalog but not fitted.}
\label{fig:atlas}
\end{figure*}

\subsection{SED fits} \label{sec:sed_cigale}

We fit the SEDs with \texttt{CIGALE} \citep{Noll2009,Boquien2019}, which models the ultraviolet to submillimeter spectrum with energy balance, nebular emission, and an AGN component on a fixed model grid.
Each fitted object has one adopted fit, at a fixed redshift chosen by the evidence and a model chosen by its population (Figure~\ref{fig:sedrule}).
The redshift is a secure spectroscopic redshift where one exists (for three quasars the systemic CO(4--3) redshift of Section~\ref{sec:specz}).
Otherwise, in this order, an emitter of the final \ha\ samples takes an assumed narrowband redshift that places \ha\ at the middle of the half-maximum window of the passband at its position (Section~\ref{sec:survey}), an NB387 \lya\ emitter the middle of the NB387 window ($z=2.177$), and an object of the QSO class its quasar-template redshift (Sections~\ref{sec:catalog_class} and~\ref{sec:zphot_z}).
All other objects take the catalog \texttt{LePhare} redshift.
Where the likelihood of a quasar-template or catalog redshift has more than one peak in its $80\%$ highest-density region, the fit takes the median of the most probable peak.
The model is a quasar grid for the spectroscopic quasars, whose spectra show broad lines, and for the QSO class.
The counterparts of the SCUBA-2 sources take a grid with the two-component attenuation of \citet{CharlotFall2000} (birth clouds and diffuse interstellar medium), and the NB387 and HETDEX \lya\ emitters the main grid without the bands that contain \lya.
Galaxies at $z<0.3$ take a grid adapted to low redshift, and all other objects the main grid.
An object of the QSO class whose spectrum shows a galaxy, or a \lya\ emitter without broad lines, is fitted with both the quasar grid and its galaxy grid and adopts the quasar grid only if that lowers $\chi^2$ by at least 10.
The quasar grid contains every model of the main grid, so only a clear improvement counts.
Of the 26 such objects (23 with a galaxy spectrum and 3 HETDEX \lya\ emitters), two adopt the quasar grid ($\Delta\chi^2=17.0$ and $29.5$), and the others keep their galaxy fits.
All grids exclude at each redshift the models older than the Universe.
Every value of a catalog row comes from its adopted fit, and two catalog columns name the redshift source and the model.
Of the $412{,}777$ fitted objects, $382{,}799$ have a successful fit: $1{,}473$ on the quasar grid (10 of them without the \lya\ bands), 9 with the two-component attenuation, 496 without the \lya\ bands, $15{,}338$ on the low-redshift grid, and $365{,}483$ on the main grid.
An auxiliary table of $455{,}192$ rows, one per object and fit, gives the adopted fits, with a flag, and the other fits we computed.
These are the fits on the main or low-redshift grid at the catalog redshift for every object, and the narrowband-redshift, \lya, quasar-grid, and two-component fits wherever they apply, so that other classifications can be applied.
The fits use the thirteen broad bands, NB387 as a continuum band, and, as fluxes or upper limits, \wise\ W1 and W2 outside the IRAC footprint, W3 and W4, MIPS 24~\micron\ \citep{SEIP2020}, SPIRE 250, 350, and 500~\micron, and SCUBA-2 850~\micron\ (Section~\ref{sec:catalog_phot}).
The fits of the \lya\ emitters leave out every band whose response at $1215.67(1+z)$~\AA\ exceeds $10\%$ of its peak ($u$ below $z\approx2.3$, and $g$ or $g$ and $r$ at $z\geq2.3$), because the models have no \lya\ line.
NB387 is left out of every fit in which $1215.67(1+z)$~\AA\ falls where its response exceeds $10\%$ of its peak.
The narrowbands NB2083 and NB2096 enter every fit at a spectroscopic or narrowband redshift, with the passband at the object's position (Section~\ref{sec:survey}), because at a fixed redshift the model predicts the narrowband flux whether or not \ha\ falls inside the passband.
Their total fluxes follow the broadband convention, the calibrated narrowband-to-continuum flux ratio of the matched $2\arcsec$ aperture (Section~\ref{sec:vas_hae}) times the total flux of $K_s$ or $K_{Xs}$.
Fits at a photometric redshift do not use NB2083 and NB2096, whose windows, about 0.04 wide in redshift, are narrower than its uncertainty.
Adding NB2096 to these fits raises the star-formation rates of the 170 emitters selected only in NB2096 by a median of $0.20$~dex and changes those of the emitters selected only in NB2083 by less than $0.01$~dex in the median.
Fits of the same emitters at their \texttt{LePhare} redshift would give stellar masses $0.05$~dex higher and star-formation rates $0.29$~dex lower in the median.
A $10\%$ model uncertainty is added in quadrature to the photometric errors.
A model enters the likelihood only if the amplitude that scales it to the fluxes is positive, and the fluxes enter at their signed values.
The main grid has $120{,}960$ models per redshift.
They combine \citet{BruzualCharlot2003} populations with a \citet{Chabrier2003} initial mass function at $0.4$ and $1~Z_\odot$ and a delayed-$\tau$ star-formation history with $\tau=0.1$, $0.3$, $0.5$, $1.5$, and $5$~Gyr, ages of $0.05$, $0.1$, $0.2$, $0.35$, $0.5$, $0.75$, $1$, $1.5$, $2$, $2.5$, $3$, $4$, $5$, $7$, $10$, and $13$~Gyr, and no burst.
The short timescales and old ages describe quiescent galaxies at $z\geq1.4$ \citep{DominguezSanchez2011}.
Nebular emission has $\log U=-2$ and no \lya\ line, whose emergent strength is set by resonant scattering that the models do not follow.
Attenuation follows the modified \citet{Calzetti2000} law of \citet{Noll2009}, with power-law slopes $\delta=0$ and $-0.3$ and a nebular color excess $\ebv=0$, $0.1$, $0.3$, $0.6$, $0.9$, $1.2$, or $1.6$, and the stellar continuum has $0.44$ times this color excess.
Dust emission follows \citet{Dale2014} with nine slopes from $\alpha=0.5$ to 4, and the AGN component uses the SKIRTOR models \citep{Stalevski2012,Stalevski2016}, as implemented in \texttt{CIGALE} by \citet{Yang2020}, at inclinations of $30\arcdeg$ and $70\arcdeg$ with AGN fractions of $0$, $0.1$, and $0.3$.
The AGN fraction $f_{\rm AGN}$ is the fraction of the total dust luminosity that the torus and polar dust of the AGN re-emit, so it excludes the emission of the accretion disk.
The quasar grid has AGN fractions of $0$, $0.1$, $0.3$, $0.5$, $0.7$, and $0.9$ ($241{,}920$ models per redshift), and its $1{,}473$ fits have a median $f_{\rm AGN}$ of $0.29$.
The SCUBA-2 counterpart grid replaces the attenuation by the two-component model of \citet{CharlotFall2000}, and adds a burst of $10$~Myr.
For the 118 extended galaxies at $i<22$ whose ratio of $i$-band Kron to $2\arcsec$ flux exceeds five, the fits use this ratio for the total fluxes, with the NEWFIRM fluxes measured on the PSF-matched images, and leave out the prior-based bands, whose total fluxes assume a point source.
For these galaxies the \texttt{EAZY} redshifts are unreliable (8 outliers among the 23 in the validation half of Section~\ref{sec:zphot_acc}, against 1 for \texttt{LePhare}) and carry \texttt{flag\_eazy\_unusable}.
The low-redshift grid adds $\tau=1$, 3, and 10~Gyr and a metallicity of $0.2~Z_\odot$ to the main grid, samples the nebular color excess at 14 values from 0 to 1.6, and has no AGN component.
The adopted stellar masses and star-formation rates are the likelihood-weighted means of $\log\mstar$ and $\log\sfr$ over the grid, and their uncertainties the corresponding likelihood-weighted standard deviations, in dex.
The linear likelihood-weighted means of \mstar\ and \sfr\ exceed these estimates by a median of $0.25$~dex in mass and $0.29$~dex in star-formation rate, because a broad likelihood pulls a linear mean toward its upper tail.
We adopt the logarithmic means because they depend less on depth.
Degrading the photometry of the COSMOS2020 galaxies of Section~\ref{sec:zphot_acc} changes their median $\log\sfr$ by $-0.03$~dex, against $+0.06$~dex for the linear means.
The catalog also gives these linear means and the best-fit values.
For the nebular \ebv\ and the AGN fraction we give the likelihood-weighted mean and standard deviation, and for every fit the $\chi^2$ and the rest-frame colors of Section~\ref{sec:sed_uvj}.
As in COSMOS2020 and the \texttt{CIGALE} fits of COSMOS2025 \citep{Weaver2022,Shuntov2025}, the fits use a fixed redshift, and the quoted mass and star-formation-rate uncertainties exclude the redshift uncertainty.
The catalog objects without a fit (Figure~\ref{fig:sedrule}) are spectroscopic stars, objects without a valid redshift, $29{,}423$ objects for which no model has a positive amplitude (almost all without a band at $\mathrm{S/N}\geq3$), and 9 HETDEX \lya\ emitters that keep too few bands once their \lya\ bands are left out.
The catalog flags the 144 fits with $\chi^2_\nu>10$ and the $68{,}555$ fits without a band at $\mathrm{S/N}\geq3$.
Figure~\ref{fig:sedfit} shows the fits of six example objects, and Figure~\ref{fig:atlas} the same objects in all 24 bands.

A fit carries the flag \texttt{sed\_reliable} if both IRAC bands enter it and at least 7 of the 13 broad bands have $\mathrm{S/N}\geq3$, or if its total $K_{Xs}$ magnitude is brighter than 22.32~AB, the $90\%$ completeness limit of the $K_{Xs}$ detections for point sources and exponential disks (Section~\ref{sec:catalog_compl}).
Every galaxy with $K_{Xs}<22.32$~AB carries the flag by construction, as do $52{,}304$ of the $382{,}799$ fits in all.
We test the flag with the far-infrared emission that the fits predict by energy balance.
For star-forming galaxies above the mass limits of the $K$-limited sample (Table~\ref{tab:sedbench}), we divide the mean background-subtracted flux of the \herschel/SPIRE maps at their positions by the mean predicted flux, in five redshift bins from $z=0.5$ to 3.
Galaxies without a far-infrared detection, whose slope the fits do not constrain, take the \citet{Dale2014} template with $\alpha=2$.
For the galaxies with $K_{Xs}<22.32$~AB, the ratio is 0.93--1.50 at 250 and 350~\micron\ in all five bins and 0.86--0.98 at 500~\micron\ below $z=2.5$.
For all galaxies with the flag, including those fainter than this limit that qualify through IRAC, it is 1.00--1.65 at 250~\micron, and for the galaxies without the flag it is at most 1.10 in the three bands.
We recommend the flagged fits wherever stellar masses, star-formation rates, or dust attenuations are used.
At $2.5\leq z<3$, the 500~\micron\ stack of the 42 galaxies with $K_{Xs}<22.32$~AB is consistent with zero (ratio 0.14, $68\%$ interval $-0.13$ to 0.39).
The SCUBA-2 850~\micron\ stacks of the galaxies with $K_{Xs}<22.32$~AB lie below the predictions, with ratios that fall from 0.69 at $0.5\leq z<1$ to 0.08 at $2.5\leq z<3$.
This is expected if the template with $\alpha=2$ is colder than these galaxies and so overpredicts their Rayleigh--Jeans emission, whereas the 250 and 350~\micron\ bands sample the dust peak that carries the energy balance.
The 850~\micron\ fluxes of individual submillimeter sources depart in the other direction.
For the 11 IRAC candidate counterparts of the SCUBA-2 sources (Section~\ref{sec:vas_submm}), the deboosted source flux is a median of 1.8 times (range 1.1--13.9) the main-grid prediction for the candidate, and the counterpart of the brightest source (Figure~\ref{fig:sedfit}) has 15.0~mJy against a predicted 9.7~mJy.
The adopted fits are on the quasar grid for the two quasar counterparts and, for the nine others, place the stars younger than 10~Myr behind birth clouds in the two-component attenuation of \citet{CharlotFall2000}.
They reproduce within $1.23\sigma$ the 850~\micron\ fluxes of the nine counterparts with a measured 850~\micron\ flux, give $\log(L_{\rm dust}/L_\odot)=11.66$--$12.84$, and lie below every upper limit except the 12~\micron\ limit of one quasar counterpart, which the model exceeds by $0.3\sigma$.
For the nine galaxies, they change the stellar masses by a median of $+0.15$~dex (range $-0.33$ to $+0.36$) against the main-grid fits.
For the counterpart of the brightest source, the adopted fit gives $14.1$~mJy and its best model $15.5$~mJy (Figure~\ref{fig:sedfit}).

\begin{figure*}[t]
\centering
\includegraphics[width=0.96\textwidth]{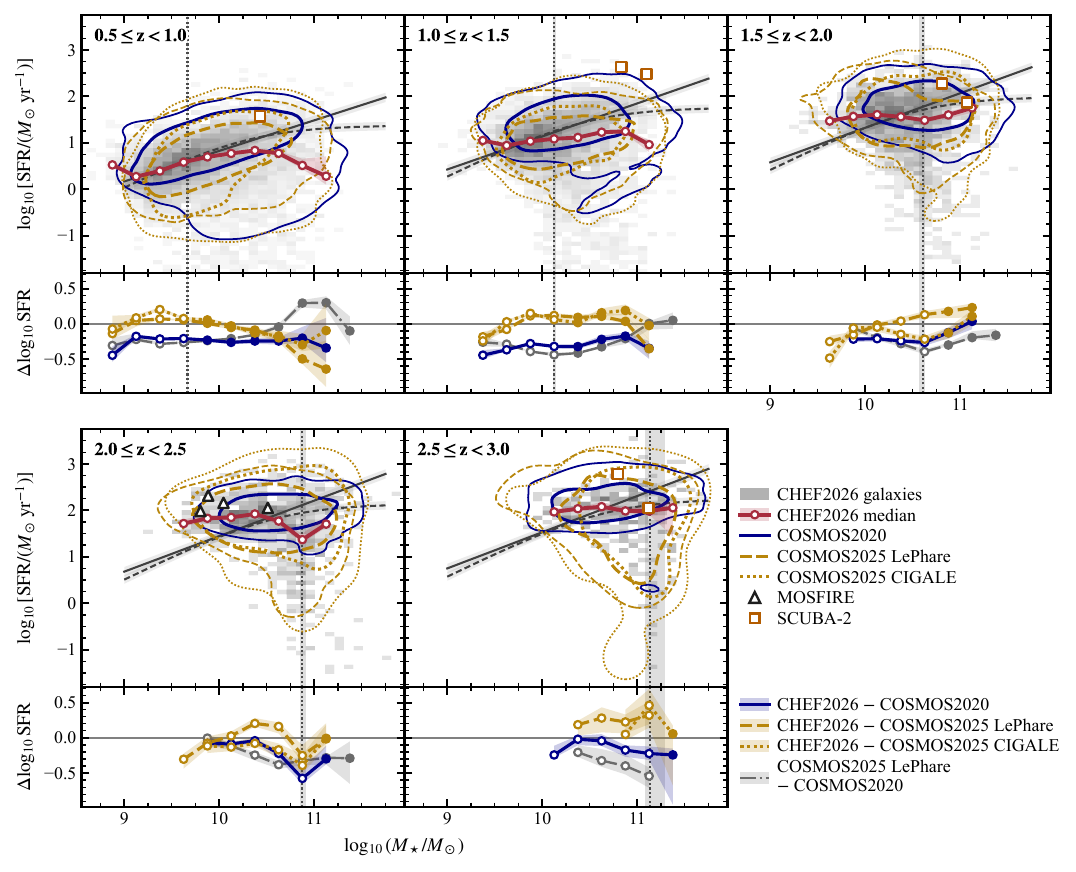}
\caption{Star-forming galaxies with an observed total $K$ magnitude below 22.32~AB in five redshift bins (upper panels) and the offsets of their median star-formation rates at fixed stellar mass (lower panels).
\textit{Upper panels:} the gray scale and the red curve are the $14{,}304$ CHEF2026 galaxies ($6{,}825$, $4{,}423$, $2{,}014$, 764, and 278 from low to high redshift) and their median star-formation rate in $0.25$~dex mass bins with at least 20 galaxies.
Of these galaxies, 400 lie below the plotted range.
The contours enclose $68\%$ and $95\%$ of the galaxies of the same selection in COSMOS2020 (dark blue solid; \citealt{Weaver2022}) and in the \texttt{LePhare} (gold dashed) and \texttt{CIGALE} (gold dotted) fits of COSMOS2025 \citep{Shuntov2025}, star-forming by their $UVJ$ and by their $NUV-r$ and $r-J$ colors, respectively.
Their counts are smoothed with a Gaussian kernel of about $0.1$~dex, wider for the smaller samples at $z>2$.
Open black triangles and open orange squares mark the four Keck/MOSFIRE galaxies with a secure redshift and the seven SCUBA-2 counterparts in the sample.
The gray solid and dashed lines are the relations of \citet{Speagle2014} and \citet{Popesso2023} at the middle of each bin on the \citet{Chabrier2003} IMF, with bands of the scatter of the compiled measurements around each relation ($0.10$ and $0.09$~dex), and the dotted vertical line and the gray band are the CHEF2026 mass limit and its $68\%$ interval.
\textit{Lower panels:} the CHEF2026 median minus the median of each COSMOS sample, in the color and line style of its contours, and the COSMOS2025 \texttt{LePhare} median minus the COSMOS2020 median (gray dash-dotted), in mass bins with at least 20 galaxies of both samples.
Filled markers are the bins of Table~\ref{tab:sedbench}, above the CHEF2026 mass limit, which contain $5{,}697$ CHEF2026 galaxies, and open markers are the other bins.
Bands are the 16th--84th percentiles of the median in $1{,}000$ bootstrap resamplings of the galaxies.}
\label{fig:sfms}
\end{figure*}
We compare the fits of the degraded COSMOS2020 galaxies of Section~\ref{sec:zphot_acc}, one realization per coverage class, with the COSMOS2020 \texttt{LePhare} masses and star-formation rates \citep{Weaver2022}.
The COSMOS2020 and COSMOS2025 masses and star-formation rates were computed for $H_0=70~\kms~\mathrm{Mpc^{-1}}$ and $\Omega_m=0.3$, and here and below we convert them to our cosmology by the squared luminosity-distance ratio at the redshift of each galaxy ($+0.02$~dex at $z\approx2.2$).
For the galaxies without an X-ray AGN, fitted at their spectroscopic redshifts, the median mass offset (this work minus COSMOS2020) is $-0.14$ to $-0.13$~dex ($0.21$~dex scatter) with near-infrared bands and $-0.02$~dex with optical bands alone.
The star-formation rates lie a median $0.14$~dex below COSMOS2020 with near-infrared bands and $0.18$~dex below with optical bands alone ($0.29$--$0.38$~dex scatter).
For massive galaxies ($\log(M_{\rm ref}/\msun)\geq10.5$, $z_{\rm ref}\geq1.25$) the median offsets are $-0.18$~dex in both quantities.
Fits of the undegraded photometry give medians within about $0.03$~dex of the degraded ones.
The offsets therefore reflect the different models and estimators of the two catalogs, and our photometric depth adds little.
In the undegraded fits, the best-fit models of 62 of the 101 massive galaxies have $\ebv\geq0.6$, and only 2 lie at the grid maximum, $\ebv=1.6$, so the grid does not truncate their attenuation.

For star-forming galaxies brighter than $K=22.32$~AB and above the CHEF2026 mass limit, the median star-formation rates at fixed stellar mass agree with those of COSMOS2025 to within $0.20$~dex from $z=0.5$ to 2.5.
We compare the star-forming galaxies of CHEF2026, COSMOS2020, and COSMOS2025 under one selection in five redshift bins from $z=0.5$ to 3 (Figure~\ref{fig:sfms} and Table~\ref{tab:sedbench}).
The selection requires an observed total $K$ magnitude below 22.32~AB ($K_{Xs}$ for CHEF2026, UltraVISTA $K_s$ for COSMOS), a star-forming class by the $UVJ$ colors ($NUV-r$ and $r-J$ colors of \citealt{Ilbert2013} for COSMOS2025), and a stellar mass above the mass limit of the $K$-limited CHEF2026 sample in each bin.
In the same bins, the central offsets of CHEF2026 from COSMOS2020 range from $-0.04$ to $-0.30$~dex.
The \texttt{LePhare} fits of COSMOS2025 lie $0.10$--$0.29$~dex below COSMOS2020 in the same bins, so the offset of CHEF2026 from COSMOS2020 has the direction and size of the difference between the two COSMOS catalogs.

\begin{deluxetable*}{lcccc}
\tabletypesize{\footnotesize}
\tablecaption{Offsets in star-formation rate at fixed stellar mass between CHEF2026, COSMOS2020, and COSMOS2025.
\label{tab:sedbench}}
\tablewidth{0pt}
\tablehead{\colhead{$z$} & \colhead{\shortstack{CHEF2026\\$-$ COSMOS2020}} & \colhead{\shortstack{CHEF2026\\$-$ COSMOS2025 LePhare}} & \colhead{\shortstack{CHEF2026\\$-$ COSMOS2025 CIGALE}} & \colhead{\shortstack{COSMOS2025 LePhare\\$-$ COSMOS2020}}}
\startdata
0.5--1.0 & $-0.24^{+0.03}_{-0.02}$ & $-0.14^{+0.03}_{-0.03}$ & $-0.11^{+0.06}_{-0.04}$ & $-0.10^{+0.08}_{-0.06}$ \\
1.0--1.5 & $-0.27^{+0.04}_{-0.02}$ & $+0.06^{+0.03}_{-0.03}$ & $+0.09^{+0.04}_{-0.04}$ & $-0.21^{+0.02}_{-0.08}$ \\
1.5--2.0 & $-0.04^{+0.05}_{-0.07}$ & $+0.20^{+0.04}_{-0.10}$ & $-0.01^{+0.05}_{-0.08}$ & $-0.19^{+0.04}_{-0.07}$ \\
2.0--2.5 & $-0.30^{+0.19}_{-0.08}$ & $-0.02^{+0.22}_{-0.07}$ & $-0.01^{+0.19}_{-0.12}$ & $-0.29^{+0.11}_{-0.21}$ \\
2.5--3.0 & $-0.24^{+0.09}_{-0.70}$ & \nodata & $+0.06^{+0.12}_{-0.73}$ & \nodata \\
\enddata
\tablecomments{Offset of the median $\log\sfr$ (dex), first sample minus second, for star-forming galaxies with an observed total $K$ magnitude below 22.32~AB (Section~\ref{sec:sed_cigale}).
Each offset is the median, over the $0.25$~dex bins of stellar mass above the CHEF2026 mass limit that contain at least 20 galaxies of both samples, of the difference between the median $\log\sfr$ of the two samples in the bin.
From $z=0.5$ to 3.0 the CHEF2026 columns rest on 6, 4, 2, 1, and 1 mass bins with $3{,}609$, $1{,}648$, 333, 84, and 23 CHEF2026 galaxies, and the last column on 7, 5, 3, and 2 bins from $z=0.5$ to 2.5.
Mass limits $\log(M_{\rm lim}/\msun)$ of the CHEF2026 selection, by redshift bin: 9.67, 10.12, 10.61, 10.87, 11.13.
Star-formation rates: \texttt{CIGALE} instantaneous rate, likelihood-weighted mean of its logarithm (CHEF2026); \texttt{LePhare} median, \texttt{lp\_SFR\_med} (COSMOS2020) and \texttt{sfr\_med} (COSMOS2025); \texttt{CIGALE} instantaneous rate, \texttt{sfr\_inst} (COSMOS2025).
Star-forming class: $UVJ$ colors (CHEF2026, COSMOS2020), $NUV-r$ and $r-J$ colors (COSMOS2025).
Super- and subscripts give the 16th and 84th percentiles of $1{,}000$ bootstrap resamplings.}
\end{deluxetable*}

\subsection{Stellar-mass limits and rest-frame colors}
\label{sec:sed_mlim}\label{sec:sed_uvj}

\begin{figure}[t]
\centering
\includegraphics[width=\columnwidth]{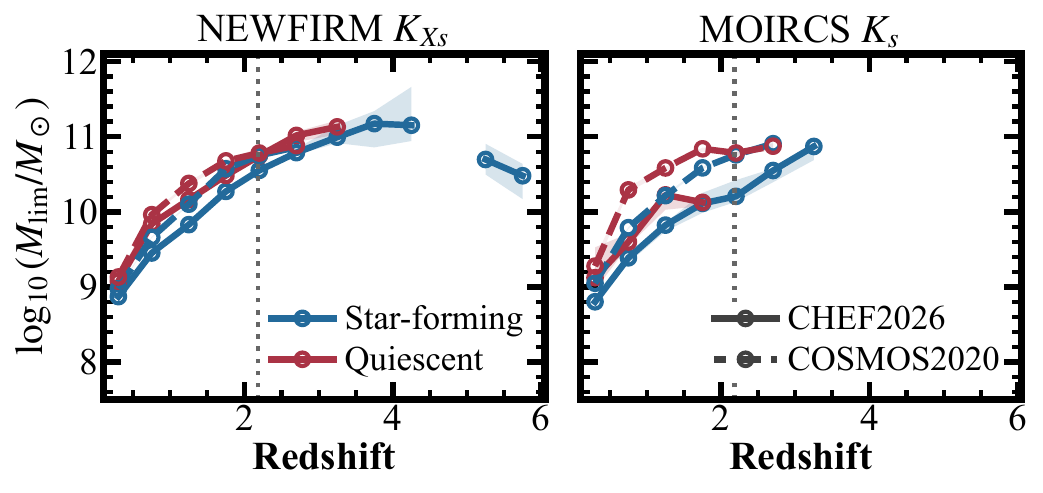}
\caption{Stellar-mass limits against redshift for star-forming (blue) and quiescent (red) galaxies, in $K_{Xs}$ (left) and $K_s$ (right), for CHEF2026 (solid) and for COSMOS2020 galaxies at our $K$ depths (dashed).
The CHEF2026 curves end at the last redshift bin with at least five galaxies in the faint tail, and the star-forming $K_{Xs}$ curve skips the bin at $4.5\leq z<5.0$, which has fewer.
The COSMOS2020 curves end at $z=3$, the upper redshift of the COSMOS2020 sample used here.
Shaded bands are the 68\% bootstrap intervals of the limits, lighter for COSMOS2020, and the dotted line marks $z=2.18$.}
\label{fig:mlim}
\end{figure}

\textit{Stellar-mass limits.}
We estimate the stellar-mass limit $M_{\rm lim}(z)$ with the empirical method of \citet{Pozzetti2010}, separately for $UVJ$ star-forming and quiescent galaxies, whose mass-to-light ratios differ.
The galaxies used in each redshift bin are detected at $\mathrm{S/N}\geq5$ in the total flux of a $K$ band, have a fit with positive \mstar\ and \sfr\ and $\chi^2_\nu\leq10$, and are not stars or quasars.
Because the depth varies across the field (Section~\ref{sec:bandqa_depth}), we rank the galaxies by $m-m_{\rm lim}$, where $m_{\rm lim}$ is the local $5\sigma$ total-flux depth.
We scale the mass of each galaxy in the faintest $20\%$ to that depth, $\log M_{\rm lim}=\log\mstar+0.4(m-m_{\rm lim})$, and adopt the 95th percentile of the scaled masses, requiring at least five galaxies in the faint tail.
The intervals are the 16th--84th percentiles of 500 bootstrap resamplings.
At $2.0\leq z<2.4$ (Figure~\ref{fig:mlim}), the $K_{Xs}$ limits are $\log(M_{\rm lim}/\msun)=10.55$ ($10.52$--$10.56$) for star-forming and $10.74$ ($10.70$--$10.78$) for quiescent galaxies, and the $K_s$ limit for star-forming galaxies is $10.20$ ($10.15$--$10.54$).
These empirical limits at the local depth lie below those of the fixed cut at 22.32~AB in Table~\ref{tab:sedbench} ($10.87$ for star-forming galaxies at $2.0\leq z<2.5$).
The same estimator applied to COSMOS2020 galaxies \citep{Weaver2022}, with their own masses and colors and our $K$ depths, gives limits $0.04$--$0.55$~dex higher than the three measured CHEF2026 limits (Figure~\ref{fig:mlim}).
Without the split by $UVJ$ class, the CHEF2026 $K_{Xs}$ limit lies $0.14$~dex below the COSMOS2020 limit, and the quiescent fractions of the $K_{Xs}$-detected galaxies at $2.0\leq z<2.4$ are similar ($10\%$ in CHEF2026 and $8\%$ in COSMOS2020).
This suggests that most of the difference in $K_{Xs}$ comes from the mass scale of the fits, which give the degraded COSMOS2020 galaxies masses $0.13$--$0.14$~dex below their COSMOS2020 values (Section~\ref{sec:sed_cigale}).
The star-forming $K_s$ limit rests on 30 galaxies in its faint tail, and the quiescent one is not measured, since its faint tail contains 4 galaxies, fewer than the five required.
We adopt the CHEF2026 limits, which describe the masses and $UVJ$ classes of the catalog.

\begin{figure}[t]
\centering
\includegraphics[width=\columnwidth]{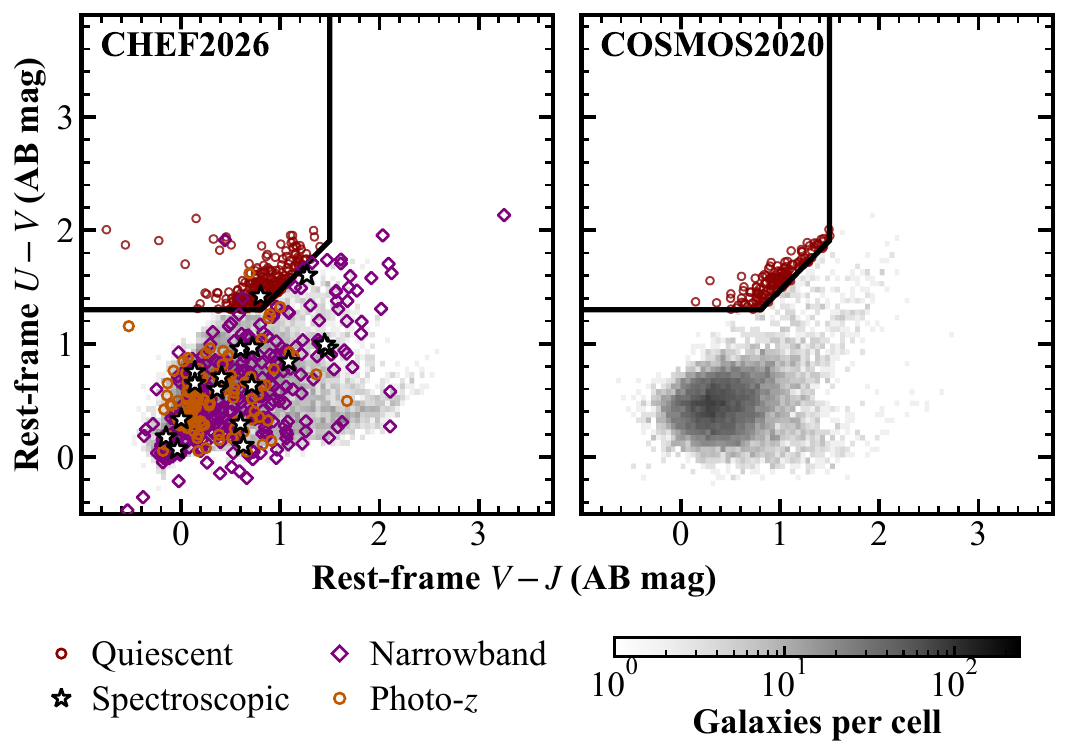}
\caption{Rest-frame $U-V$ against $V-J$ at $2.0\leq z<2.4$ for the $12{,}910$ CHEF2026 galaxies with both IRAC bands, whose rest-frame $J$ is measured (left), and their $12{,}910$ COSMOS2020 partners (right), as the number of galaxies per color cell on a shared logarithmic scale.
The partners come from the one-to-one pairing in stellar mass of all $30{,}985$ CHEF2026 galaxies at $2.0\leq z<2.4$ with COSMOS2020 galaxies of the same redshift interval.
Red circles are quiescent candidates, $1.9\%$ of the CHEF2026 galaxies and $1.1\%$ of their partners, and the black line is the quiescent boundary of \citet{Muzzin2013}.
The left panel also marks, at the redshift of their fits, the 457 members and member candidates of the structure with both IRAC bands and a fit at $2.0\leq z<2.4$.
They are 17 spectroscopic members without a quasar spectrum (open black stars), 335 candidate narrowband members (\lya\ or \ha\ emitters; purple diamonds), and 105 photometric-redshift candidates (orange circles).
Our colors are measured with \texttt{EAZY} at the redshift of the \texttt{CIGALE} fit (Section~\ref{sec:sed_uvj}).
Galaxies with fewer than two IRAC bands, whose $V-J$ is extrapolated by the templates, are not shown, and 3 CHEF2026 and 5 COSMOS2020 galaxies lie outside the plotted range.}
\label{fig:uvj}
\end{figure}

\textit{Rest-frame colors.}
With \texttt{EAZY} \citep{Brammer2008} at the redshift of the \texttt{CIGALE} fit, we measure rest-frame absolute magnitudes in the Johnson $U$ and $V$ passbands of \citet{MaizApellaniz2006}, 2MASS $J$, \galex\ NUV, HSC $r$, and VISTA $J$.
They give $U-V$ and $V-J$ with 2MASS $J$, and $NUV-r$ and $r-J$ with VISTA $J$.
\texttt{EAZY} fits its templates at that redshift and derives each rest-frame flux mainly from the observed bands nearest in wavelength.
A galaxy is quiescent by the criterion of \citet{Muzzin2013} if $U-V>1.3$, $V-J<1.5$, and $U-V>0.88(V-J)+0.59$ ($+0.69$ at $z<1$), and by that of \citet{Ilbert2013} if $NUV-r>3.1$ and $NUV-r>3(r-J)+1$.
At $z=2.18$ the rest-frame $J$ band is observed near $3.9$~\micron, between the two IRAC bands, so $V-J$ is measured only where IRAC data exist and is otherwise extrapolated by the templates.
We pair each CHEF2026 galaxy at $2.0\leq z<2.4$ with the COSMOS2020 galaxy of the same interval nearest in stellar mass within $0.2$~dex, without replacement.
Figure~\ref{fig:uvj} shows on the $UVJ$ plane the pairs whose CHEF2026 galaxy has both IRAC bands.
Of all pairs, $4.1\%$ of the CHEF2026 galaxies and $1.3\%$ of their COSMOS2020 partners lie in the quiescent region.
Our fits of 595 degraded COSMOS2020 galaxies at $2.0\leq z<2.4$, at their spectroscopic redshifts, recover $54\%$ ($45$--$61\%$) of their 22 quiescent galaxies, averaged over realizations.
They classify $1.7\%$ ($1.4$--$2.1\%$) of the star-forming ones as quiescent, which gives $3.7\%$ ($3.1$--$4.3\%$) quiescent against $3.7\%$ ($3.0$--$4.5\%$) by their COSMOS2020 colors.
With IRAC photometry, the fits recover $63\%$ ($53$--$73\%$) of the quiescent galaxies and classify $0.2\%$ of the star-forming ones as quiescent, against $6.3\%$ without IRAC.
Among the CHEF2026 galaxies without IRAC data, the quiescent fraction is $6.1\%$ ($1.4\%$ for their COSMOS2020 counterparts), against $1.9\%$ ($1.1\%$) for galaxies with both IRAC bands.
The excess of CHEF2026 over COSMOS2020 thus comes mainly from the galaxies without IRAC data, the group most contaminated in the degraded COSMOS2020 test.
We therefore recommend the $UVJ$ classes only for galaxies with both IRAC bands and a good photometric-redshift class, in which the quiescent fraction is $3.1\%$, against $3.8\%$ for their COSMOS2020 counterparts.

\section{Value-Added Samples and the \textit{Cosmic Himalayas}}
\label{sec:science}\label{sec:vas}

The catalog also contains samples selected for the structure at $z\approx2.18$ or for galaxy populations at $z\sim2$, each with its own selection function (Table~\ref{tab:vas}).

\begin{deluxetable*}{lp{2.8in}rp{1.5in}}
\tabletypesize{\footnotesize}
\setlength{\tabcolsep}{3.5pt}
\tablecaption{Value-added samples of the catalog.
\label{tab:vas}}
\tablewidth{0pt}
\tablehead{
\colhead{Sample} & \colhead{Selection} & \colhead{Number} & \colhead{Reference}
}
\startdata
\ha\ emitters, NB2083 & $\Sigma\geq3$, $\mathrm{EW_0}\geq25$~\AA, color cut (detection completeness: Table~\ref{tab:compl}) & 103 & \citet{Sobral2013} \\
\ha\ emitters, NB2096 & same & 213 & \citet{Sobral2013} \\
\lya\ emitters, NB387 & published sample (452 distinct LAEs in 465 entries), without 14 low-redshift contaminants & 438 & \citet{Liang2021} \\
Spectroscopic members & secure redshift at $2.15<z<2.22$ (Section~\ref{sec:catalog_member}) & 38 & this work \\
Photometric-redshift candidates & $\geq50\%$ of the likelihood in $2.08<z<2.28$; purity $\approx43\%$ & 216 & this work \\
BX, BM & $U_nG\mathcal{R}$ criteria in our filters (completeness: Table~\ref{tab:selfunc}) & 29{,}701; 27{,}109 & \citet{Steidel2004} \\
LBG & $u$-dropout & 9{,}596 & \citet{Sawicki2019} \\
Star-forming, passive $gzK$ & $gzK$ criteria & 7{,}390; 334 & \citet{ArcilaOsejo2013};\newline\citet{Daddi2004} \\
Spectroscopic quasars & compilation (Section~\ref{sec:specz}) & 253 & this work \\
Photometric QSO class & template $\chi^2$, morphology, $W1-W2$ (Section~\ref{sec:catalog_class}) & 1{,}415 & this work \\
AGN flag & $W1-W2$ criteria, or spectroscopic quasar & 928 & \citet{Stern2012};\newline\citet{Assef2018} \\
Red galaxies & $J-K>2.3$ (Vega), NEWFIRM; MOIRCS & 229; 11 & \citet{Franx2003} \\
SCUBA-2 peaks & $\mathrm{S/N}\geq3.5$ (7 secure at $\mathrm{S/N}\geq5$, about 26 expected real) & 61 & this work \\
IRAC counterparts & chance probability $\leq0.05$ within $8\farcs7$ & 11 & \citet{Downes1986} \\
\enddata
\tablecomments{Numbers count distinct objects in the final samples, and peaks for SCUBA-2.
Of the photometric QSO class, the 214 objects at $i<21$ are those whose purity is measured (Section~\ref{sec:catalog_class}).
The two NB2083 emitters that MOIRCS resolves $1\arcsec$ apart belong to one object, so the 104 NB2083 emitters are 103 objects.
Of the flagged AGN, 47, 164, and 770 meet the criteria of \citet{Stern2012} and of \citet{Assef2018} at $90\%$ reliability and at $75\%$ completeness.
The 9 SCUBA-2 sources with a candidate, 3 of them secure, have 11 counterpart candidates.}
\end{deluxetable*}

\textit{Visual inspection.}
The \ha\ candidates, the detections on the rims of bright stars, and random samples of the color-selected candidates were inspected on image cutouts with OpenAI's GPT-6.1, a multimodal large language model (LLM) that reads images and text.
We reviewed the \ha\ classifications (Appendix~\ref{sec:app_vi}).

\subsection{\ha\ emitters} \label{sec:vas_emitters}\label{sec:vas_hae}

The \ha\ samples are selected by the color-excess significance $\Sigma$ \citep{Bunker1995,Sobral2013} and by the rest-frame equivalent width corrected for the line's contribution to the broadband \citep{Sobral2009}.
$\Sigma$ is the narrowband excess over the continuum in the $2\arcsec$ aperture divided by the quadrature sum of the two aperture noises.
Each noise is the NMAD of source-free $2\arcsec$ apertures in cells of $1\arcmin$ (NB2083) or $3\farcm4$ (NB2096), interpolated to the source position.
For narrowband detections without a line in the band, $\Sigma$ has an NMAD of $0.91$--$0.98$ at narrowband $\mathrm{S/N}=3$--$8$.
Line fluxes and equivalent widths use the FWHMs of the manufacturers' transmission curves (Table~\ref{tab:bands}), computed for an angle of incidence of $5\arcdeg$ at 120~K for NB2083 and measured at room temperature for NB2096.
The catalog also gives, for each candidate, the passband at its position averaged over the exposures that cover it (Section~\ref{sec:survey}), with the mean angle of incidence, the half-maximum wavelengths, and the corresponding \ha\ redshift window.
The continuum is measured at the narrowband position in the broadband that contains it ($K_s$ for NB2083, $K_{Xs}$ for NB2096; Section~\ref{sec:catalog_phot}).
It is corrected for Galactic extinction and by a color term, the running median of the broadband-minus-narrowband color of continuum sources against narrowband magnitude.
For NB2096 a second, local term (the median residual color of the 200 nearest continuum sources) removes color offsets that follow the PSF matching of the two NEWFIRM bands.
Candidates require $\Sigma\geq3$ and $\mathrm{EW_0}\geq25$~\AA, the selection thresholds of the High-redshift Emission Line Survey \citep[HiZELS;][]{Sobral2013}.

\begin{figure}[t]
\centering
\includegraphics[width=\columnwidth]{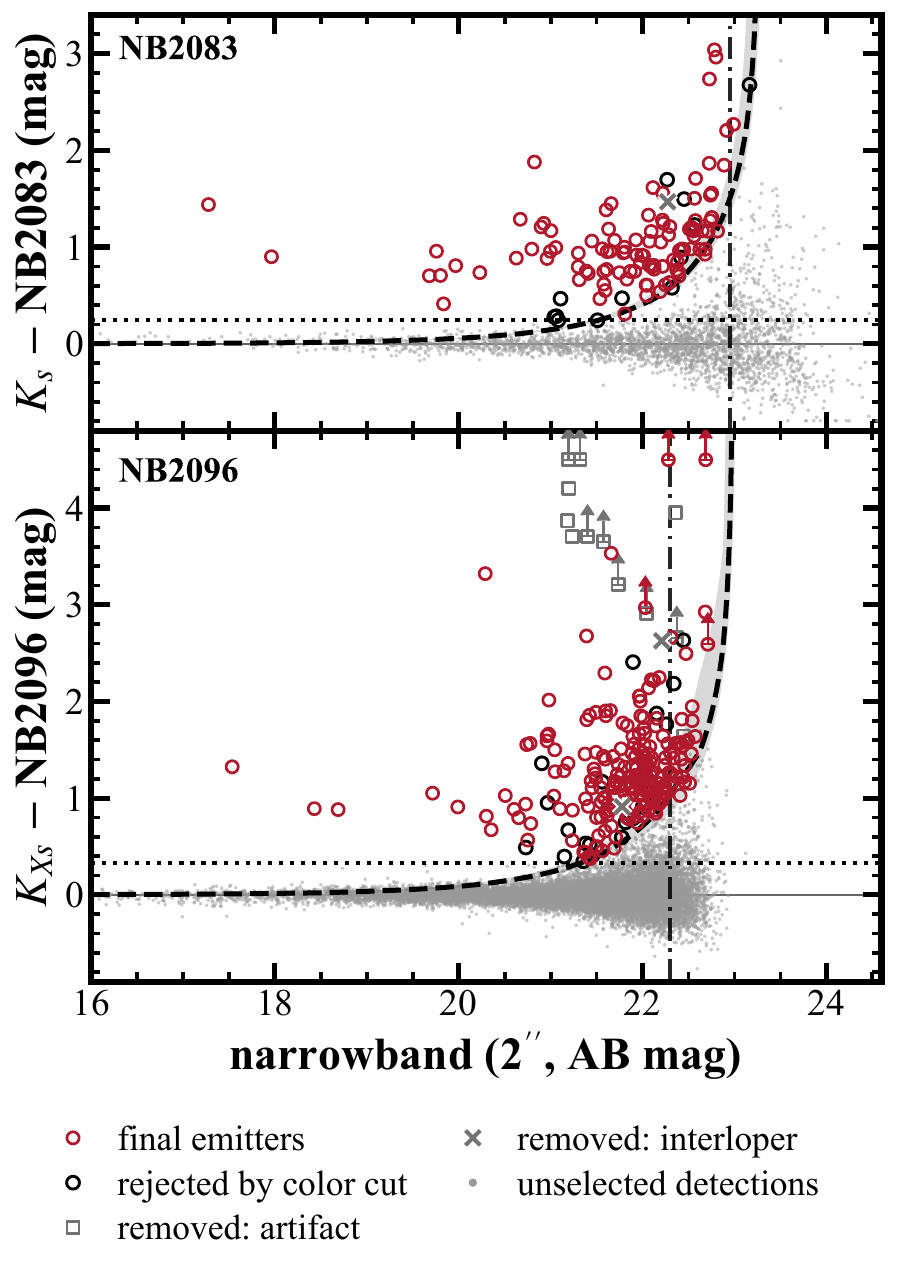}
\caption{Selection of the \ha\ emitters in MOIRCS NB2083 (top; continuum from $K_s$) and NEWFIRM NB2096 (bottom; continuum from $K_{Xs}$): broadband minus narrowband color against the $2\arcsec$ narrowband magnitude.
Each source is selected with its own local noise by $\Sigma\geq3$ and $\mathrm{EW_0}\geq25$~\AA\ (dotted line).
The dashed curve is $\Sigma=3$ at the median noise, and the shaded band spans $\Sigma=3$ at the 16th and 84th percentiles of the noise, so emitters in the deeper parts of the mosaics lie below the dashed curve.
The dash-dotted line is the median local $5\sigma$ depth.
Gray points are the detections in the selection area that were not selected.
Red circles are the final emitters, black circles are the candidates rejected by the continuum color cut, gray squares are the NB2096 candidates in the selection area removed as image artifacts after visual inspection, and gray crosses are the three candidates removed as spectroscopic interlopers.
Arrows mark colors that are lower limits or lie beyond the panel.}
\label{fig:emitters}
\end{figure}

Of the $3{,}222$ NB2083 and $17{,}545$ NB2096 detections, $117$ and $236$ candidates pass these cuts.
The NB2096 count excludes 15 candidates that visual inspection identified as image artifacts or that lie in regions of diffuse light, which are excluded from the selection area.
Low-redshift line emitters are then rejected with a ($g-i$, $i-K$) color cut, an analog of the $BzK$ criterion \citep{Daddi2004}.
Sources with $i-K<1.6~(g-i)-0.5$ in dereddened colors are removed, and sources too faint for the cut to decide are kept with a flag.
Candidates that spectroscopy identifies as stars or as line emitters at other redshifts are also removed (one star and two \oiii\ emitters at $z\approx3.2$).
The final samples contain $104$ (NB2083) and $213$ (NB2096) \ha\ emitters (Figure~\ref{fig:emitters}).

The SED fit restricted to the \ha\ and \oiii\ solutions flags $2$ NB2083 and $6$ NB2096 emitters as \oiii\ emitters at $z\approx3.16$.
Pure-noise peaks in NB2096 have an HSC counterpart only by chance, at a rate of 0.18.
The HSC $g$ and $i$ forced photometry at the positions of the NB2096 emitters other than quasars therefore bounds the fraction of pure-noise emitters in each bin of $\Sigma$.
All 22 of these emitters at $\Sigma\geq8$ have a counterpart.
If every real emitter has one as often, the purities are $0.86$, $0.89$, $0.92$, $0.95$, and $1.00$ at $\Sigma=3$--$3.5$, $3.5$--$4$, $4$--$5$, $5$--$8$, and $\geq8$.
These purities give 16 (9--28) expected pure-noise emitters among the 213, about 10 of them among the $80$ emitters with $\Sigma<4$, which carry a low-purity flag.
The catalog gives each NB2096 emitter its model probability of being pure noise (\texttt{p\_noise}).
The negated NB2083 image yields no spurious emitters.
We correct the counts with the \oiii\ fraction per narrowband magnitude measured in our Keck/MOSFIRE spectra and, in NB2096, weight each emitter by the probability that it is real.
The expected numbers of \ha\ emitters are then $91.6$ ($68\%$ interval $86.0$--$97.1$) in NB2083 and $171$ ($160$--$176$ over the purity intervals) in NB2096.
Both the narrowband-excess and the final samples are part of the catalog.

Of the candidates that pass the color cut and have a Keck/MOSFIRE, DESI, or SDSS redshift, 22 of 23 in NB2083 and 16 of 17 in NB2096 lie at $2.15<z<2.22$, and the other two are the \oiii\ emitter at $z=3.17$ and the star removed above.
The third removed interloper, an \oiii\ emitter at $z=3.20$, has a HETDEX \lya\ redshift.
Seven of the 24 confirmed \ha\ emitters are in neither sample, because their narrowband excesses fall below the threshold or they are not detected.

Inside the MOIRCS footprint, $22$ of the $34$ NB2096 emitters are also NB2083 emitters.
The NB2083 emitters are recovered less often in NB2096 because the NB2096 data are about 1~mag shallower and NB2096 covers mainly the red part of the NB2083 redshift window (Section~\ref{sec:survey}).
Of the $12$ NB2096 emitters without an NB2083 counterpart, one lies at the low-weight edge of the MOIRCS mosaic, which is masked in the NB2083 detection, and shows an NB2083 excess as well ($\Sigma\approx6$--$9$).
The other $11$ show no NB2083 excess at the level expected from their NB2096 excesses ($\Sigma\leq2.4$, against $4.9$--$26$ expected).
They lie where light reaches the MOIRCS filter at large angles of incidence (median $7.9\arcdeg$, against $4.6\arcdeg$ for the emitters selected in both bands), where the red edge of NB2083 moves to $z\approx2.19$--$2.20$.
Their \ha\ lines therefore lie at the red end of the window, where only NB2096 detects them.
NB2096 thus extends both the redshift coverage and the area of the survey.

\subsection{\lya\ emitters} \label{sec:vas_lae}

The \lya\ sample is the published sample of \citet{Liang2021}, selected on the same NB387 image.
The $465$ published entries ($452$ distinct LAEs, from two reductions of these data) are matched to our NB387 detections within $1\arcsec$.
Of these 452 LAEs, 37 have a secure spectrum within $1\arcsec$ (Section~\ref{sec:specz}), and all 17 with HETDEX or Keck/MOSFIRE spectra lie at $z\approx2.18$.
DESI DR1 and SDSS DR18, which preferentially target brighter emission-line galaxies at $z<1.6$ and quasars, confirm six quasars in the band and place 14 sources at $z=0.70$--$1.50$, which we remove, leaving 438 LAEs.
Inside the area where the LAEs are matched (Section~\ref{sec:obs_imaging}), $89.1\%$ ($68\%$ Wilson interval $87.2$--$90.7\%$) of the published entries have an NB387 detection within $1\arcsec$, and the entries outside it carry a flag.
For the entries without an NB387 detection within $1\arcsec$, the catalog gives the nearest object and its separation.
Within the area unmasked in NB387, $6$ of $102$ NB2083 and $22$ of $192$ NB2096 emitters have a published LAE within $1\farcs5$, whereas the same positions shifted by $1$ or $2\arcmin$ in right ascension have none ($0$ of $1{,}140$).
Most of these \ha\ emitters therefore have no \lya\ counterpart at the NB387 depth.
Of the 26 HETDEX \lya\ emitters at $2.15<z<2.22$ \citep{MentuchCooper2026}, the 16 at $z=2.170$--$2.199$, inside the NB387 window, coincide with NB387 LAEs within $0\farcs73$, and the other 10 lie at $z=2.200$--$2.210$, at or beyond its red edge.

\subsection{Color-selected samples} \label{sec:vas_color}

\begin{figure*}[t]
\centering
\includegraphics[width=\textwidth]{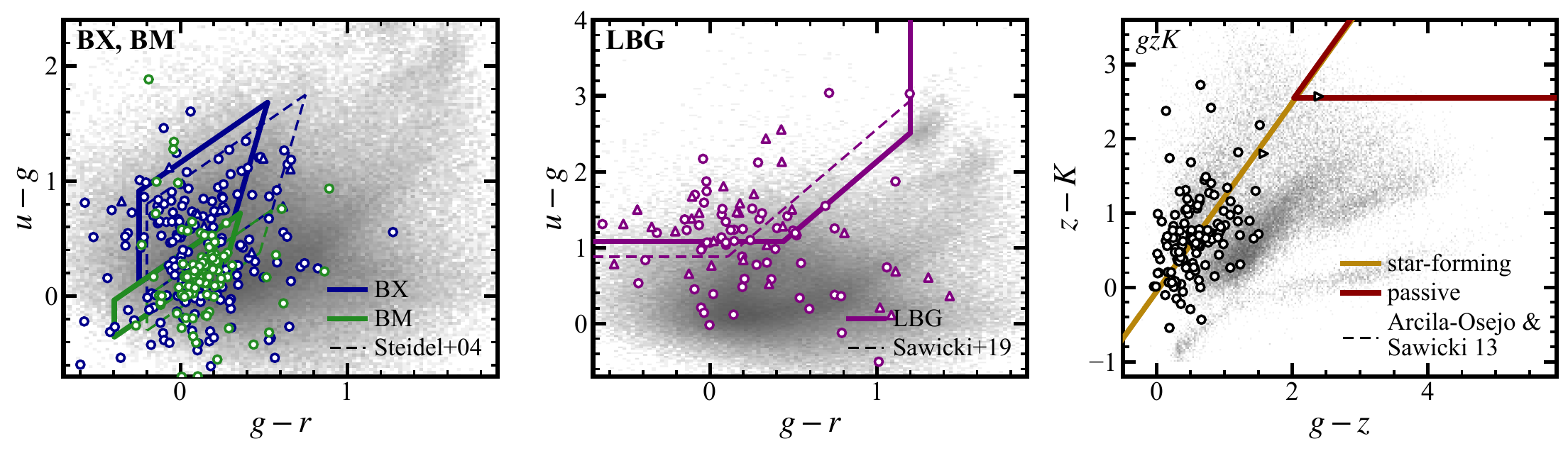}
\caption{Color selections in our filters: BX and BM (left) and LBGs (middle) in $u-g$ against $g-r$, and $gzK$ (right).
Thin dashed lines are the boundaries of \citet{Steidel2004}, \citet{Sawicki2019}, and \citet{ArcilaOsejo2013}, and thick lines the adopted ones in the same color ($gzK$ for NEWFIRM $K_{Xs}$), over the logarithmic density of the field objects that meet the requirements of each selection.
Open markers are field objects with a secure spectroscopic redshift in the target window of the selection of the same color (BM below $z=1.9$), mostly line emitters and quasars, with triangles for $2\sigma$ limits.}
\label{fig:colorsel}
\end{figure*}

The BX, BM \citep{Steidel2004}, and $gzK$ (\citealt{ArcilaOsejo2013}, adapted from the $BzK$ criterion of \citealt{Daddi2004}) criteria were defined in filters that differ from ours, and the $u$-dropout Lyman-break criterion \citep[LBG;][]{Sawicki2019} in the same CFHT $u$ and HSC $g$ and $r$ filters as ours.
We therefore move each boundary into our filter system with synthetic photometry of galaxy templates through the measured filter responses, keeping its functional form, which leaves the LBG boundary unchanged.
The transformed cut is adopted for star-forming $gzK$, and the cut of \citet{ArcilaOsejo2013} is kept, with a flag, for passive $gzK$.
For BX, BM, and the LBGs we shift the transformed boundary lines by the amounts within $\pm0.20$~mag that maximize the harmonic mean of completeness and purity in the target window on the calibration galaxies of COSMOS2020 degraded to our depth (Section~\ref{sec:zphot_acc}).
For the LBGs the $u-g$ boundary rises by $0.20$~mag and the diagonal boundary moves by $0.20$~mag along its normal toward bluer $u-g$, which raises the harmonic mean on the test galaxies from $0.20$ for the transformed boundary to $0.25$.
BM takes the lower BX edge, $u-g=(g-r)+0.361$, as its upper edge, so that the two selections share one boundary, as those of \citet{Steidel2004} do, and have no object in common.
Figure~\ref{fig:colorsel} shows the original and adopted boundaries.
The optical selections require a $\chi^2$ detection with $u$, $g$, and $r$ measurements and a combined $griz$ $\mathrm{S/N}$ of at least 5, with no magnitude limit, and colors of bands with $\mathrm{S/N}<2$ use the $2\sigma$ limit.
The $gzK$ selections require $\mathrm{S/N}\geq5$ in $K$ and $\geq2$ in $z$.
None of these selections applies the significance cut or quality flag~0 of Section~\ref{sec:catalog_flow}.
These samples are separate, overlapping catalogs with the sizes of Table~\ref{tab:vas}.
At the limit $\mathcal{R}\leq25.5$ of \citet{Steidel2004}, with $\mathcal{R}=r-0.52(r-i)$ for point-source total magnitudes, the BX surface density is $6.7$~arcmin$^{-2}$, against the $5.2$~arcmin$^{-2}$ they give after correcting for $9\%$ interlopers.
At this limit $60\%$ of the selected galaxies lie outside $1.9\leq z<2.7$.
Table~\ref{tab:selfunc} gives the completeness and the fraction of selected galaxies outside the target window, both measured on $60{,}000$ COSMOS2020 galaxies drawn at random to $i<26.5$ and degraded to our depths and coverage as in Section~\ref{sec:zphot_acc}, with the COSMOS2020 photometric redshift as the reference.
Down to $\mathcal{R}\leq24.5$, brighter samples are more complete and less contaminated.
From all galaxies to $\mathcal{R}\leq24.5$, $C$ rises from $31\%$ to $52\%$ for BX, from $23\%$ to $50\%$ for BM, and from $18\%$ to $59\%$ for the LBGs.
Over the same range $f_{\rm out}$ falls from $65\%$ to $51\%$, from $61\%$ to $52\%$, and from $56\%$ to $45\%$.
Most of the selected galaxies outside the window lie below it at every limit.
Brighter than $\mathcal{R}\approx24.5$, $f_{\rm out}$ of BX and LBG rises again as low-redshift galaxies enter: $61\%$ of the BX brighter than $\mathcal{R}=23.5$ lie at $z<0.5$, as \citet{Steidel2004} found for their bright BX candidates.
For star-forming $gzK$, $C$ reaches $80\%$ and $f_{\rm out}$ $9.5\%$ at $K\leq22.0$, and the low $C$ over all galaxies comes from the target galaxies whose $K$ flux is below its $2\sigma$ limit or missing ($3{,}515$ of $6{,}240$).

The catalog adds flag columns to the BX, BM, LBG, and $gzK$ selections without removing rows.
They are the near-star flag of the $\chi^2$ detection (Section~\ref{sec:catalog_flow}), a flag for sources at the edge of the coverage or in the outskirts of a large galaxy, and the class that the model classification of the continuum cutouts gives to detections on the rims of $G=15$--$16$ stars (Appendix~\ref{sec:app_vi}).
The near-star flag marks $9.2\%$ of the LBG, $3.1\%$ of the BX, $2.2\%$ of the BM, and $2$--$3\%$ of the star-forming $gzK$ candidates.
The classification flags $144$ LBG candidates ($1.5\%$) and $23$ BX candidates as pieces of bright-star rims.

\begin{deluxetable}{lccc}
\tabletypesize{\footnotesize}
\setlength{\tabcolsep}{3.5pt}
\tablecaption{Redshift-window selection functions of the color-selected samples.
\label{tab:selfunc}}
\tablewidth{0pt}
\tablehead{
\colhead{Selection} & \colhead{Target $z$} & \colhead{$C$} & \colhead{$f_{\rm out}$} \\
\unitcolhead{} & \unitcolhead{} & \unitcolhead{(\%)} & \unitcolhead{(\%)}
}
\startdata
BX                  & 1.9--2.7 & $31.1^{+0.6}_{-0.6}$ & $65.1^{+0.6}_{-0.7}$ \\
BM                  & 1.4--2.1 & $22.9^{+0.5}_{-0.5}$ & $61.3^{+0.7}_{-0.7}$ \\
LBG                 & 2.7--3.5 & $17.6^{+0.6}_{-0.6}$ & $56.0^{+1.3}_{-1.3}$ \\
$gzK$, star-forming & 1.4--2.5 & $13.2^{+0.4}_{-0.4}$ & $18.6^{+1.2}_{-1.1}$ \\
$gzK$, passive      & 1.4--2.5 & $14.9^{+2.7}_{-2.9}$ & $11^{+5}_{-5}$ \\
\enddata
\tablecomments{$C$: completeness, the fraction selected by the cut among the degraded COSMOS2020 galaxies that lie in the target window, belong to the targeted population, and have the required band coverage.
The targeted population is $UVJ$ star-forming for BX, BM, LBG, and star-forming $gzK$, and $UVJ$ quiescent for passive $gzK$.
$f_{\rm out}$: fraction of the selected galaxies outside the target window, with the COSMOS2020 photometric redshift as the reference.
Super- and subscripts give the $68\%$ bootstrap interval.}
\end{deluxetable}

\subsection{Infrared-selected and submillimeter sources}
\label{sec:vas_submm}\label{sec:vas_ir}\label{sec:science_obscured}

\textit{Infrared-selected sources.}
From the AllWISE photometry (Section~\ref{sec:catalog_phot}) we flag candidate AGN with three color criteria: that of \citet{Stern2012}, $W1-W2\geq0.8$ and $W2<15.05$ (Vega), and those of \citet{Assef2018} at $90\%$ reliability, $W1-W2>0.65\exp[0.153~\max(0,~W2-13.86)^2]$, and at $75\%$ completeness, $W1-W2>0.71$.
The criteria apply to sources with clean AllWISE photometry: no contamination flag, and $W2$ $\mathrm{S/N}\geq10$ for the first criterion and $>5$ for the other two.
Red galaxies are flagged with the $J-K>2.3$ (Vega) criterion of \citet{Franx2003}, which becomes $J_X-K_{Xs}>1.398$ and $J_X-K_s>1.393$ in AB through our filters, with $\mathrm{S/N}\geq3$ in $J_X$ and $\geq5$ in $K$.
Two combined flags collect these selections.
An AGN flag (Table~\ref{tab:vas}) marks non-stellar objects inside the four-band area that meet any of the AllWISE criteria or are spectroscopic quasars, and a red-galaxy flag marks objects with a red $J-K$ color.

\textit{Submillimeter sources.}
The 850~\micron\ catalog lists the $61$ peaks of the matched-filtered map with $\mathrm{S/N}\geq3.5$, selected from the 850~\micron\ map alone (Figure~\ref{fig:emittermap}, right).
Each peak carries a false-detection probability from the same peak search on the negated map \citep{Weiss2009}.
The $7$ peaks with $\mathrm{S/N}\geq5$, above the highest peak of the negated map ($\mathrm{S/N}=4.61$), are flagged as secure, and about $26$ of the $61$ are expected to be real.
Peak fluxes are deboosted with source injections into the map, weighted to the S2CLS counts \citep{Geach2017}.
For each source we search for IRAC 4.5~\micron\ sources with $\mathrm{S/N}\geq5$ within $8\farcs7$.
A source is a candidate counterpart if its chance probability $p=1-\exp(-\pi r^2 n)$ is at most $0.05$ \citep{Downes1986}, where $r$ is its separation and $n$ the local density of such sources.
Of the $61$ sources, $9$ have a candidate, and the same procedure at offset positions predicts $3.1$ ($2.9$--$3.2$) chance identifications, so the counterparts are positional candidates.
One source has a spectroscopic member of the structure among its candidates.
The source list, the candidates, and the deblended SPIRE and SCUBA-2 fluxes at the catalog positions (Section~\ref{sec:catalog_phot}) are part of the catalog.
\subsection{Structure membership} \label{sec:catalog_member}

Each source carries a class, \texttt{membership\_class}, that places it with respect to the \textit{Cosmic Himalayas} from three kinds of evidence, in order of precedence:
\begin{enumerate}
\item \textit{Spectroscopic redshift.}
A secure DESI, SDSS, or Keck/MOSFIRE redshift (Section~\ref{sec:specz}) in $2.15<z<2.22$ makes a source a spectroscopic member, and one outside that range makes it a spectroscopic interloper or star.
Single-line HETDEX redshifts do not set the class.
Sources whose spectra disagree carry a class of conflicting spectroscopy.
\item \textit{Narrowband emitter.}
A source in the \lya\ or final \ha\ samples is a candidate narrowband member.
The \oiii\ flag (Section~\ref{sec:vas_hae}) and the interloper flags, set where the broadband photometry suggests another line or a continuum galaxy, move such a source to an ambiguous narrowband class.
\item \textit{Photometric redshift.}
Eligibility requires dereddened total $i<24$, $\mathrm{S/N}\geq5$ in $i$, a local $5\sigma$ $2\arcsec$ depth of at least $24.5$ in $i$, coverage in all of $grizy$, and no near-star flag.
Galaxies that meet these cuts are eligible, and objects of the QSO class are eligible if they are also brighter than $i=21$.
An eligible object is listed as a photometric-redshift candidate if at least half of its \texttt{LePhare} likelihood, for the QSO class that of the quasar library, lies in $2.08<z<2.28$.
\end{enumerate}
The membership classes contain 38 spectroscopic members (16 quasars, among them CH-Q01 to CH-Q12 and CH-Q14, and 22 emission-line galaxies), 501 candidate narrowband members, and 216 photometric-redshift candidates.
CH-Q13, at $z=2.223$, lies just above the redshift window of the spectroscopic members, and the other three member quasars lie in the window but are not linked to CH-Q01 to CH-Q11 at the linking length of Section~\ref{sec:survey}.
All 24 confirmed \ha\ emitters lie at $2.15<z<2.22$ and make up 22 of the members.
One of them has no catalog object within $1\arcsec$, and two were observed through two slits on the same object.
On the degraded COSMOS2020 validation galaxies with $i<24$ (Section~\ref{sec:zphot_acc}), the criterion recovers $16.0\%$ ($12.4$--$19.5\%$) of the galaxies inside $2.08<z<2.28$, and $43\%$ ($35$--$51\%$) of the selected galaxies lie inside this interval.
Most contaminants are galaxies at $1.7<z<2.5$ just outside this interval.
For each eligible galaxy we also provide the fraction of its likelihood in $2.08<z<2.28$, so that users can define their own selection.

\begin{figure*}
\centering
\includegraphics[width=\textwidth]{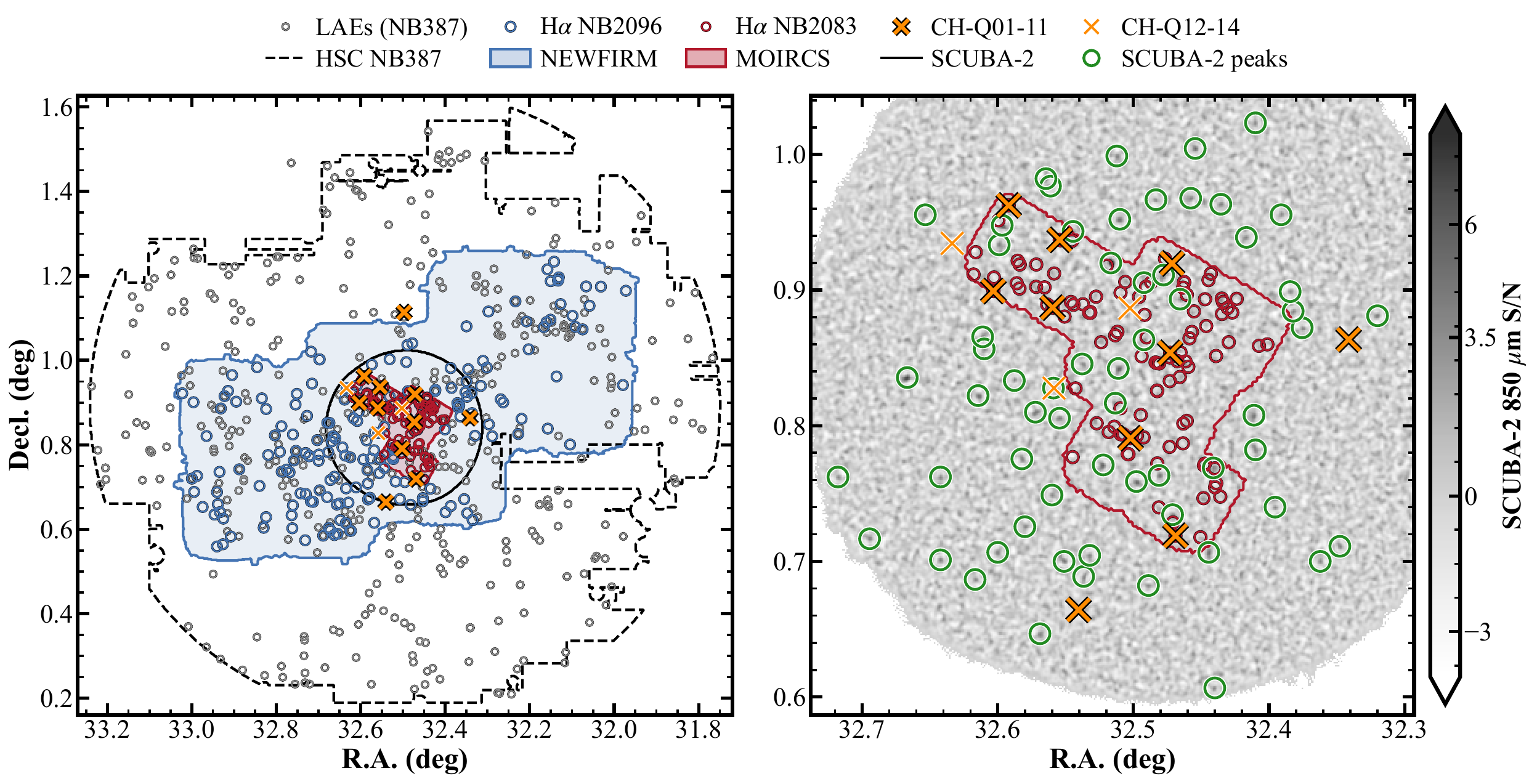}
\caption{Tracers of the \textit{Cosmic Himalayas}.
\textit{Left:} the 451 entries of the 438 LAEs (gray), the 213 NB2096 (blue) and 104 NB2083 (red) \ha\ emitters, and the quasars of the structure (orange crosses, thick with a black edge for CH-Q01 to CH-Q11 and thin for CH-Q12 to CH-Q14), with the outlines of the NB387, NEWFIRM, MOIRCS, and SCUBA-2 coverage.
\textit{Right:} the matched-filtered SCUBA-2 850~\micron\ signal-to-noise map with the 61 peaks at $\mathrm{S/N}\geq3.5$ (green circles), the NB2083 \ha\ emitters, CH-Q01 to CH-Q14, and the MOIRCS outline. The gray scale is linear in S/N, from white at $\mathrm{S/N}=-4$ to dark gray at $\mathrm{S/N}=8$.
Colors follow Figure~\ref{fig:footprints}.
R.A.\ increases to the left.}
\label{fig:emittermap}
\end{figure*}

Figure~\ref{fig:emittermap} shows the tracers of the structure with the coverage of each sample: the quasars CH-Q01 to CH-Q14 (Section~\ref{sec:survey}), the LAEs over the $1.31~\mathrm{deg}^2$ NB387 field, and the \ha\ emitters over the NEWFIRM and MOIRCS areas.
It also shows the SCUBA-2 map, which covers the quasar concentration and the MOIRCS area.

\section{Summary and Data Availability} \label{sec:summary}

\subsection{Data products and availability} \label{sec:products}

The catalog is a master catalog with one row per object and five extensions in the same row order: the photometry, the \texttt{LePhare} and \texttt{EAZY} photometric redshifts, the adopted \texttt{CIGALE} fits, and the classes with the spectroscopic redshifts, the membership classes, and the sample flags.
The redshift likelihoods of the three photometric-redshift fits are separate files.
Sample tables give the quantities that belong to a sample rather than to an object: the final \ha\ samples, the published \lya\ sample, the SCUBA-2 sources with their candidate counterparts, the quasars CH-Q01 to CH-Q14 and CH-BGQ01 to CH-BGQ34 (Section~\ref{sec:survey}), and the spectroscopic-redshift catalog of Section~\ref{sec:specz}, which lists every spectrum in the field except the Keck/MOSFIRE spectra presented in CHEF~II.
Ancillary files give the masks and detection areas, the maps of the detection limits and completeness magnitudes for random catalogs (Section~\ref{sec:catalog_compl}), the filter curves, the Galactic extinction coefficients, and an identifier crosswalk to the names and target numbers used in earlier papers, proposals, and observations.
The catalog of each detection image, the photometry of every band, method, and aperture, the fits not adopted, the PSF-homogenized mosaics, and the PSF models and matching kernels are provided on request.
Appendix~\ref{sec:app_datamodel} describes the catalog, and Table~\ref{tab:datamodel} lists the principal columns of the master catalog and of the sample tables.

For optical selections we recommend the significance cut, $\texttt{anchor\_peak\_snr}\geq4.5$ and $\texttt{anchor\_L}>0$.
Selections affected by blends should also require $\texttt{anchor\_quality\_flag}=0$, and selections affected by bright-star halos should require \texttt{anchor\_flag\_near\_star} to be false (Sections~\ref{sec:catalog_det} and~\ref{sec:catalog_flow}).
Stellar masses, star-formation rates, and dust attenuation should be taken from the fits with \texttt{sed\_reliable} (Section~\ref{sec:sed_cigale}).
Section~\ref{sec:known_issues} lists the properties of the catalog that need care in use.

The catalog products are currently available from the corresponding author upon reasonable request.
We plan a public release of the full products presented in this paper within about two years of publication.
The CHEF survey page (\url{https://cosmic-himalayas.github.io/chef/}) describes how to access the data, provides a preview of the images and the catalog, and is updated as the release proceeds.

\subsection{Summary} \label{sec:summary_points}

We present CHEF2026, the master catalog of the $\approx1~\mathrm{deg}^2$ CHEF field around the densest $z\sim2.2$ quasar overdensity in the SDSS footprint \citep{Liang2025}, imaged from the far-ultraviolet to the submillimeter.
Our main results are as follows.

\begin{enumerate}
\item Twenty-four bands cover the far-ultraviolet to 850~\micron, with three narrowbands tuned to the structure redshift (Table~\ref{tab:bands} and Figure~\ref{fig:filterset}).
The median $5\sigma$ point-source depths in $2\arcsec$ apertures are $26.3$ ($u$) to $24.2$~mag ($y$) in the optical, $22.2$--$23.5$~mag in the near-infrared, and $21.46$ and $21.54$~mag in the two IRAC bands for the fitted total fluxes (Section~\ref{sec:bandqa_depth}).
\item Six images are searched independently for sources: the optical $\chi^2$ image, MOIRCS $K_s$, NEWFIRM $K_{Xs}$, and the three narrowbands.
The $\chi^2$ catalog is $50\%$ complete at $i=26.46$, and its $285{,}537$ detections with $\texttt{peak\_snr}\geq4.5$ and $L>0$ have a false-detection fraction of $\approx0.3\%$.
Cross-matching links the six samples into $456{,}513$ objects with PSF-homogenized photometry (Section~\ref{sec:catalog}).
\item The fourteen-band photometric redshifts reach $\sigma_{\rm NMAD}=0.030$ with $2.0\%$ outliers (\texttt{LePhare}) for 942 spectroscopic galaxies at $z<1.62$.
For COSMOS2020 galaxies degraded to our depths, the outlier fraction is $19\%$ over all redshifts and $22\%$ at $2\leq z<2.5$, and it rises from $10\%$ at $i<22$ to $45\%$ at $i\geq25$.
Each \texttt{LePhare} redshift carries a reliability class based on its number of significant bands (Section~\ref{sec:zphot_acc}).
\item The \texttt{CIGALE} fits give stellar masses and star-formation rates for $382{,}799$ objects.
For the star-forming galaxies with $K_{Xs}<22.32$~AB above the mass limit, the stacked \herschel/SPIRE 250 and 350~\micron\ fluxes are 0.93--1.50 times the energy-balance predictions, and the median star-formation rates at fixed stellar mass lie within $0.20$~dex of COSMOS2025 from $z=0.5$ to 2.5.
At $2.0\leq z<2.4$, the $K_{Xs}$ mass limits are $\log(M_{\rm lim}/\msun)=10.55$ for star-forming and $10.74$ for quiescent galaxies (Sections~\ref{sec:sed_cigale} and~\ref{sec:sed_mlim}).
\item The value-added samples contain $104$ (NB2083) and $213$ (NB2096) \ha\ emitters with $\mathrm{EW_0}\geq25$~\AA, of which 21 and 16 are spectroscopic members of the structure, and $438$ of the $452$ published NB387 \lya\ emitters.
The color selections give $29{,}701$ BX, $27{,}109$ BM, $9{,}596$ LBG, $7{,}390$ star-forming and $334$ passive $gzK$ candidates, and the $61$ SCUBA-2 peaks with $\mathrm{S/N}\geq3.5$ include 7 secure sources.
Every source carries a structure-membership class (Section~\ref{sec:science}).
\end{enumerate}

ALMA observations of the molecular gas and dust continuum in the hosts of the 11 quasars in Bands~4 and~7 already give the CO(4--3) redshifts of three of them (Section~\ref{sec:specz}), presented in a paper of the series in preparation.
Approved ALMA programs also target CO(3--2) emission at $z\approx2.2$ across the structure in Band~3 mosaics.
\textit{Chandra} X-ray (340~ks) and radio continuum observations are underway.

The catalog is the photometric basis of later papers of the series and may provide target lists for wide-field spectroscopic campaigns, for example with the Subaru Prime Focus Spectrograph \citep{Takada2014} or Magellan/IMACS \citep{Dressler2011}.
Most of its objects lie outside the structure, from low redshift to $z>3$, and are measured in the same way as the members.

\section*{Acknowledgments}
This work was supported by JSPS KAKENHI Grant Numbers JP24K17084, JP25H00663, JP25H00674, and JP22K21349, and by the JSPS Core-to-Core Program, grant number JPJSCCA20210003.
YL acknowledges the International Leading Research program ``Comprehensive understanding of the formation history of structures in the Universe'' and SUPER-IRNET 1.0 and 2.0.
SC gratefully acknowledges support from the Italian Ministry of University and Research (MUR) under the FARE 2020 program CosmicNodes (grant no. R20S99FS3J, CUP H45E21000490001) and from the European Research Council (ERC) Consolidator Grant 864361 (CosmicWeb).
This work is sponsored by the Chinese Academy of Sciences (CAS), through grant Y329011V01 to the CAS South America Center for Astronomy (CASSACA). WW is supported by the CASSACA Key Research Project E52H540101 / E52H540201 / E52H540301 and the NAOC Research Program E5TG540101.
FXA acknowledges the support from the National Natural Science Foundation of China (12303016), the Natural Science Foundation of Jiangsu Province (BK20242115), and the Yunnan Revitalization Talent Support Program Young Talent Project.

This paper is based in part on data collected at the Subaru Telescope, which is operated by the National Astronomical Observatory of Japan (NAOJ), including data of the Hyper Suprime-Cam (HSC) Subaru Strategic Program \citep[PDR3;][]{Aihara2022} retrieved from the HSC data archive system.
The HSC collaboration includes the astronomical communities of Japan and Taiwan and Princeton University.
The HSC instrumentation and software were developed by NAOJ, Kavli IPMU, the University of Tokyo, KEK, ASIAA, and Princeton University, and the HSC pipeline uses software developed for the Vera C. Rubin Observatory.
We are honored and grateful for the opportunity of observing the Universe from Maunakea, which has the cultural, historical, and natural significance in Hawai{\okina}i.

This work is also based on observations at NSF Cerro Tololo Inter-American Observatory, NSF NOIRLab (Prop.~IDs 2024B-182778 and 2025B-897187; PI: Y.~Liang), which is managed by AURA under a cooperative agreement with the U.S. National Science Foundation.
It is also based on observations at the Canada-France-Hawai{\okina}i Telescope, operated by the National Research Council of Canada, CNRS/INSU, and the University of Hawai{\okina}i, at the W.~M.~Keck Observatory, and at the James Clerk Maxwell Telescope (program M23BP028), operated by the East Asian Observatory, with additional funds for SCUBA-2 from the Canada Foundation for Innovation.
We also use observations collected at the European Southern Observatory under ESO programme 179.A-2010.
This paper makes use of the following ALMA data: ADS/JAO.ALMA\#2025.1.01243.S. ALMA is a partnership of ESO (representing its member states), NSF (USA) and NINS (Japan), together with NRC (Canada), NSTC and ASIAA (Taiwan), and KASI (Republic of Korea), in cooperation with the Republic of Chile. The Joint ALMA Observatory is operated by ESO, AUI/NRAO and NAOJ.
The National Radio Astronomy Observatory and Green Bank Observatory are facilities of the U.S. National Science Foundation operated under cooperative agreement by Associated Universities, Inc.

This research has made use of the NASA/IPAC Infrared Science Archive and the MAST archive, and of data from \spitzer\ \citep[SpIES;][]{Timlin2016}, \wise\ and NEOWISE \citep[AllWISE;][]{Cutri2013}, \galex\ \citep{Martin2005}, \herschel\ \citep[HerMES and HerS;][]{Oliver2012,Viero2014}, an ESA space observatory with science instruments provided by European-led Principal Investigator consortia and with important participation from NASA, and the ESA mission \textit{Gaia} \citep[DR3;][]{GaiaDR3}, processed by the \textit{Gaia} Data Processing and Analysis Consortium.

Spectroscopic redshifts are taken from DESI \citep[DR1;][]{DESI2025}, SDSS \citep{Lyke2020,Almeida2023}, and HETDEX \citep{MentuchCooper2026}.
DESI is managed by the Lawrence Berkeley National Laboratory and supported by the U.S. Department of Energy Office of Science (Contract No. DE--AC02--05CH11231), the U.S. National Science Foundation, and the DESI Member Institutions.
The DESI collaboration is honored to be permitted to conduct scientific research on I'oligam Du'ag (Kitt Peak), a mountain with particular significance to the Tohono O'odham Nation.
Funding for SDSS has been provided by the Alfred P. Sloan Foundation, the U.S. Department of Energy Office of Science, the Heising-Simons Foundation, the U.S. National Science Foundation, and the Participating Institutions.
HETDEX is led by the University of Texas at Austin McDonald Observatory and Department of Astronomy.

Data analysis was carried out in part on the Multi-wavelength Data Analysis System and with the cooperation of the Center for Computational Astrophysics, both at NAOJ.

The visual inspection used the GPT-6.1 model of OpenAI through the Codex command-line interface (Appendix~\ref{sec:app_vi}).
The authors used Claude (Anthropic), ChatGPT (OpenAI), and Gemini (Google) to help write the analysis and figure code and to support proofreading and text checks.
All code, results, and text were checked by the authors, who take full responsibility for them.
Access to ChatGPT was provided through the OpenAI ChatGPT for Academic Researchers program.

\facilities{Subaru (HSC, MOIRCS), CFHT (MegaCam), Blanco (NEWFIRM), Spitzer (IRAC), GALEX, WISE, Herschel (SPIRE), JCMT (SCUBA-2), ALMA, ESO:VISTA (VIRCAM), UKIRT (WFCAM), Sloan, Mayall (DESI), Keck:I (MOSFIRE), HET (VIRUS), Gaia, IRSA, MAST}

\software{Astropy \citep{Astropy2022}, SExtractor \citep{Bertin1996}, SWarp \citep{Bertin2002}, MCSRED2 \citep{Tanaka2011}, SciPy \citep{Virtanen2020}, EAZY \citep{Brammer2008}, LePhare \citep{Arnouts1999,Ilbert2006}, CIGALE \citep{Boquien2019}, SMURF \citep{Chapin2013}, Codex CLI (OpenAI; visual inspection, Appendix~\ref{sec:app_vi})}

\appendix

\section{Catalog Data Model} \label{sec:app_datamodel}

\begin{sloppypar}
The master catalog has one row per cross-matched object, identified by \texttt{chef\_id} and by its coordinate designation CHEF~J\textit{hhmmss.ss}$\pm$\textit{ddmmss.s}, from its J2000 position on the \textit{Gaia} DR3 frame truncated to $0.01$~s in right ascension and $0\farcs1$ in declination.
Its five extensions share this row order, and the objects without a fit have empty entries in the photometric-redshift and SED extensions.
The position of an object is that of the first available of its detections in the order of Section~\ref{sec:catalog_assoc}.
The total fluxes are those of Section~\ref{sec:catalog_phot}, and the coverage bits and the near-star flag summarize for each object the masks of Section~\ref{sec:bandqa_mask}.
Objects detected only in the MOIRCS $K_s$ or NEWFIRM $K_{Xs}$ image are rows of the same catalog, and those that enter the fits are fitted on their forced photometry (Section~\ref{sec:zphot_acc}).
The sample tables carry the \texttt{chef\_id} of their objects, and the spectroscopic-redshift table (Section~\ref{sec:specz}) has one row per spectrum, with the identifier of the matched object where one lies within $1\arcsec$.
The rows of our Keck/MOSFIRE spectra are presented in CHEF~II.
The data model distributed with the catalog gives the unit, description, and source of every column.
\end{sloppypar}

\begin{deluxetable*}{lcp{0.56\textwidth}}
\tabletypesize{\scriptsize}
\tablecaption{Principal columns of the master catalog and the sample tables.
\label{tab:datamodel}}
\tablewidth{0pt}
\tablehead{
\colhead{Column} & \colhead{Units} & \colhead{Description}
}
\startdata
\cutinhead{PHOTOMETRY}
\texttt{chef\_id}, \texttt{iau\_name} & \nodata & Identifier, used in every extension and in the sample tables to link their objects, and designation CHEF~J\textit{hhmmss.ss}$\pm$\textit{ddmmss.s} \\
\texttt{ra}, \texttt{dec}    & deg     & Position (J2000) on the \textit{Gaia} DR3 frame \\
\texttt{anchor\_peak\_snr}, \texttt{anchor\_L} & \nodata & Significance and $L$ of the $\chi^2$ detection (Section~\ref{sec:catalog_det}) \\
\texttt{anchor\_quality\_flag}, \texttt{anchor\_flag\_near\_star} & \nodata & Quality and near-star flags of that detection (Sections~\ref{sec:catalog_det} and~\ref{sec:catalog_flow}) \\
\texttt{flag\_newfirm\_astrometry} & \nodata & NEWFIRM astrometry offset from \textit{Gaia} (Section~\ref{sec:known_issues}) \\
\texttt{f\_X}, \texttt{e\_X}, \texttt{lim\_X} & $\mu$Jy & Total flux, uncertainty, and limit flag in band X (Section~\ref{sec:catalog_phot}) \\
\cutinhead{LEPHARE}
\texttt{z\_lephare}, \texttt{z\_lephare\_p16}, \texttt{z\_lephare\_p84} & \nodata & Median and percentiles of the likelihood (Section~\ref{sec:zphot_z}) \\
\texttt{z\_lephare\_p16\_cal}, \texttt{z\_lephare\_p84\_cal} & \nodata & Calibrated $68\%$ interval (Section~\ref{sec:zphot_acc}) \\
\texttt{reliability\_lephare}, \texttt{z\_qso} & \nodata & Reliability class, and quasar-template redshift for the QSO class \\
\cutinhead{EAZY}
\texttt{z\_eazy}, \texttt{flag\_eazy\_unusable} & \nodata & Photometric redshift, and the extended galaxies where it is not to be used (Section~\ref{sec:sed_cigale}) \\
\cutinhead{CIGALE}
\texttt{z\_sed}, \texttt{sed\_fit\_z\_source}, \texttt{sed\_fit\_model} & \nodata & Redshift, its source, and the model of the adopted fit (Figure~\ref{fig:sedrule}) \\
\texttt{log\_stellar\_mass}, \texttt{log\_sfr} & dex & Stellar mass and star-formation rate (Section~\ref{sec:sed_cigale}) \\
\texttt{sed\_reliable}       & \nodata & Reliable fit (Section~\ref{sec:sed_cigale}) \\
\texttt{UV}, \texttt{VJ}, \texttt{quiescent} & mag, \nodata & Rest-frame colors and UVJ class (Section~\ref{sec:sed_uvj}) \\
\cutinhead{CLASS}
\texttt{class\_adopted}, \texttt{z\_spec} & \nodata & Star, galaxy, or QSO (Section~\ref{sec:catalog_class}), and spectroscopic redshift (Section~\ref{sec:specz}) \\
\texttt{membership\_class}   & \nodata & Structure-membership class (Section~\ref{sec:catalog_member}) \\
\texttt{is\_hae\_nb2083}, \texttt{is\_lae}, \ldots & \nodata & One flag per value-added sample of Table~\ref{tab:vas}: \ha\ in NB2083 and NB2096, \lya, BX, BM, LBG, star-forming and passive $gzK$, and SCUBA-2 counterparts (Section~\ref{sec:science}) \\
\texttt{flag\_agn}, \texttt{flag\_red\_jk} & \nodata & AGN and red $J-K$ galaxies (Section~\ref{sec:vas_ir}) \\
\cutinhead{Sample tables}
\texttt{p\_noise}            & \nodata & NB2096 \ha\ sample: probability that the emitter is pure noise (Section~\ref{sec:vas_hae}) \\
\enddata
\tablecomments{The five extensions of the master catalog have the same rows in the same order, and the data model distributed with the catalog describes every column. The sample tables are \texttt{halpha\_nb2083}, \texttt{halpha\_nb2096}, \texttt{lya\_nb387}, \texttt{scuba2\_sources}, \texttt{scuba2\_counterparts}, \texttt{chq\_quasars}, \texttt{chq\_bgq}, and \texttt{specz\_spectra}. X is one of the band keys \texttt{galex\_fuv}, \texttt{galex\_nuv}, \texttt{cfht\_u}, \texttt{hsc\_g}, \texttt{hsc\_gdeep}, \texttt{hsc\_r}, \texttt{hsc\_i}, \texttt{hsc\_z}, \texttt{hsc\_y}, \texttt{hsc\_nb387}, \texttt{moircs\_ks}, \texttt{moircs\_nb2083}, \texttt{newfirm\_jx}, \texttt{newfirm\_kxs}, \texttt{newfirm\_nb2096}, \texttt{irac\_ch1}, \texttt{irac\_ch2}, \texttt{wise\_w1} to \texttt{wise\_w4}, \texttt{mips\_24}, \texttt{spire\_250}, \texttt{spire\_350}, \texttt{spire\_500}, and \texttt{scuba2\_850}.
The deep $g$ stack (\texttt{hsc\_gdeep}) is not used in the fits, and \texttt{wise\_w1} and \texttt{wise\_w2} enter only the SED fits outside the IRAC footprint.}
\end{deluxetable*}

\section{Visual Inspection} \label{sec:app_vi}

The \ha\ candidates, the detections on bright-star rims, and random samples of the color-selected candidates were inspected with the multimodal LLM GPT-6.1 of OpenAI.
The \ha\ candidates were classified on an earlier detection run, and we reviewed these classifications.
They are carried to the catalog detections by position, and candidates we classified as artifacts are removed (Section~\ref{sec:vas_hae}).
Six NB2096 candidates are new or changed relative to the earlier detection run, and two GPT-6.1 runs accepted all six.
For the detections on the rims of $G=15$--$16$ stars, GPT-6.1 accepted $0.83$ of injected sources and $\geq0.99$ of real sources away from stars, and it rejected $0.92$ of rim fragments.
These classifications, also made on the earlier detection run, are carried as flags to the catalog detections within $0.5\arcsec$, which holds for $78\%$ of them.
No row is removed because of a model classification.

\begin{figure*}[t]
\centering
\includegraphics[width=\textwidth]{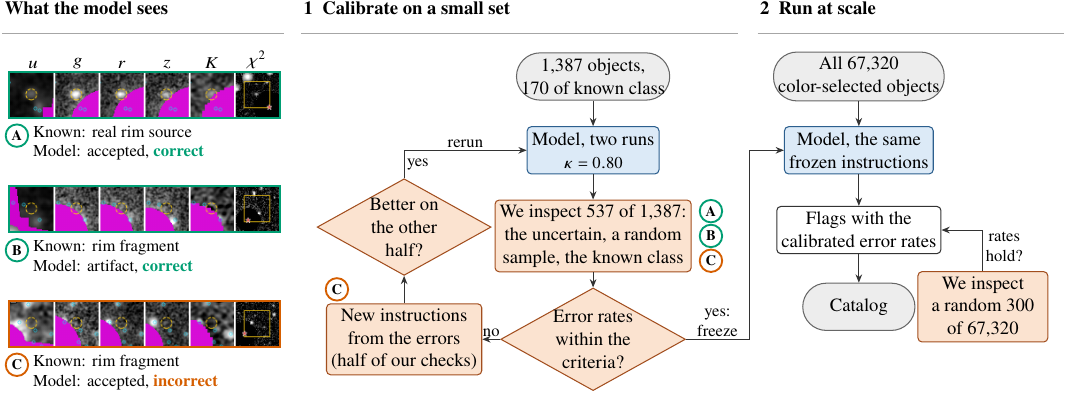}
\caption{Visual inspection of the color-selected samples.
The left part shows a real source (A) and two fragments (B, C) on bright-star rims as the model saw them.
Each panel shows the central $8\arcsec$ of its cutout, and $20\arcsec$ for the $\chi^2$ detection image.
Dashed circles mark the $2\arcsec$ apertures and the box the $12\arcsec$ cutout, and magenta marks masked pixels or pixels without data.
Frames and letters are green where the model was correct and vermilion where it was not.
The middle part shows the calibration on 1,387 objects, and the right part the run on all 67,320 color-selected objects.
Steps of the model are blue, our inspection by eye orange, the data gray, and the flags white.}
\label{fig:viflow}
\end{figure*}

\textit{Color-selected samples.}
The model saw only image cutouts in $u$, $g$, $r$, $z$, $K$, and the $\chi^2$ detection image, with the catalog photometry of each object.
Each object was classified as a real source with clean or contaminated photometry, a star, an artifact, or uncertain, and only a real source counts as accepted.
The targets were 200 random candidates from each of the BX, BM, LBG, and star-forming $gzK$ color regions, all 334 passive $gzK$ candidates (one of them also drawn as star-forming), and the 89 rim candidates without a carried classification.
The cutouts, the instructions, and the model were fixed before the first run.
The workflow calibrates the inspection on these 1,387 objects and, once its error rates meet the criteria, runs it unchanged on all 67,320 selected objects (Figure~\ref{fig:viflow}).

\textit{Reference objects.}
The targets included 170 unmarked reference objects of known class, 5 of which were also drawn candidates.
They are 45 spectroscopic galaxies at $1.4<z<3.5$, 25 objects far from bright stars, 30 rim detections with a $K_{Xs}$ counterpart, 25 spectroscopically confirmed stars, and 45 rim fragments.
Both runs correctly classified $0.82^{+0.05}_{-0.06}$ of the spectroscopic galaxies, $0.96^{+0.02}_{-0.06}$ of the objects far from stars, and $0.77^{+0.07}_{-0.09}$ of the rim detections with a $K_{Xs}$ counterpart (68\% intervals).
They classified all 25 stars as stars and $0.62\pm0.07$ of the rim fragments as artifacts, and they classified none of the 45 spectroscopic galaxies and 25 objects far from stars as an artifact.

\textit{Repeat run.}
Every object was classified twice, the second time on cutouts in a new layout.
An object is accepted only if both runs accept it.
The two runs agree on acceptance for $0.95$ of the objects, with Cohen's $\kappa=0.80\pm0.02$.

\textit{Check by eye.}
The check covers 537 of the objects, inspected visually and blind to the model's classes and to which objects are reference objects.
They are every object the runs do not both accept, 30 random accepted objects per selection, the accepted objects that a catalog class or flag marks for a check, and the reference objects.
For each selection the check gives the fraction of stars and artifacts among the accepted and among the rejected objects, and from both the contamination of the selection.
It also gives our own accuracy on the spectroscopic reference objects.

\textit{Use.}
The model classes of the color-selected samples are not used in the catalog, and these samples carry only the flags of Section~\ref{sec:vas_color}.
The workflow revises the instructions from the errors of the model on one half of the objects checked by eye and keeps the revision only if it does better on the other half.
It then applies the fixed instructions unchanged to all selected objects and checks a random 300 of them by eye to confirm the rates.
In our experience, the results of current LLMs depend less on the prompt than on the workflow that constrains and verifies them, often called a harness.
We therefore show the workflow (Figure~\ref{fig:viflow}) rather than the prompt, which is available from the corresponding author on request together with the cutouts.

\bibliography{refs}
\bibliographystyle{aasjournal}

\end{document}